\documentclass[journal]{IEEEtran}
\IEEEoverridecommandlockouts

\usepackage{amsfonts,amssymb}
\usepackage{array}
\usepackage{ifpdf}
\usepackage{cite}
\usepackage{stfloats}
\usepackage{bm}
\usepackage{verbatim}
\usepackage{color,xcolor}
\usepackage{subfigure}
\usepackage{cite}
\usepackage{subfloat}
\usepackage{graphicx}
\usepackage[cmex10]{amsmath}
\usepackage{algorithm}
\usepackage{algorithmic}
\usepackage{epsfig,epsf,color,balance,cite}
\usepackage{epstopdf}
\usepackage{makecell}
\usepackage{longtable}
\usepackage{multirow}
\usepackage{bbding}
\usepackage{lipsum} %%
\usepackage{threeparttable}
\usepackage[hyphens]{url}
\usepackage{hyperref}
\usepackage{url}

\begin{document}

\title{Intelligent Surfaces Empowered Wireless Network: Recent Advances and The Road to 6G}

\author{\IEEEauthorblockN{Qingqing Wu, Beixiong Zheng, Changsheng You, Lipeng Zhu, Kaiming Shen, Xiaodan Shao,  Weidong Mei, Boya Di, Hongliang Zhang, Ertugrul Basar, \emph{Fellow, IEEE}, Lingyang Song, \emph{Fellow, IEEE},  Marco Di Renzo, \emph{Fellow, IEEE}, Zhi-Quan Luo, \emph{Fellow, IEEE}, and Rui Zhang, \emph{Fellow, IEEE}
\thanks{Qingqing Wu is with the Department of Electronic Engineering, Shanghai Jiao Tong University Shanghai, Shanghai 200240, China (e-mail: qingqingwu@sjtu.edu.cn). Beixiong Zheng is with the School of Microelectronics, South China University of Technology, Guangzhou 511442, China (e-mail: bxzheng@scut.edu.cn). Changsheng You is with the Department of Electronic and Electrical Engineering, Southern University of Science and Technology (SUSTech), Shenzhen, China (e-mail: youcs@sustech.edu.cn). Lipeng Zhu is with the Department of Electrical and Computer Engineering, National University of Singapore, Singapore 117583 (e-mail: zhulp@nus.edu.sg). Kaiming Shen is with the School of Science and Engineering, The Chinese University of Hong Kong, Shenzhen 518172, China (e-mail: shenkaiming@cuhk.edu.cn). Xiaodan Shao is with the Institute for Digital Communications (IDC), Friedrich Alexander University Erlangen–Nuremberg, 91058 Erlangen, Germany (e-mail: xiaodan.shao@fau.de). Weidong Mei is with the National Key Laboratory of Wireless Communications, University of Electronic Science and Technology of China, Chengdu 611731, China (e-mail: wmei@uestc.edu.cn). Hongliang Zhang, Boya Di, and Lingyang Song are with School of Electronics, Peking University, Beijing 100871, China (email: hongliang.zhang@pku.edu.cn; boya.di@pku.edu.cn; lingyang.song@pku.edu.cn). Ertugrul Basar is with the Communications Research and Innovation Laboratory (CoreLab), Department of Electrical and Electronics Engineering, Koc¸ University, Sariyer, Istanbul 34450, Turkey (e-mail: ebasar@ku.edu.tr). Marco Di Renzo is with Université Paris-Saclay, CNRS, CentraleSupélec, Laboratoire des Signaux et Systèmes, 91192 Gif-sur-Yvette, France (e-mail: marco.di-renzo@universite-paris-saclay.fr). Zhi-Quan Luo is with The Chinese University of Hong Kong, Shenzhen 518172, China, also with Shenzhen Research Institute of Big Data, Shenzhen 518172, China, and also with Peng Cheng Laboratory, Shenzhen 518071, China (e-mail: luozq@cuhk.edu.cn). Rui Zhang is with School of Science and Engineering, Shenzhen Research Institute of Big Data, The Chinese University of Hong Kong, Shenzhen, Guangdong 518172, China (e-mail: rzhang@cuhk.edu.cn), and also with the Department of Electrical and Computer Engineering, National University of Singapore, Singapore 117583 (e-mail: elezhang@nus.edu.sg).}
\thanks{{\it (Corresponding author: Rui Zhang.)}}
}
}

\maketitle
%\thanks{S. Zhang is with the Department of Electronic and Information Engineering, The Hong Kong Polytechnic University, Hong Kong SAR, China (email: shuowen.zhang@polyu.edu.hk). She was with the Department of Electrical and Computer Engineering, National University of Singapore.}
%\thanks{B. Zheng, C. You, and R. Zhang are with the Department of Electrical and Computer Engineering, National University of Singapore, Singapore 117583 (e-mail:\{elezbe, eleyouc, elezhang\}@nus.edu.sg). This work is supported in part by the National University of Singapore under Research Grant R-261-518-005-720.}
%\thanks{Corresponding author: Rui Zhang. }
%

\begin{abstract}
Intelligent surfaces (ISs) have emerged as a key technology to empower a wide range of appealing applications for wireless networks, due to their low cost, high energy efficiency, flexibility of deployment and capability of constructing favorable wireless channels/radio environments. 
Moreover, the recent advent of several new IS architectures further expanded their electromagnetic functionalities from passive reflection to active amplification, simultaneous reflection and refraction, as well as holographic beamforming. 
%In recent years, ISs have received extensive attention from academics and industries, promoting ISs evolving into more advanced architectures. 
However, the research on ISs is still in rapid progress and there have been recent technological advances in ISs and their emerging applications that are worthy of a timely review.  
Thus, we provide in this paper a comprehensive survey on the recent development and advances of ISs aided wireless networks. Specifically, we start with an overview on the anticipated use cases of ISs in future wireless networks such as 6G, followed by a summary of the recent standardization activities related to ISs. 
Then, the main design issues of the commonly adopted reflection-based IS and their state-of-the-art solutions are presented in detail, including  reflection optimization, deployment, signal modulation, wireless sensing, and integrated sensing and communications. Finally, recent progress and new challenges in advanced IS architectures are discussed to inspire futrue research.
\end{abstract}

%, as a collective of intelligently controllable metasurfaces enabling manipulation of electromagnetic waves,


\section{Introduction}
\subsection{Overview of 6G}

The global mobile traffic in the next decade is envisaged to increase explosively and reach over 5000 exabytes per month in 2030 \cite{WeiJiang2021OJCS}. 
Such a demand will pose a huge challenge to the current fifth-generation (5G) wireless network in terms of energy consumption, cost, complexity, etc. 
To address this challenge and foster new applications, worldwide academic and industrial communities, together with standardization organizations, have been devoting their efforts to embrace the next/sixth-generation (6G) wireless network in the future. 
Starting from the first global 6G research program named ``6Genesis" in 2018 \cite{Finland6Genesis}, extensive preliminary projects, works, and standard proposals have been pursued to define/identity the 6G framework, usage scenarios, key technical requirements, and enabling technologies \cite{rajatheva2020white}. 
A very recent significant progress is that in June 2023, a new and complete 6G vision has been proposed  by the International Mobile Telecommunications for 2030 and beyond (IMT-2030) promotion group during the 44th meeting of the international telecommunication union radiocommunication sector working party 5D (ITU-R WP5D) \cite{6GIMT2030,RuiQiArxiv}. 
Evolving from 5G, 6G is envisioned to provide global seamless coverage, expand the utilized spectrum resources beyond the millimeter wave (mmWave) frequency bands, and establish the physical-digital world mapping to support a wide range of new applications and services intelligently, which will drive the next wave of digital economic growth, as well as sustainable far-reaching societal changes and digital equality.

%Moreover, 6G will upgrade the security and integrated sensing ability to an unprecedented level.

\begin{figure}[t]
  \centering
  \includegraphics[width=0.75\linewidth]{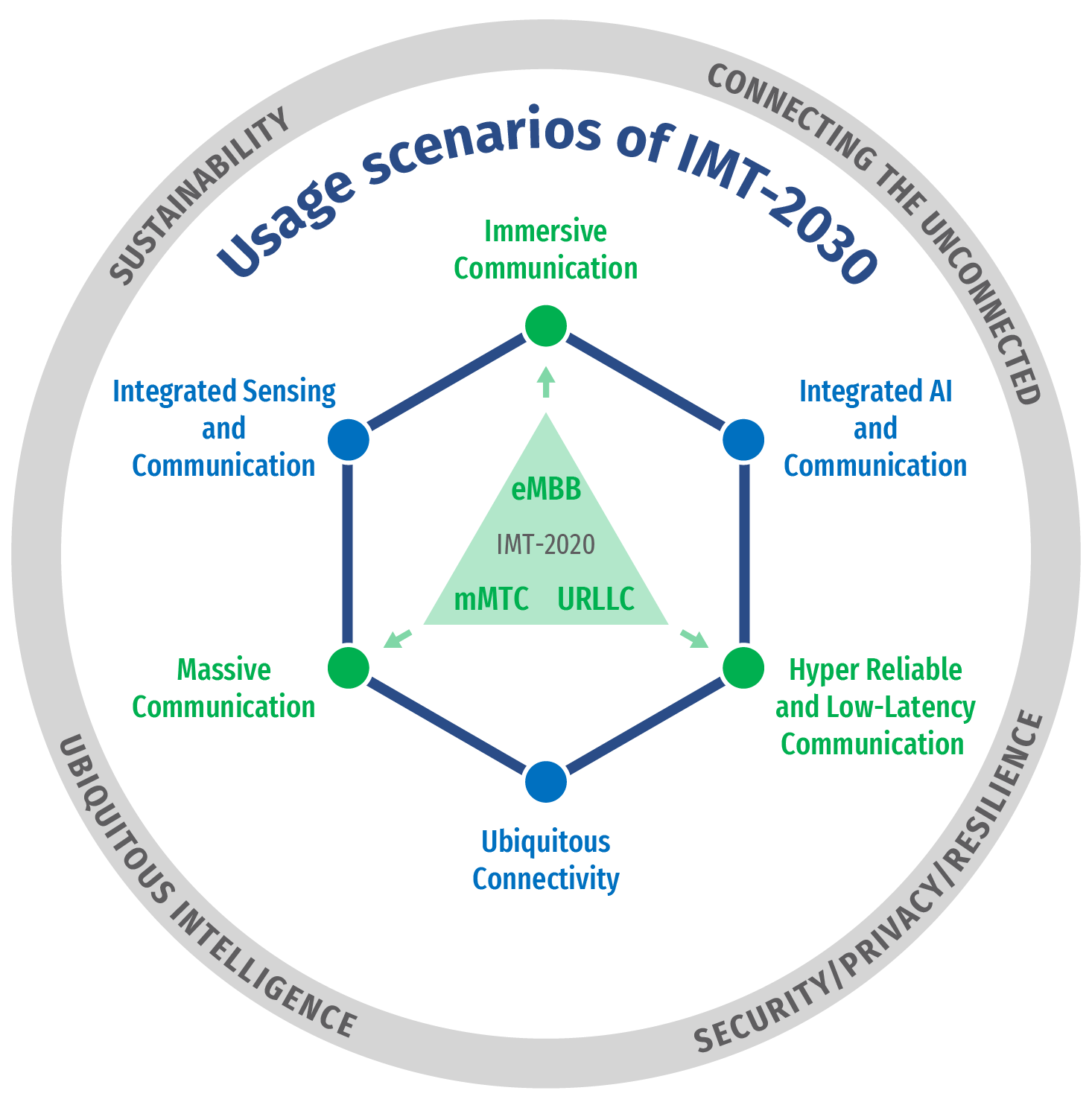}
  \caption{Illustration of 6G usage scenarios of IMT-2030 \cite{6GIMT2030}.} \label{6Gusagecases}
\end{figure}

To specifically characterize 6G visions, six usage scenarios with required network capabilities are suggested by IMT-2030, as illustrated in Fig. \ref{6Gusagecases} \cite{6GIMT2030}. Particularly, the three 5G scenarios, i.e., enhanced mobile broadband (eMBB), massive machine type communications (mMTC), and ultra-reliable and low-latency communications (uRLLC) have evolved to immersive communication, massive communication, and hyper-reliable and low-latency communication in 6G, respectively. Specifically, immersive communication covers application scenarios offering an interactive and immersive experience for users, such as the extended reality and holographic communications, which generally require 10 Gbps user experienced data rate and 1 Tbps peak data rate \cite{6GRoadCOMST,6GIMT2030}. Massive communication aims at supporting massive number of connections to facilitate Internet-of-things (IoT) applications like smart cities, smart home, health and environment monitoring, which necessitates the requirement of 30-50 Mbps/$\text{m}^{\text{2}}$ area traffic capacity and 10$^{\text{6}}$-10$^{\text{8}}$ devices/$\text{km}^{\text{2}}$ connection density \cite{6GIMT2030}. Hyper-reliable and low-latency communication scenarios include smart factory, smart health, and smart transportation to support intelligent industrial control, remote precise surgery, and fully autonomous driving, which may require 0.1 ms latency and 99.99999$\text{\%}$ reliability \cite{WeiJiang2021OJCS,6GRoadCOMST}.

Furthermore, three new usage scenarios are identified in 6G, namely integrated sensing and communication (ISAC), integrated artificial intelligence (AI) and communication, and ubiquitous connectivity. Specifically, ISAC enables services requiring sensing capabilities such as high-precision localization and tracking, radio imaging and map generation, and activity pattern recognition, which rely on centimeter-level positioning accuracy and millimeter-level sensing/imaging resolution. Integrated AI and communication will support numerous intelligent applications, including human-machine interaction, digital twin-based prediction and decision-making, and heavy computation tasks, which call for AI-oriented communication and computing support. Ubiquitous connectivity targets at filling in the blanks like scarcely covered areas (such as sea,
desert, forest, etc.) and complicated indoor environments by maintaining consistent user experience over them.
%while providing IoT and broadband communication. 
In summary, 6G visions are expected to be fulfilled by the above six usage scenarios which call for unprecedentedly high network performance as compared to 5G.

\subsection{Overview of Intelligent Surfaces (ISs)}

\subsubsection{Historical Development of ISs}
\begin{figure}[t]
  \centering
  \includegraphics[width=\linewidth]{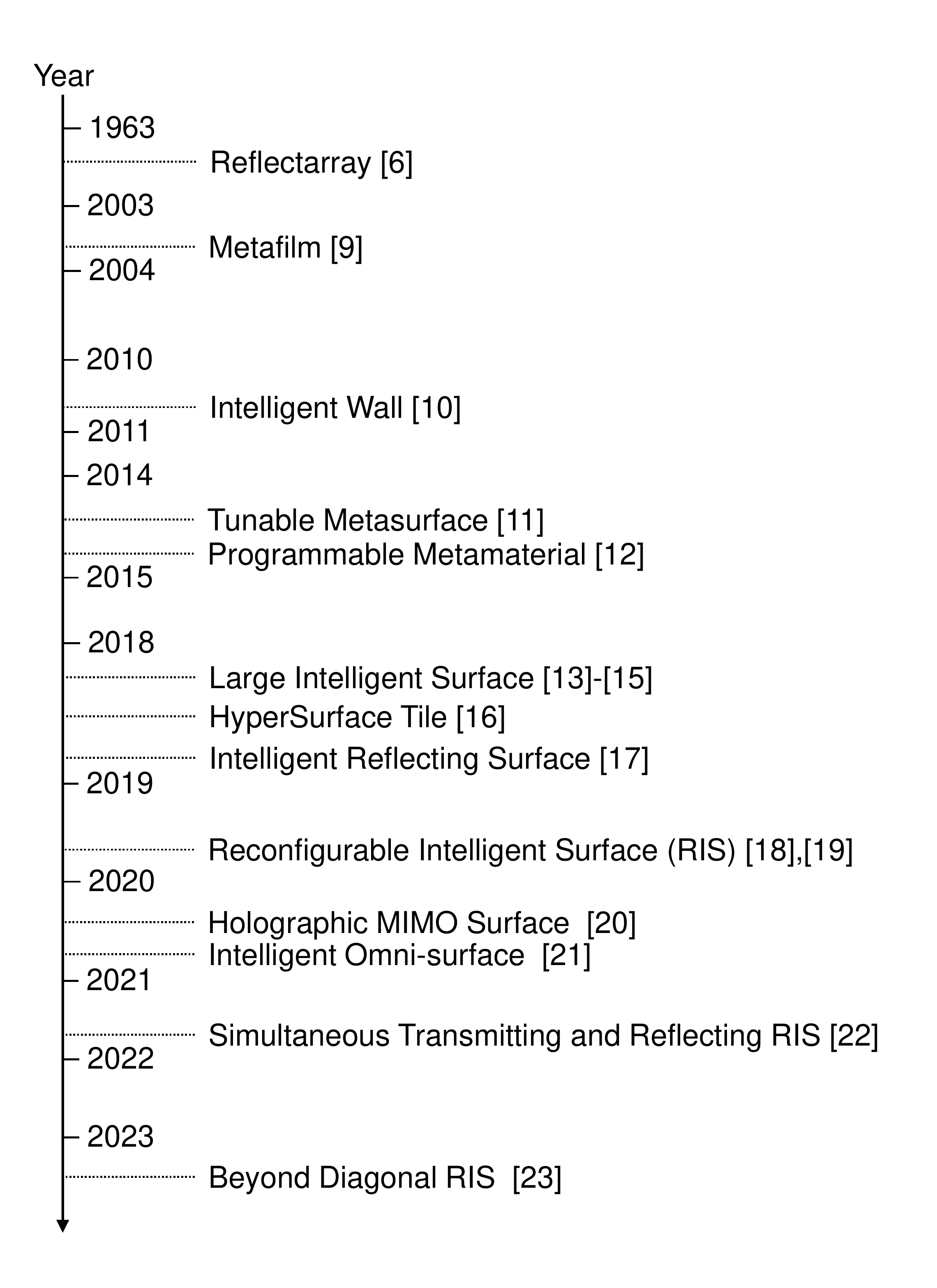}
  \caption{Historical development of ISs.} \label{IS-Timeline}
\end{figure}

As shown in Fig. \ref{IS-Timeline}, ISs can be traced back to the early study on a special type of antenna array called reflectarray \cite{Reflect-array}.
Reflectarray operates with an active source radiating radio-frequency (RF) signals towards a nearby reflective surface formed by arrays of passive antennas to achieve desired radiation patterns, which was aimed at reducing the hardware cost of conventional active antenna arrays. 
Subsequently, the research on investigating the more sophisticated interaction between electromagnetic (EM) waves and metamaterials was also pursued (e.g., \cite{VeselageMeta,DavidRSmith}). 
In particular, the transition conditions for the average EM fields across a surface with a type of metamaterial called metafilm were studied in 2003, showing that the EM wave can be controlled by electrically changing the properties of a metafilm \cite{Metafilm2003}. 
This result inspired the new idea of deploying metamaterial-made IS in the environment to alter wireless signal propagation, which thus opened up a new direction for IS beyond the traditional reflectarray.
Subsequently, an intelligent wall equipped with a certain type of metamaterial called active frequency selective surface was proposed to manipulate the propagation environment inside a building \cite{IntelligentWall2010}. 
Afterwards, the controllability of metamaterials over EM waves underwent further development. In 2014, tunable metasurfaces were developed to construct a spatial microwave binary phase modulator \cite{TunableMS2014}. 
Meanwhile, programmable metamaterials were proposed to manipulate EM waves by digitally coding the metamaterials with a field-programmable gate array (FPGA) \cite{Programmable2014}, which greatly extended the flexibility of controlling the EM wave. 

The aforementioned development of programmable metamaterials/surfaces promoted the formation of the new IS concept and its applications in wireless communications. 
In 2018,  the large intelligent surface (LIS) was proposed for enhancing wireless sensing and communication performance in \cite{ShaHu2018LISPositioning,ShaHu2018LISData,HCW2018LIS}. 
As an extension of the conventional massive multiple-input-multiple-output (MIMO) technology, LIS, like massive MIMO, is also based on the active array architecture, which incurs higher hardware cost and power consumption with increasingly more antenna elements employed. 
In the same year, a software-controlled programmable metamaterial architecture named HyperSurface Tile was proposed in \cite{Akyildiz2018HyperS} for proactively controlling the radio propagation environment to enhance the wireless communication performance. 
In addition, a similar concept called intelligent reflecting surface (IRS) was proposed independently in \cite{qqw2018IRS}, which, for the first time, investigated the joint active and passive beamforming design problem in an IRS-aided wireless system for maximizing the communication spectral efficiency (SE).  
In particular, a quadratic power scaling law with the increasing number of IRS elements was revealed in \cite{qqw2018IRS}, which unveiled the more cost-effective performance gain of IRS passive beamforming as compared to the conventional active beamforming (via e.g., massive MIMO and LIS) with only a linear power gain over the number of antenna elements. 
IS structures that are not limited to reflection, i.e., reconfigurable intelligent surface (RIS) were discussed in \cite{CWH2019RIS} and \cite{renzo2019-smart}. Also, holographic MIMO surface \cite{huang2020-holographic} was proposed to encompass relay-type and transmitter-type, as well as discrete and continuous implementations. 
To extend the signal coverage of reflection-based IRS, the intelligent omni-surface (IOS) \cite{SHZ2020IOS} and the Simultaneous Transmitting and Reflecting RIS (STAR-RIS) \cite{JQX2021STARRIS} were subsequently proposed to enable both signal reflection and refraction of ISs. 
A more recent IS development was the general Beyond Diagonal RIS (BD-RIS) architecture \cite{HSB-Architectures-2023}, which includes the above-mentioned IS architectures as special cases by enabling more sophisticated signal processing across IS elements.  Another architecture based on multi-layer ISs is refereed to as stacked intelligent metasurface \cite{an2023stacked}. 
ISs have put forth the emerging concept of smart radio environment that can provide uninterrupted wireless connectivity, in an energy sustainable manner by recycling existing radio waves \cite{MarcoX,MarcoY,basar2023reconfigurable}. 

\subsubsection{Classifications of ISs}
As a collective term for programmable metasurfaces with various functionalities, ISs can be categorized from the following perspectives.\footnote{We will use IS/ISs consistently in the sequel of this paper unless specified otherwise.}
\begin{itemize}
  \item \textbf{\emph{Continuous/Discrete ISs:}} Discrete ISs refer to ISs composed of discrete elements with certain spacing between adjacent elements, while continuous ISs are formed by a continuous-aperture surface.
  \item \textbf{\emph{Reflection-/Refraction-based ISs:}} The signals impinging on an IS can be either reflected or refracted completely (i.e., IRS and intelligent refracting surface, respectively) or partially reflected and refracted at the same time to achieve a full-space signal coverage. 
  \item \textbf{\emph{Active/Passive ISs:}} The passive IS only reflects/refracts the incident signal with controllable phase shift and/or amplitude without amplification. 
  In contrast, the active IS is able to amplify the impinging signal while manipulating its phase shift at the same time.\footnote{In the rest of this paper, the term passive refers to ISs with no power amplification capabilities.}
\end{itemize}

\begin{figure*}[!t]
  \centering
 \includegraphics[width=1.0\textwidth]{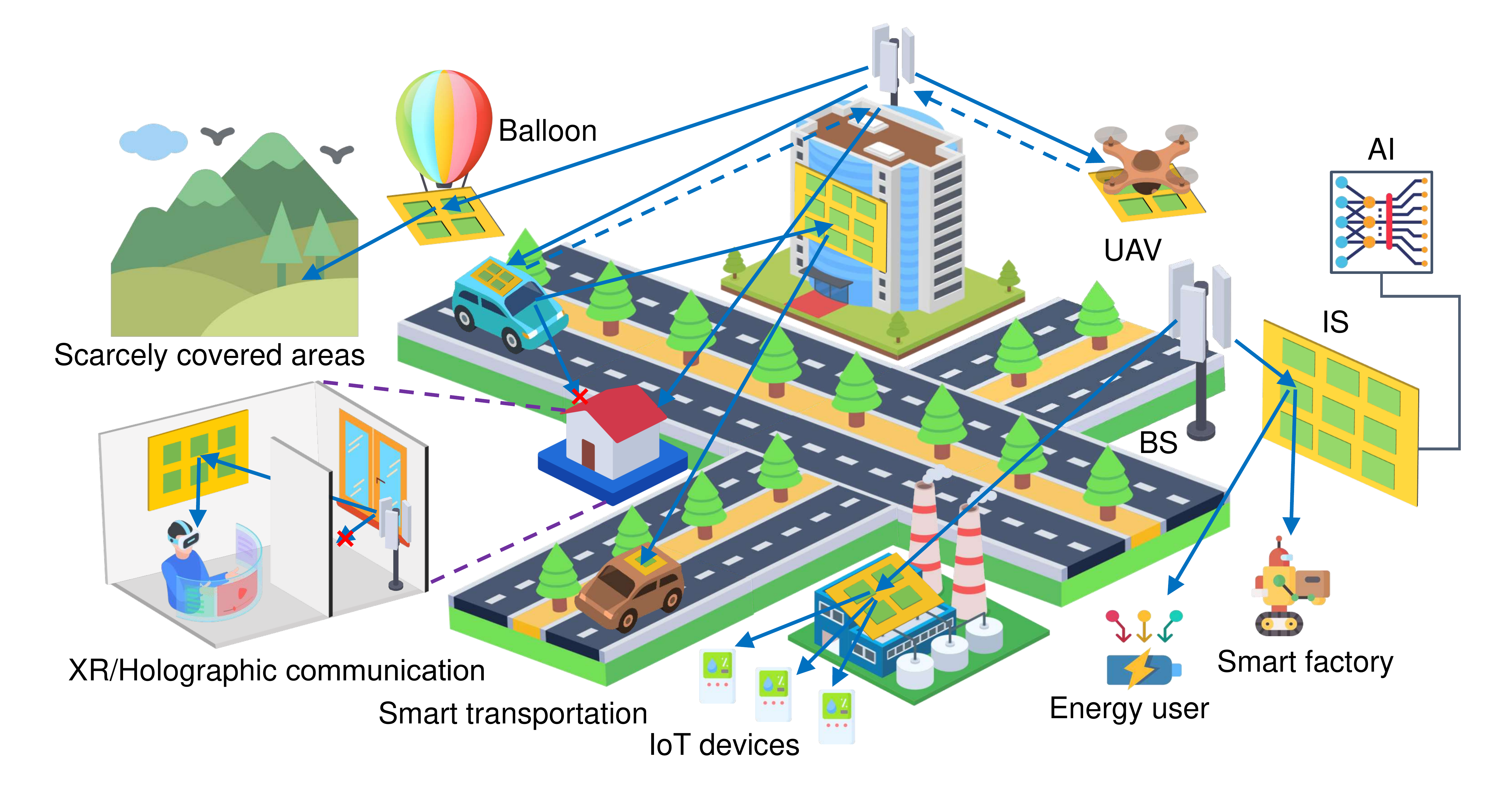} 
  \caption{Illustration of IS applications in 6G.} \label{ISapp6G}
\end{figure*}

\subsubsection{Applications of ISs in 6G}

As shown in Fig. \ref{ISapp6G}, ISs are envisioned to facilitate and enable a wide range of applications for future 6G usage scenarios. 
For example, to enable immersive communication applications, such as extended reality (XR) and ultra-high definition (UHD) video services usually supported by high-frequency wideband signal transmission, ISs can help overcome the severe path loss and bypass obstacles by providing high beamforming gains and additional propagation paths, respectively \cite{chen2023intelligent}. 
Additionally, to facilitate hyper-reliable and low-latency communication applications, such as industrial control, telemedicine and emergency services, ISs can be utilized to improve the channel quality and transmission reliability by creating the dominant line-of-sight (LoS) path to mitigate the undesired multipath fading.
Particularly, for services with stringent requirements on reliability and latency, especially in high-mobility scenarios such as fully autonomous driving, ISs can be either deployed at the roadside to enhance the signal strength for the vehicles (or passengers inside vehicles) on the road, or mounted on the vehicles to mitigate the fast channel fading arising from their high mobility \cite{huang2022-transforming}.
Moreover, to support massive communications in e.g., smart cities, environmental monitoring and other IoT applications, ISs can be used to improve the system throughput and fairness for densely deployed devices by jointly designing reflections/refractions and their non-orthogonal multiple access (NOMA) scheme \cite{RIS-NOMA,Zheng2020NOMA}. 

In addition to the aforementioned three scenarios, ISs can  effectively facilitate applications in the newly introduced scenarios for 6G. 
Under the integrated AI and communication scenario, especially the edge AI scenario where edge servers aggregate data from mobile devices over wireless links to perform AI tasks such as federated learning, ISs can be utilized to provide favorable wireless channel environment for real-time data aggregation and exchange, distributed computing, remote monitoring 
 and controlling, etc., thereby improving AI performance and scalability \cite{YapengZhao2023TCOM,YapengZhao2023JSTSP}. 
On the other hand, AI algorithms can also be promisingly applied to solve non-convex, highly nonlinear and large-scale design and optimization problems in IS-aided wireless networks, such as passive beamforming, channel estimation, optimal deployment and resource allocation \cite{AIRIS-china2021}. 
Besides, the use of massive and  flexibly deployed ISs in proper locations is appealing to achieve the ubiquitous connectivity even in a complex propagation environment with rich scatterers/obstacles. 
In addition, ISs can be deployed on maritime or aerial mobile platforms (e.g., ships, unmanned aerial vehicles (UAVs), balloons, airships, satellites, and so on) to provide wireless coverage for scarcely covered areas \cite{qingqing2021JSAC-UAV,You2021WC-IRS-UAV,Toka2024arxiv}. 
Furthermore, the flexibility of ISs to reflect or refract wireless signals enables coverage for dead zones in conventional terrestrial networks \cite{SHBYMZHL-Beamforming-2022}.

Apart from communication applications, ISs can also be employed to improve both the sensing and communication performance in ISAC applications. 
For instance, to enable ISAC services such as navigation, high-precision positioning, user activity detection and tracking, ISs can be flexibly deployed on the surface of walls, buildings, and vehicles as additional reference nodes to enhance the positioning accuracy of the future dual-functional radar and communication (DFRC) BS \cite{RISTA-Whitepaper,KaitaoISAC-MutualAssist}, as well as mounted on targets like automobiles or UAVs to enlarge their radar cross section (RCS) for high-resolution detection and highly accurate real-time tracking \cite{huang2022-transforming,KaitaoISAC-MutualAssist}. 
In addition to the above, ISs can assist in many other applications, such as enhancing the physical-layer security \cite{xrgPLSWCL2020,9524501,9606864,9496108} and boosting both the communication rate and the energy transmission efficiency for simultaneous wireless information and power transfer (SWIPT) as well as wireless powered communications \cite{qqwSWIPTPIEEE2022,9531372,9982493,9896755,10384328,9849458,9647934}. 

\subsubsection{Prototypes/Field Trials of ISs}
{The versatile promising applications of ISs in wireless networks have spurred extensive worldwide endeavors in experimentation, measurement, and prototyping. Some representative results are summarized in Table \ref{Prototypes}. The usefulness of ISs has been demonstrated in  wireless systems operating at frequencies ranging from sub-6 GHz to mmWave, Terahertz (THz), and even optical frequency bands \cite{IS-OF-optical-1,IS-OF-optical-2,fang2023reconfigurable,IS-OF-optical-3}.
}

\begin{table*}[!t]
        \centering
        {
	\caption{List of Representative IS Prototypes/Field Trials}\label{Prototypes}
    \renewcommand\arraystretch{1.8}
	\begin{tabular}[!t]{|m{0.6cm}<\centering|m{16.4cm}|}      
		\hline
        \bf Year & \bf Description of IS prototypes/field trials \\
        \hline 
        2018 & A meta-structure-based reflectarray facilitating 28GHz-band 5G systems was developed by NTT DOCOMO and Metawave \cite{IS-OF-mmWave-1}. \\
        \hline
        2019 & A programmable metasurface-based 8-phase shift-keying transmitter operating at the 4.25 GHz band was presented \cite{Tang_2019}. \\
        \hline
        2020 & A surface with over 3200 elements operating at the 2.4 GHz band was developed, improving signal strength \cite{Arun_2020_RFocus}. \\
        \hline
        2020 & A 36-element array of inexpensive antennas operating at the 2.4 GHz band was developed, improving channel capacity \cite{LZQ-2020-MCC-PRE}. \\
        \hline
        2020 & A high-gain yet low-cost RIS was developed with 256 elements and 2-bit phase shifting operating at the 2.3 GHz or 28.5 GHz band \cite{DLL-2020-AC-Exp}. \\
        \hline
        2020 & A smart surface prototype was developed with measurements showing throughput improvement over commercial MIMO systems \cite{Dunna-2020-ScatterMIMO}. \\
        \hline
        2021 & An RIS path loss model was verified by experimental measurements conducted at the 4.25 GHz and the 10.5 GHz band \cite{TWK-2021-TWC}. \\
        \hline
        2021 & An RIS prototype with 1100 elements working at the 5.8 GHz band was developed, achieving high performance in various scenarios \cite{1bit_Emil_TC2021}. \\
        \hline
        2022 & An IRS prototype operating at the 2.6 GHz band was developed by Huawei and its collaborators, showing superiority to 5G systems \cite{IS-OF-Huawei}. \\
        \hline
        2022 & A prototype of IOS-based wireless communications was developed with the functionality of IOS verified by experiments \cite{HSBYMMZHL-Implementation-2022}. \\
        \hline
        2022 & An RIS prototype with the ability to continuously control the phase shifts was developed with its properties characterized \cite{Fara-2022-MWC-Continue}. \\
        \hline
        2022 & An RF-switch-based RIS prototype operating at the 5.3 GHz band was developed with dataset characterization \cite{Rossanese-2022-WINTECH}. \\
        \hline
        2022 & RIS prototypes covering sub-6 GHz to 28 GHz bands were developed with field tests conducted by ZTE \cite{IS-OF-ZTE}. \\
        \hline
        2023 & Field trials on RIS operating at the sub-6 GHz band were conducted, achieving 40 dB indoor received power improvement \cite{JR-2023-TAP}. \\
        \hline
        2023 & An electronically almost-$360^{\circ}$ steerable metamaterial surface that operates above the 24 GHz band was developed \cite{CWK-2023-NSDI-mmWall}. \\
        \hline
        2023 & A multi-RIS prototype was developed to facilitate Wi-Fi and commercial 5G networks operating at the 3.4 GHz and the 5.8 GHz band. \cite{xiong2023multirisaided}. \\
        \hline
        2023 & A reconfigurable metasurface operating at the 0.34 THz band was developed, achieving $1^{\circ}$ angular precision of beam scanning \cite{FL-2023-LSA}. \\
        \hline
        2023 & An RIS prototype operating at the 5G FR2 frequency band was developed by Greenerwave and verified by Rohde \& Schwarz \cite{IS-OF-Greenerwave}. \\
        \hline 
        2023 & Field trials on  throughput enhancement with RIS were conducted in a 5G mmWave commercial network at the  frequency of 27 GHz \cite{YH-2023-MCOM}. \\
        \hline 
        2024 & Experiments on using RIS to enhance wireless coverage were conducted in a 5G commercial network operating at the 2.6 GHz band \cite{JS-2024-MWC}. \\
        \hline 
		\end{tabular}
        }
\end{table*}

\subsection{Standardization Activities}

As an innovative radio access technology, IS has already been included in the ITU-R IMT-2030 report on future technology trends, as a candidate technology for enhancing the radio interface of future telecommunication standards \cite{ITU-R,9679804}. 
During the last two years, IS has been under evaluation for its inclusion in future telecommunication standards, and pre-standarization and standardization bodies are currently working on it. 
In this subsection, we briefly overview the most relevant standarization activities on IS. Also, we put them in context with other standarization activities focusing on technologies that have strong ties with IS, notably THz communications and ISAC.

\subsubsection{Third Generation Partnership Program (3GPP)} The mission of 3GPP is the creation of the mobile broadband standard, with an increasing emphasis towards connecting the internet of things -- whether the need is for uRLLC at one end of the scale or for energy efficiently low-cost, low-power devices at the other \cite{3GPP}. 3GPP unites seven telecommunications standard development organizations (ARIB, ATIS, CCSA, ETSI, TSDSI, TTA, TTC), known as ``Organizational Partners'' providing their members with a stable environment to produce the reports and specifications that define technologies. 3GPP specifications cover cellular telecommunications technologies, including radio access, core network and service capabilities, which provide a complete system description for mobile telecommunications. The 3GPP specifications also provide hooks for non-radio access to the core network, and for interworking with non-3GPP networks.

As far as IS is concerned, several proposals have been made during the 3GPP release 18 (Rel-18) radio access network (RAN), including:
\begin{itemize}
\item New radio (NR) repeaters and IS from KDDI Corporation (RWS-210300) -- Evaluation of the performance benefits of ISs assuming compatibility with legacy user equipments (Rel-15/16/17).
\item NR smart repeaters from China Mobile Communications Corporation (CMCC) (RWS-210339) -- Study candidate side control information such as timing, beam, and bandwidth information for smart repeaters (RAN1, RAN4); and specification of RF requirements considering side control information (RAN4).
\item Support of IS for 5G-Advanced from Zhongxing Telecommunication Equipment (ZTE) Corporation, Sanechips (RWS-210465) -- Technical enhancements and specification support for IS, including beam management, channel state information enhancement, control interface between ISs and BSs, interference coordination.
\item Introducing IS for 5G-Advanced from Sony Europe (RWS-210306) -- Study phase in Rel-18 to investigate IS entrance impact on specifications including initial access, beam management, and channel models.
\item Motivation of IS requirements in Rel-18 from China Unicom (RWS-210390) -- Study of ISs including deployment scenarios and channel modeling, potential impact on beam management and interference coordination.
\item NR smart repeaters for Rel-18 from Qualcomm (RWS-210019) -- Work in Rel-18 on smart repeaters, including the study of side control information (RAN1), protocol support (RAN2), and RF and electromagnetic compatibility requirements (RAN4).
\item Smart repeaters enhancements from MediaTek Inc. (RWS-210099) -- Smart repeater enhancements including support for beamforming (RAN1), interference management (RAN1), operation and integration (RAN2), necessary backward compatibility with legacy user equipments.
\item IS from Rakuten Mobile Inc. (RWS-210247) -- Study item on ISs for Rel-18 considering channel modeling, use cases and deployment scenarios.
\end{itemize}

As far as 3GPP is concerned, Rel-18 will include, on the other hand, a study item (RP-213700 \cite{3GPP-NCR-RP}, TR-38.867 \cite{3GPP-NCR-TR}) on network controlled repeaters (NCRs). 
These repeaters are intended to address the limitations of conventional RF repeaters as well as to provide enhanced low-cost coverage.  
Conventional amplify-and-forward repeaters are considered which operate with minimal processing and do not require signal decoding and reconstruction. 
Therefore, they operate at a lower complexity and latency than decode-and-forward repeaters. 
The price to pay for the reduction in complexity lies in a reduced link quality and increased network interference, which is due to the interference/noise amplification. 
When applied to high frequency bands (above 24 GHz), amplify-and-forward repeaters provide limited coverage. 
In fact, static directional repeaters cannot adapt to the channel conditions and user mobility, which is necessary in communication systems where directional communications and high beamforming gains are necessary. 
NCRs will be designed to include an in-band control interface to configure and manage their behavior, with the objective of reducing the amplification of noise and supporting communications with a better spatial directivity. 
A detailed description of NCRs in the context of 3GPP standardization can be found in \cite{NCR}. NCRs are often considered a stepping stone for ISs, since the control channel may be reused. Recent research works have also compared NCRs and ISs, as summarized in \cite{NCR}.

\subsubsection{European Telecommunication Standards Institute (ETSI)} ETSI is a leading organization for the development of standards on information and communication technologies, fulfilling European and  global market needs \cite{ETSI}. ETSI has over 900 members from 64 countries over 5 continents. ETSI produces specifications, standards, reports and guides, enabling technologies in a multi-vendor, multi-network, multi-service environment. Standards are created in an open approach, thanks to the direct participation of members and contribution-driven consensus-based working procedures. The standardization work is carried out in different technical groups, which include industry specification groups (ISG). ISGs are the perfect tool for developing ``early'' standardization activities, resulting from fundamental and applied research projects, targeting for the publication of deliverables where technological aspects are streamlined for consideration by standards organizations. 

In September 2021, ETSI launched ISG-IS, with the objective of reviewing and establishing global
standardization activities for the IS technology, which focused on the following aspects:
\begin{itemize}
\item Definition of use cases, key performance indicators, and deployment and operational scenarios; 
\item Radio-frequency aspects, including surface models, channel characterization, radiation characterization, 
and radiation exposure limits;
\item IS-aided air-interface technologies, mechanics, and requirements;
\item System and network level control signaling aspects;
\item System and network architecture considerations; 
\item Baseline evaluation methodology and performance analysis (link-level and system-level);
\item IS microelectronics, enabling technologies, and proof-of-concepts (prototyping);
\item IS verification and validation.
\end{itemize}

In the first phase from 2021 to 2023, the activities of ISG-IS were focused on three main work items (WIs) \cite{ISG-RIS-WI}, including 
\begin{itemize}
\item \textit{WI-1: Use cases, deployment scenarios and requirements} -- WI-1 focused on identifying relevant use cases for IS, with the corresponding general key performance indicators, deployment scenarios, and operational requirements for each identified use case. This includes system/link performance, spectrum, co-existence, and security. 
\item \textit{WI-2: Technological challenges and impact on architecture and standards} -- WI-2 focused on  technological challenges to deploy IS as a new network entity, the internal architecture, framework and required interfaces for IS, and the potential recommendations and specifications to standardization groups for supporting IS as a network entity.
\item \textit{WI-3: Communication models, channel models, and evaluation methodology} -- WI-3 focused on communication models striking a suitable trade-off between electromagnetic accuracy and simplicity for performance evaluation and optimization at different frequency bands; channel models (deterministic and statistical) that include path-loss and multipath propagation effects, as well  as the impact of interference for application to different frequency bands; channel estimation, including reference scenarios, estimation methods, and system designs; and key performance indicators and the methodology for evaluating the performance of IS for application to 
wireless communications, including the coexistence between different network operators, and for fairly 
comparing different transmission techniques, communication protocols, and network deployments.
\end{itemize}

The second phase (initial specifications) of ISG-IS kicked off in September 2023. The group is currently working on the following three WIs \cite{ISG-RIS-WI-NEW}:
\begin{itemize}
\item \textit{WI-4: Implementation and practical considerations} -- The scope of WI-4 is to investigate relevant implementation and practical considerations for IS in a wide range of frequency bands and deployment scenarios and provide possible solutions and prototyping results. 
\item \textit{WI-5: IS-aided air-interface technologies and mechanics} -- The scope of WI-5 is to identify use cases, deployment scenarios, and transmission and reception schemes for diversity and multiplexing in IS-aided channels.
\item \textit{WI-6: Multi-functional IS -- Modeling, optimization, and operation} -- The scope of WI-6 is to identify technological challenges and solutions for multi-functional IS, incorporating transmission, reflection, sensing, computation, and other potential functions, as well as corresponding appropriate deployment scenarios and resource allocation schemes.
\end{itemize}

The activities of ISG-IS during the second phase will be tailored to providing an initial specification framework for the technology.
A major objective of ISG-IS is to establish collaborations with other relevant ISGs, which are focused on emerging technologies that have strong ties with IS. The most relevant and related ISGs are ISG-THz on THz communications \cite{ISG-THZ} and ISG-ISC \cite{ISG-ISC} on ISAC. Specifically, ISG-THz concentrates on channel sounding and modeling for IS-aided channels and ISG-ISC concentrates on how IS can be exploited for improving the performance of ISAC, as well as how ISAC can be exploited for improving the operation and deployment of IS.
In summary, standardization activities for IS and its applications in THz communications and ISAC are under a vivid debate by standards organizations \cite{ISG-RIS}, \cite{ISG-THZ}, \cite{ISG-ISC}.

\begin{figure}[!t]
	\centering
	\includegraphics[width=0.5\textwidth]{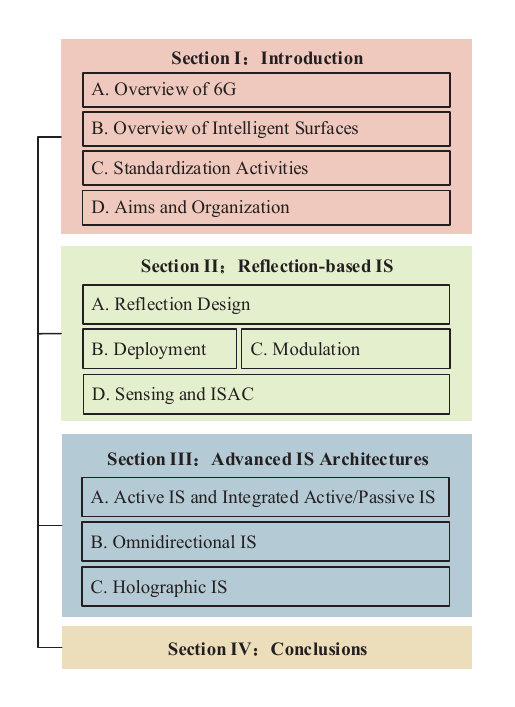}
	\caption{Organization of this paper.}
	\label{org}
\end{figure}

\begin{table*}[!t]
	\centering
	\caption{List of Representative Overview/Survey/Tutorial Papers on ISs}\label{contributions}
    \renewcommand\arraystretch{1.5}
	\begin{tabular}[!t]{|m{1.5cm}<{\centering}|m{1.5cm}<{\centering}|m{9.8cm}|m{0.8cm}<{\centering}|m{1.8cm}<{\centering}|}      
		\hline
		%   Title
		\multicolumn{2}{|m{3cm}|}{\centering \bf Contents / Topics } & \multicolumn{1}{c|}{\bf Major Contributions} & {\bf Ref.} &{\bf Publication Year} \\ \hline		
		%   Channel Modeling & Characterization 
		\multicolumn{2}{|m{3cm}|}{\centering Channel Modeling \& Characterization} & {Offer an overview of RIS-based channel measurements/experiments, large-scale path loss models and small-scale multipath fading models. } & {\cite{huang2022-reconfigurable}} & {2022}  \\ \hline		
      \multicolumn{2}{|m{3cm}|}{\centering Communication Models} & {Offer a comprehensive tutorial on electromagnetically consistent communication models for IS, including discrete-type and continuous-type IS. } & {\cite{m1}} & {2022}  \\ \hline
      \multicolumn{2}{|m{3cm}|}{\centering Fundamentals of Metamaterial-based IS} & {Offer a comprehensive tutorial on the theory behind metamaterial-based RIS, including near-field channel models. } & {\cite{m2}} & {2020}  \\ \hline
		%   System Design & Challenges
		\multicolumn{2}{|m{3cm}|}{\multirow{4}{3cm}[-0.6cm]{\centering {System Design \& \\ Challenges}}} & {Discuss the up-to-date solutions and practical challenges of the IRS focusing on the channel estimation and the passive beamforming design.} & {\cite{zheng2022-survey}} & {2022}  \\ \cline{3-5}
		\multicolumn{2}{|m{3cm}|}{} & {Provide an overview of RIS and explore the theoretical performance limits of RIS-aided wireless communications} & {\cite{BasarAC2019}} & {2019}  \\ \cline{3-5}
		\multicolumn{2}{|m{3cm}|}{} & {Provide a comprehensive overview of signal processing techniques on RIS/IRS-aided wireless systems.} & {\cite{CunhuaPan2022JSTSP}} & {2022}  \\
        \cline{3-5}
		\multicolumn{2}{|m{3cm}|}{} & {Give a tutorial for IRS-aided wireless communications, and elaborate on technical challenges and potential directions.} & {\cite{wu2021-intelligent}} & {2021}  \\ \hline		
		%   System Design & Performance Analysis   
		\multicolumn{2}{|m{3cm}|}{\multirow{2}{3cm}[-0.2cm]{\centering System Design \& \\ Performance Analysis}} & {Discuss the optimization frameworks and performance analysis methods for large intelligent surfaces.} & {\cite{alghamdi2020-intelligent}} & {2020}  \\ \cline{3-5}
		\multicolumn{2}{|m{3cm}|}{} & {Explain the fundamental principles of RISs based on electromagnetic waves, and provide a survey of RIS focusing on the performance and applications.} & {\cite{liu2021-reconfigurable}} & {2021}  \\ \hline		
		%   System Design & Applications   
		\multicolumn{2}{|m{3cm}|}{\multirow{3}{3cm}[-0.2cm]{\centering System Design \& \\ Applications}} & {Provide a technical study on the IRS, with emphasis on the performance enhancement of different scenarios induced by the reconfigurability of IRS.} & {\cite{gong2020-smart}} & {2020}  \\ \cline{3-5}
		\multicolumn{2}{|m{3cm}|}{} & {Explore the up-to-date solutions and challenges of the RIS in view of CSI acquisition, passive information transfer and low-complexity robust system design.} & {\cite{yuan2021-reconfigurableintelligentsurface}} & {2021}  \\	\cline{3-5}
		\multicolumn{2}{|m{3cm}|}{} & {Provide an overview on using IS for low-complexity data modulation.} & {\cite{m3}} & {2021}  \\ \hline	
		%   Applications & Challenges
		\multicolumn{2}{|m{3cm}|}{\multirow{3}{3cm}[-0.2cm]{\centering Applications \& \\ Challenges}} & {Present a comprehensive yet concise overview of the IRS technology and explores the main design challenges.} & {\cite{wu2020-smart}} & {2020}  \\ \cline{3-5}
		\multicolumn{2}{|m{3cm}|}{} & {Elaborate on the concept of smart radio environments (SREs) empowered by reconfigurable AI meta-surfaces.} & {\cite{renzo2019-smart}} & {2019}  \\ 
            \cline{3-5}
		\multicolumn{2}{|m{3cm}|}{} & {Discuss the concept, applications, challenges and future research directions of RIS/IRS.} & {\cite{CunhuaPan2021CM}} & {2021} \\ \hline		
		%   Deployment
		\multicolumn{2}{|m{3cm}|}{\multirow{2}{3cm}[-0.6cm]{\centering Deployment}} & {Deliver a tutorial of multi-IRS-aided wireless networks, emphasizing on the aspects of IRS reflection design and channel acquisition.} & {\cite{mei2022-intelligent}} & {2022}  \\ \cline{3-5}
		\multicolumn{2}{|m{3cm}|}{} & {Envision the IRS-empowered 6G wireless networks from the perspective of its architecture, deployment strategy, and applications.} & {\cite{Naeem2022IRS-Empowered}} & {2022}  \\ \cline{3-5}
		\multicolumn{2}{|m{3cm}|}{} & {Review the IRS deployment design in wireless networks including the BS/User-side deployment and the hybrid deployment strategies.} & {\cite{you2022-how}} & {2022}  \\ \hline			
		%   Architectures             
		\multirow{2}{1cm}[-0.4cm]{\centering Architec-tures} & {Holographic Surfaces} & {Introduce the Holographic MIMO Surfaces in terms of the fundamental basics, applications and design challenges.} & {\cite{huang2020-holographic}} & {2020} \\ \cline{2-5}
		~ & {Intelligent Omni Surfaces} & {Present a survey of IOS with the emphasis on design principles, system model and design, hardware implementation and experiments, as well as  potential research directions.} & {\cite{zhang2022-intelligent}} & {2022} \\ \hline
		%   Hardware Implementation
		\multicolumn{2}{|m{3cm}|}{\multirow{2}{3cm}[-0.2cm]{\centering Hardware \\ Implementation}} & {Overview the different implementations and channel models for RIS as well as the RIS-assisted optimization.} & {\cite{elmossallamy2020-reconfigurable}} & {2020} \\ \cline{3-5}
		\multicolumn{2}{|m{3cm}|}{} & {Elaborate on the mechanisms of RISs in the view of digital metasurfaces, and presents several hardware implementations and architectures.} & {\cite{cheng2022-reconfigurable}} & {2022} \\ \hline		
		%   Applications             
		\multirow{2}{1cm}[-1.5cm]{\centering Applica-tions} & {IRS-NOMA} & {Discuss the performance improvements induced by the IS, including the channel gains, fair power allocation, coverage range and energy efficiency.} & {\cite{sena2020-what}} & {2020} \\ \cline{2-5}
		~ & {Radio Localization and Mapping} & {Provide an overview of RIS-based radio localization and mapping, highlighting the challenges and open issues left behind.} & {\cite{wymeersch2020-radio}} & {2020} \\ \cline{2-5}
		~ & {Wireless Transceivers} & {Explain two paradigms of IS as RF chain-free transmitter and space-down-conversion receiver.} & {\cite{tang2020-wireless}} & {2020} \\ \cline{2-5}
		~ & {Wireless Energy Transfer} & {Deliver a tutorial overview on IRS-aided WPT/WIPT systems, with a special focus on system designs and open issues.} & {\cite{wu2022-intelligent}} & {2022} \\ \hline  		
		\end{tabular}
\end{table*}

\subsection{Aims and Organization}
In contrast to the existing overview, survey, and tutorial papers on ISs listed in Table \ref{contributions} and two recent books \cite{wu2023intelligent,wu2023intelligentswipt}, this paper aims to provide a comprehensive overview of the latest development and advancement of ISs, with an emphasis on recent research results as well as new IS architectures and their design issues.
Specifically, this paper first provides an in-depth overview of the state-of-the-art results on the design of the commonly adopted reflection-based IS including its reflection optimization, deployment, signal modulation, wireless sensing, and ISAC. 
Next, we delve into the recent progress in advanced IS architectures and their associated new design issues and solutions. 
Moreover, we highlight the important problems that remain unsolved to motivate future research on ISs.   
We hope that this paper will help unveil the immense potential of ISs for applications in future wireless networks such as 6G, and provide useful and up-to-date guidance for the future research and development for ISs.   

As depicted in Fig. \ref{org}, the rest of this paper is organized as follows. Section II presents a detailed overview of state-of-the-art results on the reflection-based IS design. Section III discusses the recent progress and new issues of advanced IS architectures. Finally, Section IV concludes this paper.

%\caption{IS beam training design.}

\section{Reflection-based IS: Design Issues and State-of-the-art Results}

In this section, we overview the main design issues and state-of-the-art results for reflection-based ISs, including reflection optimization, deployment, modulation, sensing as well as ISAC.

\begin{table*}[!t]
  \centering
  \renewcommand{\arraystretch}{1.3}
  {{\caption{ A toy example of blind beamforming when $N=4$, $K=2$, and $T=6$.}
  \small
  {\begin{tabular}{|m{2.1cm}|m{1.4cm}|m{1.4cm}|m{1.4cm}|m{1.4cm}|m{1.4cm}|m{1.4cm}|}
    \hline
  {\bf Sample Index}  & 1 & 2 & 3 & 4 & 5 & 6 \\ 
  \hline
  { $\bm{(\theta_1,\theta_2,\theta_3,\theta_4)}$ }          & $(0,\pi,0,0)$  & $(0,0,0,0)$  &  $(\pi,\pi,\pi,0)$  & $(\pi,0,\pi,\pi)$ & $(\pi,\pi,0,\pi)$ & $(0,0,\pi,\pi)$ \\
  \hline
  { \bf Utility Value }  & 2.8 & 1.0  &  1.5  & 3.3  & 0.3 & 0.4 \\
  \hline
    {\bf Is $\bm{\theta_1=0}$? }  & \Checkmark & \Checkmark &  \XSolidBrush  & \XSolidBrush  & \XSolidBrush & \Checkmark \\
  \hline
    {\bf Is $\bm{\theta_1=\pi}$? }  & \XSolidBrush & \XSolidBrush &  \Checkmark  & \Checkmark  & \Checkmark & \XSolidBrush \\
  \hline
  \end{tabular} } } }
  \label{tab:blind_beamforming}
  \end{table*}

\subsection{Reflection Design}
In order to fully exploit ISs' potential, it requires judicious design of the reflection coefficients of all IS elements, thus giving rise to passive reflection/beamforming optimization problem \cite{wu2019-intelligent,8930608,9950724,10186460,10159991,10430366}. Most of the existing design methods for IS passive reflection/beamforming are based on a two-stage procedure, i.e., the cascaded BS-IS-user channel is first estimated via downlink/uplink pilots \cite{Zheng2020OFDM,Zheng2020OFDMA,Zheng2021Fast} and the reflection coefficients are then optimized accordingly \cite{wu2021-intelligent,zheng2022-survey,Swindlehurst2022estimat,you2020channel,You2021Wireless}. However, under the current cellular protocols, the pilots are dedicated to the use of direct channel estimation between the BS and users without IS reflection. Thus, the estimation of IS channels requires significant modifications of the current cellular protocols. Moreover, due to massive reflective elements (REs) of IS, a large number of additional pilots are needed to estimate the high-dimensional BS-IS, IS-user, or cascaded BS-IS-user channels \cite{Zheng2020OFDM,Zheng2020OFDMA,Zheng2021Fast,Zheng2021Efficient}, which entail excessively high overhead for the existing wireless systems.

\begin{figure}[t]
	\centering
	\includegraphics[width=8cm]{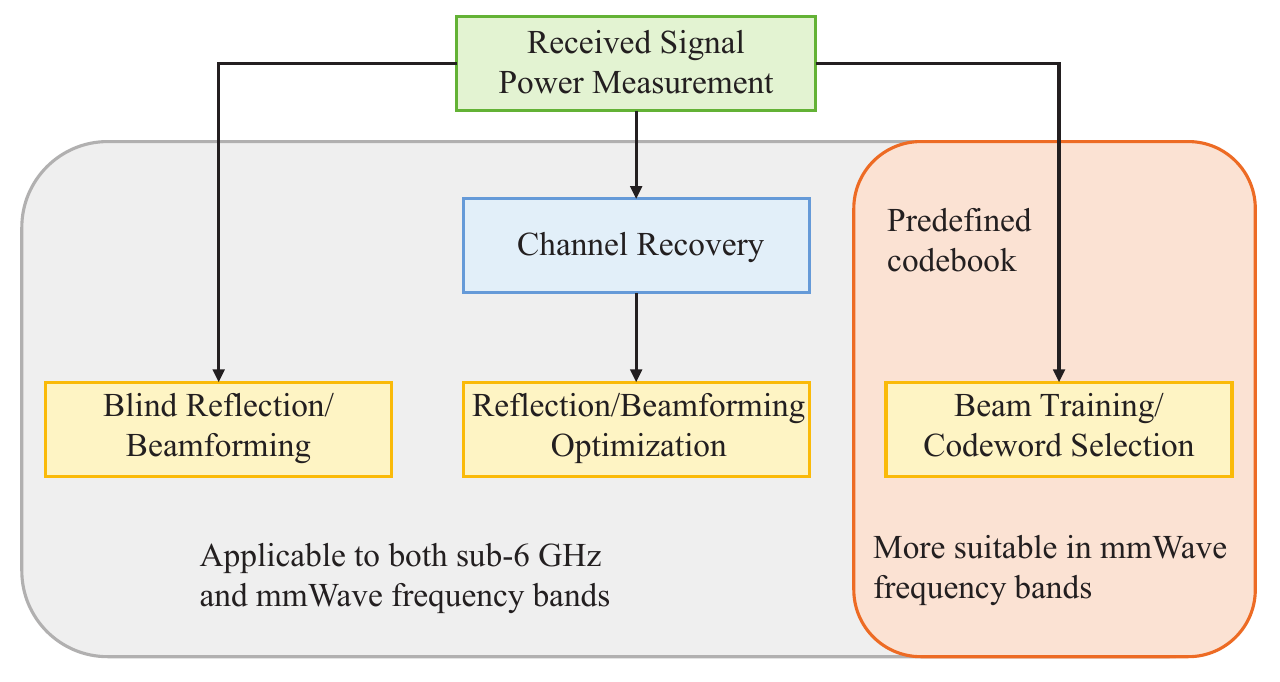}
	\caption{Three approaches for IS reflection/beamforming design based on received signal power measurement.}
	\label{fig:IS_reflection_method}
\end{figure}

To enable the seamless application of ISs in current and future wireless communication systems, recent research efforts have been devoted to developing more practical approaches for IS passive reflection/beamforming design. Specifically, instead of relying on the baseband complex-valued pilots, the IS reflection coefficients can be optimized based on the \emph{received signal power measurement} at the user terminals \cite{blind_beamforming_twc, Arun_2020_RFocus,yan2023channel,sun2023user,r6,r4}, which are easily accessible in the existing cellular networks, e.g., from the reference signal received power (RSRP). Thus, the power measurement-based IS reflection/beamforming design does not require additional pilots and is fully compatible to the current cellular protocols. Currently, such designs can be mainly classified into three categories, i.e., blind reflection/beamforming, channel recovery-based reflection/beamforming, and beam training, as shown in Fig. \ref{fig:IS_reflection_method}. 

Specifically, the blind reflection/beamforming approach optimizes the phase shift of each RE based on the user's power measurement without the need of any explicit channel state information (CSI) estimation \cite{blind_beamforming_twc, Arun_2020_RFocus,Psomas_WCOM,Nadeem_WCOM}. 
In contrast, the channel recovery-based approach first estimates the cascaded BS-IS-user channel based on the user's power measurement and then optimizes the IS reflection/beamforming accordingly \cite{yan2023channel,sun2023user}. 
Note that the above two approaches are practically applicable to both sub-6 GHz bands with rich multi-path channels and mmWave frequency bands with LoS-dominant channels. 
However, they are generally less efficient in mmWave frequency bands due to the channel/path sparsity and severe path loss\footnote{For each IS element, the channel condition (e.g., channel sparsity and number of multi-paths) is generally determined by the signal propagation environment and carrier frequency only, which is not affected by the aperture size of the whole IS. Nonetheless, the end-to-end effective channel of the BS-IS-user link is highly dependent on the IS aperture and beamforming/reflection design for both the instantaneous channels and statistical channel distribution.}. 
In particular, the received signal power may not be sufficient for CSI recovery if the IS reflection/beamforming vector is not aligned with the dominant channel path. 
Furthermore, the training overhead of these methods may be unnecessarily large because they do not fully exploit the inherent channel sparsity in mmWave frequency bands. 
To overcome the above issues, beam training serves as a more efficient approach for IS passive beamforming design in mmWave systems by selecting the best directional beam from a predefined beamforming codebook \cite{r6,r4}. In the following, the above three power-measurement-based approaches for IS passive reflection/beamforming design are discussed in more details.

\subsubsection{Blind Reflection/Beamforming}
{The blind IS reflection/beamforming approach applies to the discrete phase shift of REs.} Let us start with a naive algorithm called \emph{Random-Max Sampling (RMS)} for blind reflection/beamforming. Its idea is fairly simple: just try out a sequence of random samples of the phase shifts of IS and choose the one with the best performance thus far. The following toy example illustrates how RMS works. Assume that IS has four REs; denote by $\theta_n$ the phase shift of the $n$th RE, $n=1,2,3,4$; assume also that each phase shift $\theta_n$ takes on value from the binary set $\{0,\pi\}$. Suppose that a total of six random samples of the phase shift array $(\theta_1,\theta_2,\theta_3,\theta_4)$ have been tested as shown in Table \ref{tab:blind_beamforming}. The performance of each random sample is evaluated by the utility value, e.g., we can take the received signal-to-noise ratio (SNR) (which is in direct proportion to the RSRP) as the utility for the single-antenna transmission. Observe that RMS gives the solution $(\theta_1,\theta_2,\theta_3,\theta_4)=(\pi,0,\pi,\pi)$ in this example because the fourth random sample has the highest utility value. Clearly, RMS can attain the global optimum so long as all the possible combinations of $\{0,\pi\}$ have been tested.

However, a real-world IS has far more than four REs as assumed in the above toy example, e.g., there are 256 REs integrated into the IS prototype in \cite{blind_beamforming_twc}, so it is already intractable to search through the solution space even if each phase shift has only two choices. Then a natural idea is to partially explore the solution space. But how much portion of the entire solution space should RMS explore in order to achieve ``good'' performance?
To formalize the above question, we denote by $N$ the number of REs, $T$ the number of random samples, and $K$ the number of choices for each phase shift. For ease of discussion, we focus on the single-antenna transmission and use the SNR as the utility. A fundamental problem is how fast the SNR scales with $N$ if $T$ is much smaller than the solution space size $K^N$, which is termed as the \emph{SNR boost}\footnote{Note that the SNR boost refers to the scaling order of the SNR with respect to $N$, which is identical for both discrete and continuous phases shifts of IS.}.

The above fundamental problem has been answered in \cite{blind_beamforming_twc} for RMS. Roughly speaking, the authors in \cite{blind_beamforming_twc} showed that the SNR boost behaves like
$\mathrm{SNR}\propto N\log T$
by the RMS method. Moreover, it has been shown in \cite{wu2019-intelligent} that the scaling rate is at most quadratic in $N$, and this upper bound is tight in the ideal case where every reflected channel is perfectly aligned with the direct channel. Thus, in order to achieve a quadratic boost of SNR, we must set $T$ to be exponentially large in $N$ for the RMS method. However, $T$ can only be polynomially large in $N$ in practice. As a result, the actual performance of RMS is $\mathrm{SNR}\propto N\log N$. One may then have a follow-up question: Is it possible to strike a quadratic boost of SNR (which is the theoretical upper-bound as revealed in \cite{wu2019-intelligent}) by using a polynomial number of random samples, $T$?
%Furthermore, it is worth mentioning that RMS can be recognized as a beam-sweeping or beam-training method \cite{r6,r4,beamtraining_LiH_2022,beamtraining_Zhangw_2021} with the codebook generated uniformly. Hence, $N\log T$ is also the fundamental limit of beam sweeping or beam training, where $T$ now refers to the codebook size. But is it possible to strike a quadratic boost of SNR by using a polynomial number of random samples?

Two recent works \cite{blind_beamforming_twc,Arun_2020_RFocus} provide a positive answer to the above question by devising the so-called \emph{Conditional Sample Mean (CSM)} method; a more recent work \cite{XuFan2024TSP} further extends CSM to the joint configuration of multiple ISs. In a nutshell, CSM first calculates the mean utility conditioned on each possible phase-shift choice for a particular RE, and then chooses the phase shift with the highest conditional mean utility for this RE. We still use the toy example in Table \ref{tab:blind_beamforming} to illustrate CSM. Let us see how $\theta_1$ is decided. Recall that there are two possible choices, $0$ and $\pi$, for the phase shift; we first average out all the random samples with $\theta_1=0$ so as to obtain the mean utility conditioned on $\theta_1=0$, i.e.,
$$
\mathbb E\big[\text{utility}\big|\theta_1=0\big] = \frac{2.8+1.0+0.4}{3}=1.4,
$$
and similarly find the mean utility conditioned on $\theta_1=\pi$ as
$$
\mathbb E\big[\text{utility}\big|\theta_1=\pi\big] = \frac{1.5+3.3+0.3}{3}=1.7.
$$
Comparing the above two conditional mean utility values, the CSM method chooses $\theta_1=\pi$. Likewise, the other phase shifts are determined as $\theta_2=0$, $\theta_3=\pi$, and $\theta_4=0$. Intuitively, CSM aims at the average goodness of setting $\theta_1$ to a particular possible choice while the rest phase shifts are still pending; with every phase shift optimized in this fashion independently, CSM enables an efficient configuration of IS.

\begin{figure}[t]
\centering
\includegraphics[width=8cm]{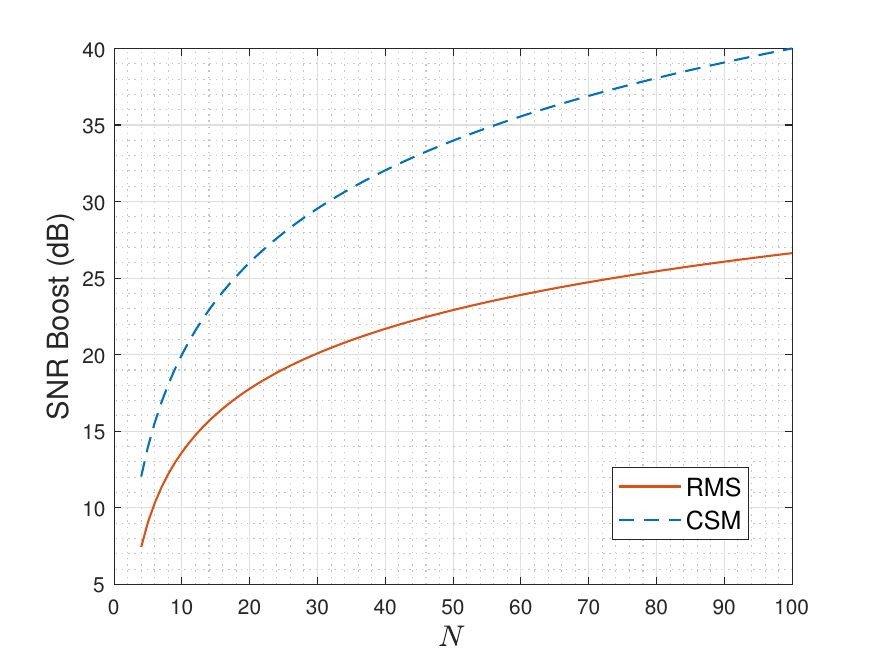}
\caption{RMS versus CSM given a polynomial number of random samples.}
\label{fig:blind beamforming}
\end{figure}

More importantly, as shown in \cite{blind_beamforming_twc}, the CSM method yields $\mathrm{SNR} \propto N^2$ for polynomially large $T$. It can be observed from Fig.~\ref{fig:blind beamforming} that CSM outperforms RMS significantly in terms of the scaling rate of SNR in $N$. In particular, according to the field tests at 2.6 GHz in the 5G network, the SNR of CSM can be approximately 8 dB higher than that of RMS when $N=256$, $K=2$, and $T=10N$. The huge advantage of CSM over RMS can be perceived from two standpoints. First, RMS solely relies on the single ``best'' random sample with the highest utility value, whereas CSM makes full use of the whole random sampling data. Second, the solution of RMS is restricted to a small space of the solution space that has been explored by the random samples, whereas the solution of CSM can be beyond this space; as shown in the previous toy example, the solution $(\pi,0,\pi,0)$ given by CSM does not appear in the six random samples tested.

\subsubsection{Channel Recovery-Based Reflection/Beamforming}
The blind reflection/beamforming design in general requires a large number of random samples (i.e., IS phase-shift realizations) as well as corresponding received signal power measurements to achieve the upper bound on the SNR boost in IS-aided communication systems. This is fundamentally because the explicit CSI between the BS, IS, and user is not exploited, which results in the underutilization of power measurements as well as high training overhead. To overcome this limitation, a new channel recovery-based reflection/beamforming approach was proposed in \cite{yan2023channel,sun2023user} based on the received signal power measurement as well. Specifically, by collecting the received signal power under different IS reflections, the cascaded BS-IS-user channel (or its autocorrelation matrix) is first estimated. Then, based on the obtained CSI, the IS reflection/beamforming coefficients are optimized more efficiently for improving the communication performance. {Note that the channel recovery-based approach applies to both discrete and continuous phase shifts of REs.}

As the phase information of the received signal is missing in its measured power, a fundamental question for the power measurement-based channel estimation is whether the complex-valued channel vector can be recovered by only exploiting the received signal powers which are positive real numbers. The results in \cite{yan2023channel} revealed that if the number of power measurements is sufficiently large, the autocorrelation matrix of the cascaded BS-IS-user channel can be uniquely determined. As such, the cascaded BS-IS-user channel can be recovered with an unknown phase ambiguity that is common to all IS REs. A special case occurs if the REs have only two phase-shift choices, i.e., 0 and $\pi$. In this case, the channel autocorrelation matrix cannot be distinguished with its conjugate. Fortunately, since the passive beamforming vector of IS has real values only in this binary phase-shift case, such a phase ambiguity does not affect the SNR boost at the user's receiver \cite{yan2023channel}. 

In general, the channel recovery can be formulated as an optimization problem which aims to find a channel vector satisfying the power measurement constraints under different IS reflections. Since the power measurement constraints are non-convex, one viable approach is by lifting up the channel/beamforming vectors into matrices by calculating their autocorrelation matrices \cite{yan2023channel}. As a result, the power measurement constraints become linear in the channel autocorrelation matrix. Since the channel autocorrelation matrix has the rank of one, the optimal solution can be found by minimizing its rank, subject to the power measurement constraints. By applying proper relaxations, the rank-minimization problem can be solved by using the semidefinite programming (SDP)-based method \cite{yan2023channel}. On the other hand, to reduce the computational complexity as well as improve the robustness to noise, a neural network (NN)-based approach was proposed in \cite{sun2023user} to estimate the cascaded BS-IS-user channel vector. Specifically, it was revealed that the received signal power can be viewed as a single-layer NN, with its input and output corresponding to the real and imaginary parts of the IS reflection vector and the expected received signal power, respectively. The weights of the single-layer NN correspond to the cascaded channel, which can be estimated by training the NN via minimizing the loss function, i.e., the mean square error (MSE) between the NN's output and the measured signal power. For both SDP-based and NN-based channel recovery, the IS passive beamforming vector can be further optimized based on the estimated CSI. Various optimization techniques proposed in the literature can be utilized, such as semidefinite relaxation (SDR) \cite{ma2022cover}, block coordinate descent (BCD) \cite{xu2020resource}, successive convex approximation (SCA) \cite{yu2020robust}, the penalty-based method \cite{xu2022optimal}, and the projected gradient ascent/descent method \cite{yan2023discrete}.

\begin{figure}[t]
	\centering
	\includegraphics[width=8cm]{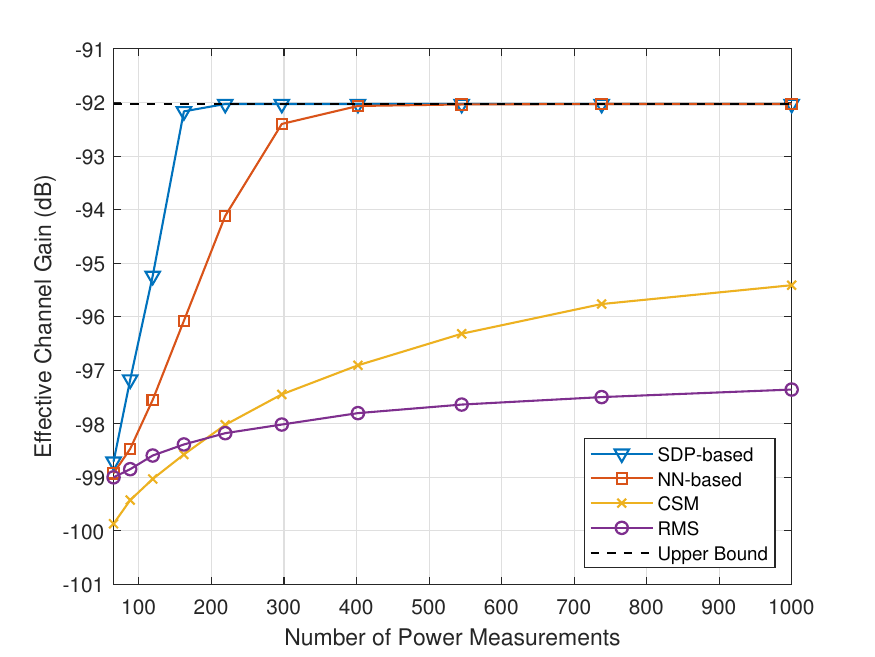}
	\caption{Effective channel gain between the BS and user aided by an IS under different IS reflection/beamforming approaches.}
	\label{fig:IS_reflection_gain}
\end{figure}

In Fig. \ref{fig:IS_reflection_gain}, we show the effective channel gains between the BS and a single user aided by an IS, employing the RMS, CSM, SDP-based and NN-based IS reflection/beamforming designs. The size of IS is $N=64$ and the phase of each IS element can only be selected from the binary set $\{0, \pi\}$. The details of the system setup, channel model, and parameter settings can be found in \cite{sun2023user}. The computational complexities for the RMS, CSM, SDP-based and NN-based methods are $\mathcal{O}(1)$, $\mathcal{O}(N)$, $\mathcal{O}(N^{4.5})$, and $\mathcal{O}(N)$, respectively. As can be observed, the SDP-based method requires the least power measurements for approaching the upper bound on the effective channel gain assuming perfect CSI, while it requires the highest computational complexity for channel recovery and IS beamforming optimization. In contrast, the RMS method has the lowest computational complexity, while its gap to the performance upper bound is large if the number of power measurements is small. The CSM and NN-based methods achieve a tradeoff between the communication performance and computational complexity. 

\subsubsection{Beam Training}

\begin{figure*}[t]
	\centering
	\includegraphics[width=16cm]{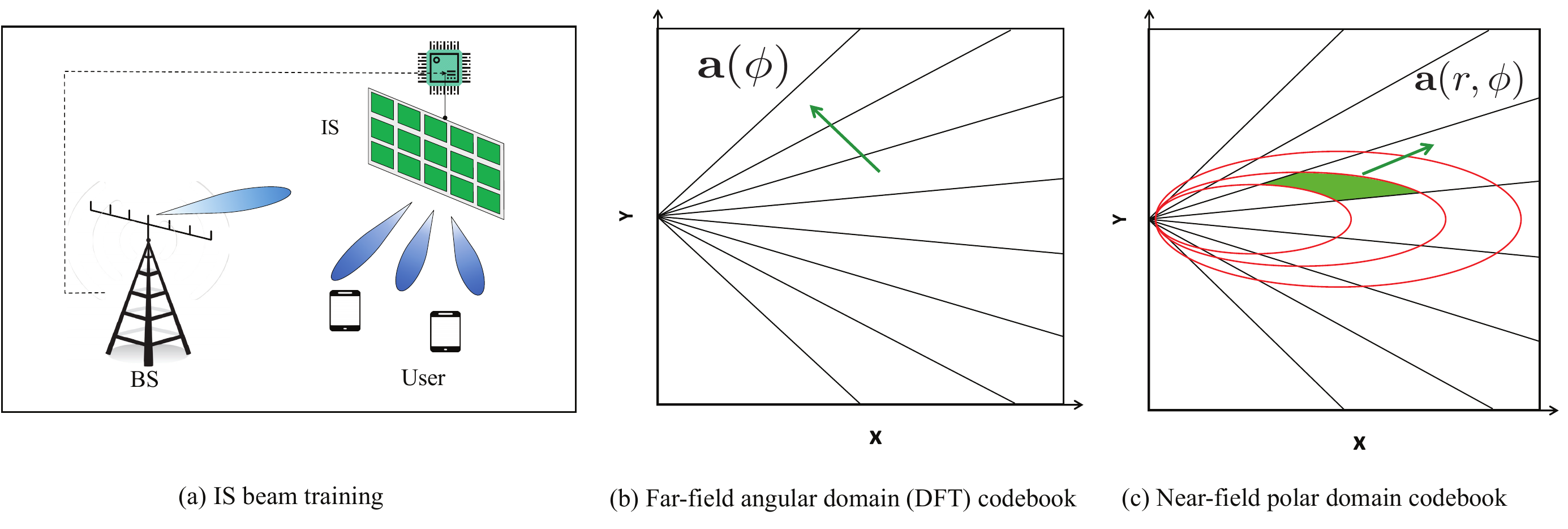}
	\caption{Illustration of IS beam training in far- and near-field channels.}
	\label{fig:IRS-beam-training}
\end{figure*}

\begin{table*}[]
	\caption{IS beam training design.}
	\label{bt_table}
	\centering
	\renewcommand{\arraystretch}{1.3}
	\setlength{\LTcapwidth}{\textwidth}
	\begin{tabular}{|l|l|l|}
		\hline
		\textbf{ }                    & \textbf{System} & \textbf{Proposed IS beam training design method}                                                                                                                                                                                                                                                                    \\ \hline
		\multirow{9}{*}{Far-field}                & \multirow{7}{*}{Narrow-band}  & Propose a quadratic phase-shift-based codeword design with wide beamwidth \cite{r2}                                                                                                                                                                                                                       \\ \cline{3-3} 
		&                              & Propose a SCA-based codebook design with beamwidth-variable codewords accommodating IS discrete phase shift \cite{r3}                                                                                                                                                                                    \\ \cline{3-3} 
		&                              & Propose a ternary-tree hierarchical beam training method for IS-aided multi-user systems \cite{r4}                                                                                                                                                                                                       \\ \cline{3-3} 
		&                              & Propose a binary hierarchical beam training with two multi-mainlobe codewords in each layer of the codebook \cite{r5}                                                                                                                                                                                     \\ \cline{3-3} 
		&                              & Propose a multi-beam training method for IS-aided multi-user systems with small training overhead \cite{r6}    
		\\ \cline{3-3} 
		&                              & Propose a computer vision-based approach to aid IS for dynamic beam tracking \cite{r8}                                                                                                                                                                                             \\ \cline{3-3} 
		&                              & Propose a simultaneous beam training and target sensing scheme at the BS for IS-aided multi-user systems \cite{r7}                                                                                                                                                                                                                                                                                                                                                                                                       \\ \cline{2-3} 
		& \multirow{2}{*}{Wide-band}    & Mitigate the beam squint effect via sum-achievable rate maximization over all subcarriers \cite{r9}                                                                                                                                                                                                       \\ \cline{3-3} 
		&                              & Mitigate the beam squint effect via mounting time-delay units for each sub-surface of IS \cite{r10}                                                                                                                                                                                                      \\ \hline
		\multirow{8}{*}{Near-field} & \multirow{6}{*}{Narrow-band}  & Propose a Cartesian-domain near-field codebook \cite{r2_1}                                                                                                                                                                                                                                               \\ \cline{3-3} 
		&                              & Propose a polar-domain near-field codebook \cite{r2_2}                                                                                                                                                                                                                                                   \\ \cline{3-3} 
		&                              & Propose to first estimate the user angle via the DFT codebook then determine the optimal range \cite{r2_3}                                                                                                                                                                                       \\ \cline{3-3} 
		&                              & Propose to estimate the user angle and range via the DFT codebook \cite{wu2023near}  
		\\ \cline{3-3} 
		&                              & Propose to first estimate the user range via a omnidirectional codebook then determine the user angle \cite{r2_4}  
		\\ \cline{3-3}   
		&                              & Propose to first estimates a coarse user angle and then refine the user angle-and-range pair  \cite{wu2023twostage}                                                                                                                                                                      \\ \cline{3-3} 
		&                              & Propose a beamwidth-variable IS reflection design to generate multi-layer codebook in the Cartesian domain \cite{r2_5}                                                                                                                                                                \\ \cline{3-3} 
		&                              & \begin{tabular}[c]{@{}l@{}}Propose 1) a deep residual network-based beam training scheme by using far-field codewords for near-field beam training \\               2) a deep residual network-based beam training scheme by using partial near-field codewords \cite{r2_6}\end{tabular} \\ \cline{2-3} 
		& \multirow{2}{*}{Wide-band}    & Employ the delay adjustable metasurface to compensate for the frequency dispersion for XL-IS-aided OFDM system \cite{r2_7}                                                                                                                                                                              \\ \cline{3-3} 
		&                              & Propose a two-phase beam training scheme by exploiting the beam split effect for XL-IS-aided OFDM systems \cite{r2_8}                                                                                                                                                                                    \\ \hline
	\end{tabular}
\end{table*}

Note that the above blind and channel recovery-based reflection/beamforming approaches can be applied in real time for improving the instantaneous channels or offline for improving the channel statistical distribution in sub-6 GHz bands. For mmWave and higher frequency bands, high beamforming gain is required to compensate for the severe path loss. Due to the channel/path sparsity, codebook-based beam training with directional beams is a more efficient approach to design IS beamforming without the need of explicit CSI estimation as illustrated in Fig. \ref{fig:IRS-beam-training}. {In the following, we discuss efficient methods for IS beam training in the far- and near-field channel scenarios, respectively, under the assumption of continuous RE phase shifts; while these methods can be readily extended to the case of discrete RE phase shifts.}

\emph{3.1) Far-Field Beam Training:}
First, consider the case where both the BS and users are located in the far-field region of the IS as shown in Fig. \ref{fig:IRS-beam-training}(b). Instead of performing joint BS and IS beam training that incurs high training overhead, a more effective approach is conducting an offline BS beam training to establish a high-gain BS-IS link by exploiting the fixed locations of the BS and IS. Then, only the IS beam training needs to be performed in the real-time phase 
%gFor IS-aided wireless communication systems, both the BS transmit beamforming and IS reflect beaFirst, consider the case where the BS-IS beam training has been 
%channel measurements indicate that their corresponding channels exhibit sparse scattering characteristics.
%Therefore, instead of obtaining the full CSI of IS that incurs excessive channel training overhead, an efficient approach is to identify the dominant channel path, and establish an initial high-SNR link from the transmitter to the receiver through the IS, referred to as the IS beam training/alignment.
%Specifically, the IS beam training scheme 
for selecting the best IS beam in a predesigned codebook that yields the strongest signal power at the user's receiver. 
For narrow-band systems, sequential \emph{single-beam} training is a straightforward method that exhaustively searches over all possible beam directions (i.e., codewords) at the IS. As revealed in \cite{r1,r2}, there generally exists a fundamental tradeoff between maximizing the achievable throughput and minimizing the beam training overhead, i.e., a more accurate beam training leads to a higher passive beamforming gain, while at the cost of higher training overhead as well. To reduce the IS beam training overhead, one line of research has developed efficient IS beam codebooks to reduce the beam-search space. For example, the authors in \cite{r2} proposed a small-size IS codebook with wide beamwidth, which is achieved by a quadratic phase-shift design for enabling higher-order variations at IS phase shifts. 
%designing IS codewords with wide beamwidth, which improves the average beam power efficiency compared with that of narrow-beam codebooks. 
In addition, under the  IS discrete phase-shift constraint,  an SCA-based algorithm was proposed in \cite{r3}  to design the IS codebook with beamwidth-adjustable codewords, while at the expense of decreased beamforming gain. Alternatively, hierarchical IS codebook design and beam search is also an efficient way to reduce the beam training overhead without compromising much the beamforming gain \cite{r4,r5}. Specifically, a hierarchical IS beam training method was proposed in \cite{r4}, which first identifies the best beam sector with wide IS beams and then refines the IS beam in the sector using narrow beams.
However, the IS codewords may not be orthogonal in practice due to IS hardware imperfections, hence resulting in a reduced beam training success rate. In addition, another hierarchical IS beam search method was proposed in 
% instead of refining the IS beam by narrowing the angle range of beam sweeping layer by layer, 
\cite{r5}, which uses two multi-mainlobe codewords in each layer for beam sweeping and exploits the beam discrimination in each layer to effectively increase the success rate.
However, hierarchical beam training approaches generally require frequent user feedback to refine the beam selection, thus incurring high feedback/training overhead that scales with the number of users.
To circumvent this issue, the authors in \cite{r6} proposed a new \emph{multi-beam} training method, where IS REs are divided into multiple sub-arrays, which allows for simultaneous beam sweeping over different directions. Then, each user can determine its optimal IS beam direction by comparing the received signal power/SNR over time, thus greatly reducing the training overhead yet without the need for successive user feedback in IS-aided multi-user systems.
Besides pilot-based beam training, other approaches for IS beam training in narrow-band systems have also been recently proposed. For instance, the authors in \cite{r8} proposed a new binocular camera-mounted IS architecture, where the visual information is collected to aid the IS in locating the BS/access point (AP) as well as dynamic beam tracking with the users, thus eliminating the need for dedicated training symbols and signal feedback. In  \cite{r7}, the authors considered an ISAC system with IS and proposed a simultaneous beam training and target sensing scheme. Specifically, the BS simultaneously senses the targets and performs the beam training for both the users and IS, based on which the LoS IS-user channel is determined by exploiting their geometry relationship.

For wide-band systems,  IS beam training designs become more challenging due to the lack of  \emph{frequency selectivity} in  IS phase shifts; thus they remain the same over all frequencies. This results in the so-called \emph{beam squint} effect, where the beams at different frequencies are generally dispersed in different directions.
To address this issue, the authors in \cite{r9} proposed to first obtain the optimal IS phase shifts for each subcarrier and then determine the common phase shift by maximizing the upper bound of the sum achievable rate over all subcarriers in orthogonal frequency-division multiplexing (OFDM) systems.
Alternatively,  the authors in \cite{r10} proposed to equip each IS sub-surface with a time-delay unit to enable frequency-dependent phase shifts over the entire frequency band,
% to compensate for the frequency-dispersion-induced phase shift, 
thus offering a flexible tradeoff between wide-band communication performance and hardware design complexity.

\emph{3.2) Near-Field Beam Training:}
Next, we consider the case where an extremely large-scale IS (XL-IS) is equipped with a huge number of REs to compensate for the product-distance path-loss. In this case, the BS and users are more likely to be located in the \emph{near-field} region of the XL-IS as shown in Fig. \ref{fig:IRS-beam-training}(c). We still assume that the BS-IS link has been aligned offline thanks to the fixed BS/IS locations, and thus focus on the IS beam training with users only. Note that directly applying the existing IS beam training methods for far-field scenarios will inevitably incur degraded beam training performance due to the channel mismatch. Thus, it is necessary to develop new near-field IS beam training methods accounting for (radiative) near-field  \emph{spherical wave-fronts} \cite{you2023nearfield,lu2023tutorial,cong2023near}.

% Owing to low cost and low power consumption, more and more reflecting elements are expected to be deployed for each IS to compensate for the passive reflection-induced product-distance path loss. Hence, IS is more likely to be developed into extremely large-scale IS (XL-IS) for future 6G communications.
%With the increasing number of IS reflecting elements from IS to XL-IS, the array aperture of XL-IS is practically large, and the Rayleigh distance increases accordingly. The scatterers are more likely to be in the near-field region of XL-IS, where the near-field (near-field) channel model should be applied in XL-IS-assisted systems.
%The existing beam training designs mainly consider the far-field (FF) channel model. As a result, the corresponding FF beam training may cause significant performance degradation in the XL-IS-assisted systems, due to the mismatches between the FF and near-field channel models.

To this end, several new codebooks catering to near-field beam training have been proposed, which can be applied to designing the near-field IS beam training. For example, a Cartesian-domain codebook was devised in \cite{r2_1}, which covers the entire three-dimensional (3D) space by performing uniform sampling in the Cartesian coordinate system. However, the dimension of the Cartesian-domain codebook is determined by the product of the numbers of sampled points at the $x$/$y$/$z$ axis, which is practically very large and thus incurs excessive beam training overhead.
To address this issue, a more efficient polar-domain codebook design was proposed in \cite{r2_2}, where the near-field region is sampled with non-uniform angle-range grids by minimizing adjacent codeword correlation.
%and requires a significantly reduced number of codewords for near-field channel representation as compared with the Cartesian-domain codebook. 
Based on this polar-domain codebook, one straightforward near-field IS beam training method is by exhaustively searching over all codewords, which, however,  requires a large number of training symbols proportional to the product of the numbers of sampled angles and ranges. To tackle this challenge, one efficient approach is by applying the \emph{two-phase} near-field beam training method proposed in \cite{r2_3} for IS. Specifically, the XL-IS/array first estimates the user angle by performing conventional DFT-codebook based beam sweeping and exploiting the ``angle-in-the-middle" effect in the received beam pattern. Then, given a set of estimated user angles, the XL-IS adopts the polar-domain beam training in the range domain to determine the user range. Numerical results showed that this two-phase method can significantly reduce the required near-field beam training overhead for XL-array/IS with negligible rate loss. This work was further extended in \cite{wu2023near} that directly utilizes the conventional DFT-codebook to resolve the user angle and range by more effectively exploiting the received signal powers.
%  polar-domain codeword is adopted to determine the user's distance in the second phase.
Alternatively, the authors in \cite{r2_4} proposed to first determine the user range via a designed omnidirectional IS codeword, and then perform the near-field beam sweeping in the angle domain to obtain user angle information.

%show that the near-field beam patterns under far-field (DFT) beamforming are broadened in the near-field region, resulting in inaccurate angle estimation with the DFT codebook. To address this issue, \cite{r2_4} proposed to first determine the user distance via an omnidirectional codeword, and then perform near-field beam sweeping in the angle domain to obtain the angle inear-fieldormation by exploiting the pencil-like near-field beam.

It is worth mentioning that the training overhead of the two-phase XL-IS beam training scheme is still proportional to the number of IS REs, which is still very large for practical implementation. 
This thus motivates various lower-overhead beam training designs customized for XL-IS-assisted communication systems \cite{r2_5,r2_6}.
For example, the authors in \cite{r2_5} proposed a beamwidth-variable IS reflection design for generating the multi-layer near-field codebook in the Cartesian domain, which shares a similar idea with that for the far-field hierarchical beam training. Besides, a new hierarchical near-field beam training method was proposed \cite{wu2023twostage}, which can be applied for XL-IS that first estimates a coarse user angle by activating partial IS elements and then refines the user angle-and-range pair by a dedicated near-field angle-and-range search.  
Moreover, the authors in \cite{r2_6} proposed a deep-learning (DL) based XL-IS beam training method, where the XL-IS applies the far-field codewords with wide beamwidth and the BS collects the received signals to feed in a deep residual network for determining the optimal polar-domain codeword. This can be achieved by exploiting the implicit relationship between the received signals of far-field codewords and the optimal near-field beamforming by DL methods. 
To further reduce the training overhead,  a partial near-field beam-based beam training was proposed in  \cite{r2_6}, where the XL-IS  adopts a small-size polar-domain codebook.  The corresponding received signals at the BS are fed into a deep residual network for extracting the optimal polar-domain codeword at the expense of lower beamforming gain.
%, which offers a flexible tradeoff between the training overhead and average achievable rate performance.

For wide-band XL-IS systems, the near-field beams generated at different frequencies with spherical wavefronts are generally dispersed at different locations (instead of directions in the far-field case) due to IS frequency non-selectivity. 
% IS reflecting elements that induce the same phase shift over all frequencies.
It is worth noting that the near-field beam squint/split effect is more severe than that in the far-field case, since the near-field beam splits in both the angular and range domains.
To tackle this challenge, the authors in \cite{r2_7} proposed to employ the delay adjustable metasurface to compensate for the frequency dispersion, in which the IS reflection and time delay are jointly optimized to enforce the near-field beam focused on the desired location at all the subcarriers in OFDM systems. Alternatively,
%However, the above design for near-field beam split mitigation requires accurate near-field CSI. This thus motivates the work
the authors in \cite{r2_8} proposed to exploit the beam split effect for near-field beam training, where a frequency domain multi-beam near-field focusing pattern at a fixed range is generated at each time for angle estimation and then the focused range of multi-beam pattern is adjusted over time for resolving the user range.

In Table \ref{bt_table}, we summarize the main approaches for the IS beam training designs under both the far-/near-field channel models. 
Despite these existing works, there are several practical challenges that need to be tackled in IS beam training.
For example, it is practically important to analyze the impact of hardware constraints/imperfections (e.g., discrete phase shift, amplitude-phase dependent IS reflection) on the IS beam training performance and develop robust beam training schemes. Besides, accurate synchronization between the BS and IS is also important to achieve joint beam training at the BS and IS. To fully exploit the potentials of IS in 6G systems, future research may also consider the joint optimization of IS deployment and beamforming, coordinated beamforming of multiple ISs, statistical channel-based IS reflection optimization for coverage enhancement, as well as machine learning-aided IS channel recovery and reflection optimization.

{{\bf{Summary:}} In this subsection, we have provided an overview of the state-of-the-art approaches for IS reflection/beamforming design based on received signal power measurements. 
For sub-6 GHz bands, the blind reflection/beamforming approach and channel recovery-based approach are both applicable. 
The blind reflection/beamforming approach does not require explicit CSI and has a low computational complexity. 
However, to achieve a satisfactory performance gain, this approach usually needs a sufficiently large number of power measurements for extracting the useful statistics in CSI, which inevitably incurs high training overhead. 
In contrast, the channel recovery-based approach requires less power measurements but higher computational complexity for channel estimation and IS reflection design. 
Both approaches can be applied to real-time IS reflection design for improving the instantaneous channels between the BS and users, under the condition of slow-fading channels. 
While for fast-fading channels, the two approaches are still feasible because they can exploit the statistical CSI or the deterministic part of time-varying wireless channels. 
As such, the corresponding IS reflection design can improve the spatial power distribution of channels between the BS and users over a long period. 
For high-frequency bands (e.g., mmWave), beam training is a more efficient approach for designing IS beamforming by effectively exploiting the channel paths sparsity in the spatial domain. 
Since this approach does not require the explicit CSI of IS-involved links and only selects the best directional beam from a designed codebook, its required training overhead and computational complexity are much lower than those of the blind reflection/beamforming approach and channel recovery-based approach. 
In addition to the pencil-like beamforming based on instantaneous channels, the wide beam of IS can also be designed based on the statistical channel knowledge/deterministic channel components to improve the long-term coverage performance over a target region.}

\subsection{Deployment}
The deployment of ISs has a significant impact on their reflected channels and the performance in IS-aided communication systems \cite{you2022-how,wu2021-intelligent,Naeem2022IRS-Empowered,zheng2022-survey}. In this subsection, we present some practical approaches to deploy ISs and compare their performance advantages in different use cases. 
\begin{figure}[t]
\centering
\includegraphics[width=7.5cm]{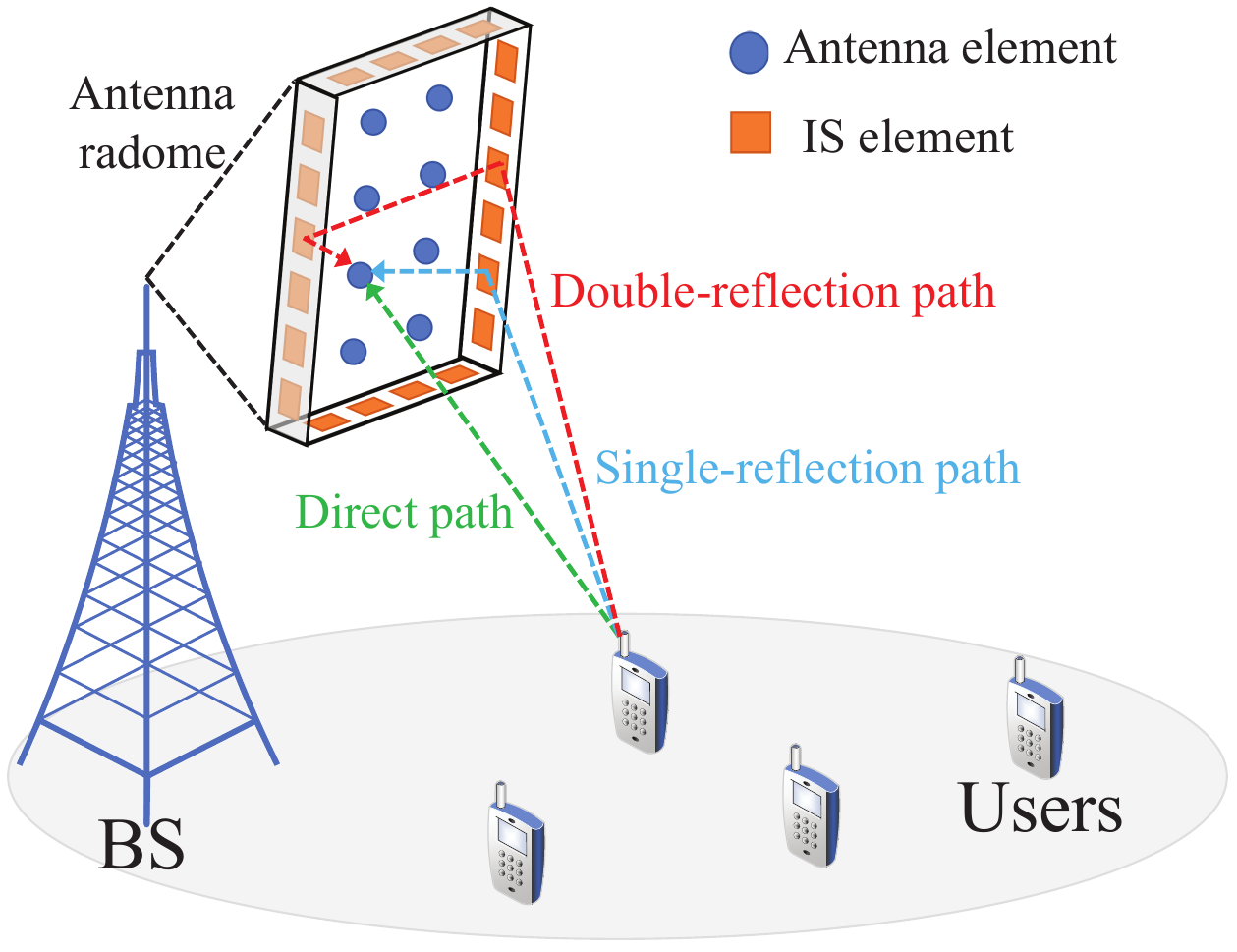}
\caption{An illustration of IS-integrated BS, where the IS elements and BS antennas are integrated within the same antenna radome.}
\label{fig:integrated_IS_BS}
\vspace{-6pt}
\end{figure}
\subsubsection{BS-Side IS}
The existing literature has mainly considered deploying ISs near user terminals to improve their performance. However, this approach may incur a high cost since ISs need to be densely deployed in the environment to cater to random user locations. Moreover, even with dense IS deployment, each user can only enjoy the passive beamforming gain brought by its nearby IS (as its signals reflected by other far-away ISs are generally much weaker due to more significant path loss). In \cite{Zhang2021Intelligent}, the authors have characterized the capacity regions achievable by two IS deployment strategies with the IS/ISs deployed near the BS and each of the distributed users, respectively, and showed the superiority of the former over the latter under the same total number of IS REs. 

Motivated by this result, the authors in \cite{huang2022empowering} proposed a co-site-IS empowered BS architecture, where ISs are deployed in the vicinity of the BS to assist in its communications with distributed users regardless of their locations. To further reduce the product-distance path loss of cascaded BS-IS-user channels, a new \emph{IS-integrated BS} architecture was proposed in \cite{huang2023integrat}, where the IS REs are integrated into the internal surfaces of the antenna radome at the BS, as shown in Fig. \ref{fig:integrated_IS_BS}. Due to the shorter distance between the IS REs and BS antennas, the IS reflection channels' path loss can be more significantly reduced as compared to that in \cite{huang2022empowering}. Moreover, since the closely spaced IS REs have different orientations, multiple reflections among them may further improve the path diversity and the effective channel gain between the BS and user. Last but not least, with the IS being installed at the BS, this architecture can help reduce the signaling overhead and provide the power supply for IS as compared to the user-side or co-site ISs which require dedicated wireless/wired signaling links and power supply.

However, the channel estimation for the IS-integrated BS is more challenging compared to the user-side and co-site IS systems because of the near-field channel response between the IS REs and BS antennas as well as more significant multi-reflection effects among IS REs. To circumvent this difficulty, the authors in \cite{huang2023integrat} proposed to employ the random phase algorithm (RPA) for IS reflection design without explicit CSI. Specifically, a large number of IS reflection patterns are generated with a random phase for each RE and their resulting communication performance is compared at the BS/user; and the one which achieves the best performance is then chosen as the IS reflection for data transmission. To further improve the performance of RPA, an iterative RPA (IRPA) method was proposed by alternately optimizing the reflection coefficients of each sub-IS \cite{huang2023integrat}.

Since both the RPA and IRPA methods require a sufficiently large number of IS reflection patterns for performance measurement, the high training overhead may render them practically inefficient. To address this issue, a codebook-based passive reflection design for the IS-integrated BS was proposed in \cite{huang2023passive}. In particular, a small number of codewords are designed offline to enhance the coverage performance of multiple sectors in a cell. Then, the BS performs online training by testing all codewords sequentially and selecting the best one for data transmission. This solution significantly reduces the training overhead for IS reflection design and it is also robust to user mobility requiring IS slow adaptation to wireless channels in a long term.

{Note that a BS-side IS is similar to a reflectarray as they both reflect signals from active antennas to attain the passive beamforming gain. 
Nonetheless, there are also differences between them. 
On one hand, compared to the reflectarray, the deployment of ISs is more flexible, which can be placed in the vicinity of the BS or integrated into the antenna radome of the BS. 
In contrast, a conventional reflectarray is usually placed at the backside of active antenna. 
On the other hand, reflectarrays are usually composed of a single planar surface mounted with reflecting elements. 
However, BS-side ISs can deploy multiple reflective surfaces with different positions and orientations around the BS \cite{Zhang2021Intelligent,huang2022empowering}, which can exploit their single and double reflection gains to further improve communication performance. 
As such, the conventional reflectarray can be regarded as a special case of the BS-side IS by deploying a single reflective surface at the backside of active antenna only.}

\subsubsection{Multi-IS Deployment}
\begin{figure}[t]
\centering
\includegraphics[width=8.5cm]{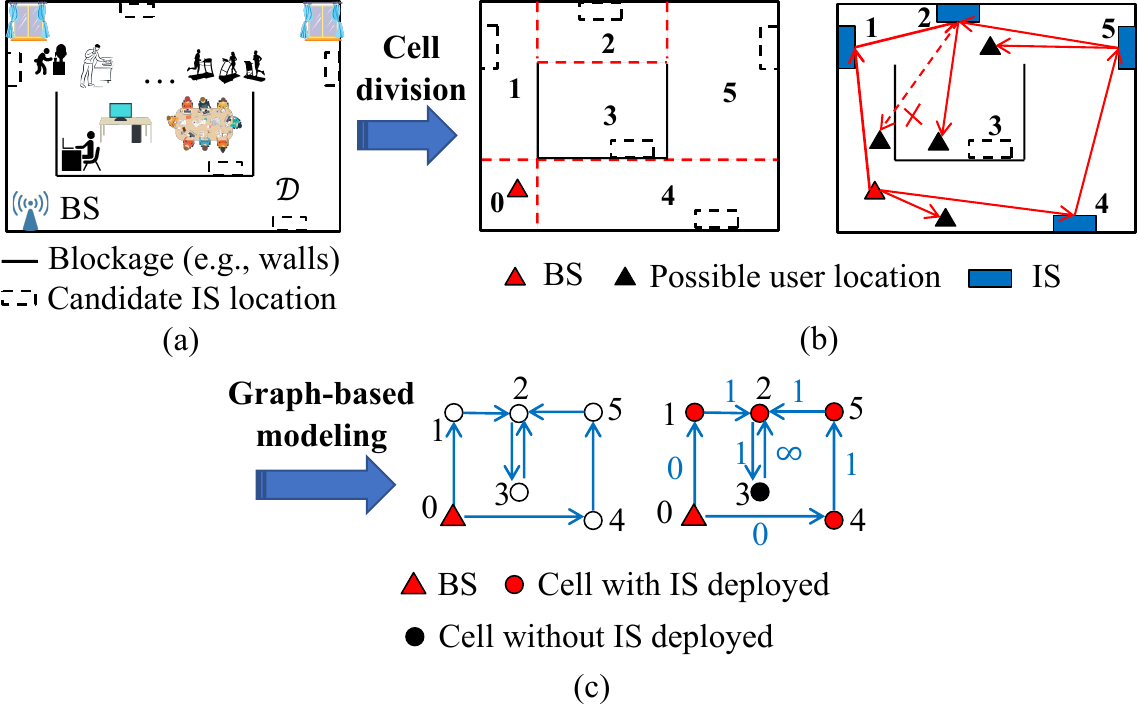}
\caption{(a) Multi-IS-reflection aided wireless network in a typical indoor environment, (b) illustration of the cell division for the region and LoS paths created by ISs deployed at selected candidate locations, and (c) an example of the graph-based model.}
\label{ISDeploy}
\vspace{-6pt}
\end{figure}
Despite the above appealing advantages of the BS-side IS, a blockage-free BS-user link via any BS-side IS may not be available in practice, since the reflection paths by BS-side ISs only may fail to bypass the dense and scattered obstacles in complex environments (e.g., an indoor environment shown in Fig. \ref{ISDeploy}(a)). To overcome this limitation, recent works\cite{mei2022-intelligent,Zheng2021Double,
mei2020cooperative,mei2021multi,mei2021distributed,mei2022split,r3_7,Zheng2021Efficient} have proposed a multi-IS-reflection aided wireless system, where multiple distributed ISs are deployed in the environment to create more available cascaded LoS paths between the BS and remote users, thereby significantly enhancing the wireless coverage, as shown in Fig. \ref{ISDeploy}(b). To achieve this purpose, how to properly deploy multiple ISs to cater to the practical environment is a critical problem of high importance.

To address this issue, the authors in \cite{mei2023joint} proposed a {\it graph-based} system modeling and optimization approach for multi-IS deployment. In particular, the network coverage area is first divided into multiple non-overlapping cells containing a set of predetermined candidate locations for deploying ISs, as shown in Fig. \ref{ISDeploy}(b). Then, based on the LoS path availability between any two candidate locations, a graph-based model was established to characterize the total deployment cost and communication performance of any IS deployment solution as shown in Fig.\ref{ISDeploy}(c), which also unveils an inherent trade-off between deployment cost and communication performance. Finally, optimal and suboptimal multi-IS deployment designs were proposed in \cite{mei2023joint} to reconcile this tradeoff.

However, the communication performance in \cite{r3_9} was only evaluated in terms of the number of signal reflections per multi-IS link, which may not accurately indicate the actual performance of each user. To tackle this issue, the authors in \cite{r3_9} proposed an enhanced graph-based system modeling approach which is able to characterize the SNR performance at any user location with a given multi-IS deployment solution. Moreover, to more effectively overcome the multiplicative path loss over the multi-reflection LoS links, the joint use of active and passive ISs was studied in \cite{r3_9}. The results in \cite{r3_9} revealed that incorporating active ISs is beneficial to reduce the total deployment cost for achieving a given SNR target in the target region by dramatically reducing the number of passive ISs required. 

\begin{figure}[t]
\centering
\includegraphics[width=0.4\textwidth]{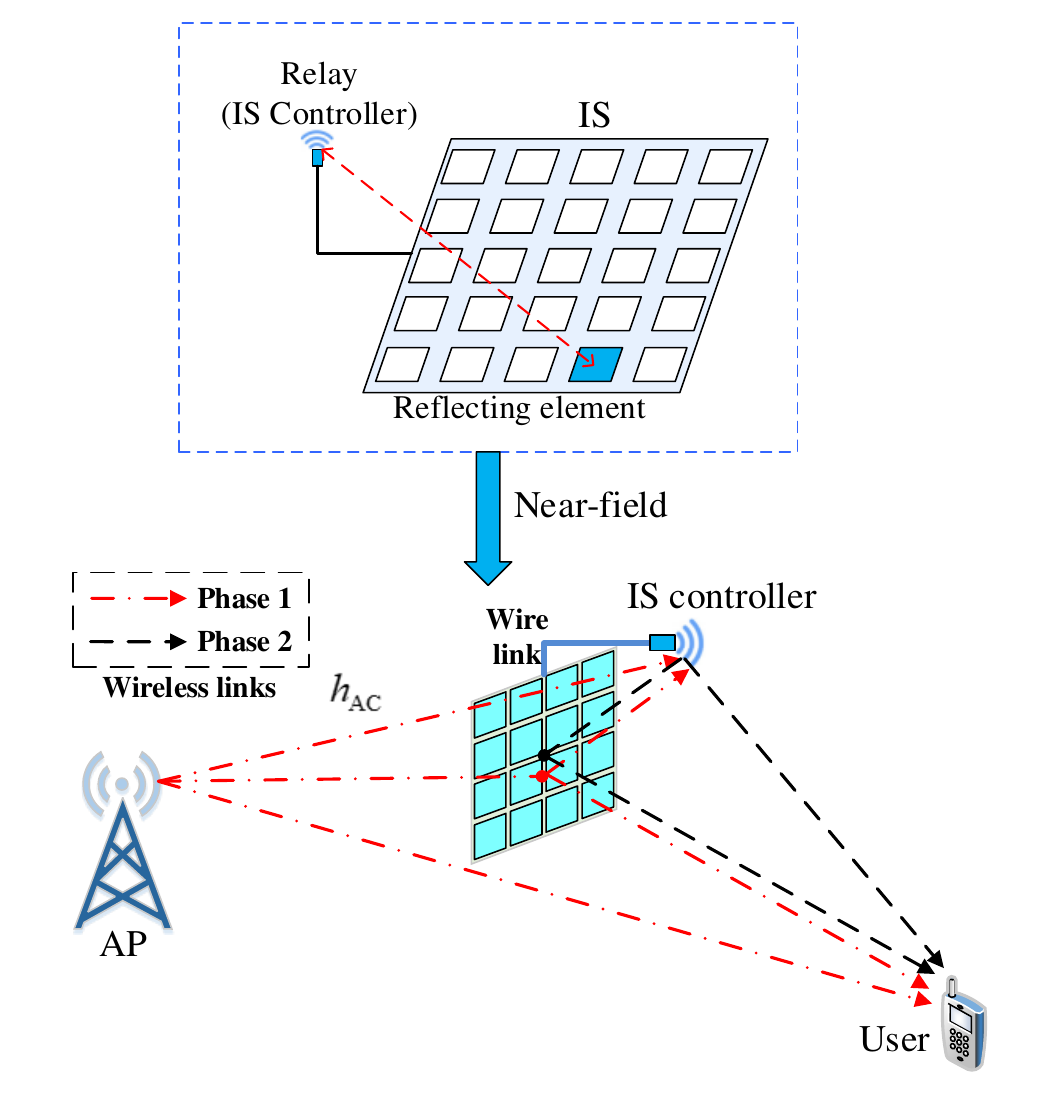}
\caption{A relaying IS-assisted communication system \cite{zheng2021-irs}.}
\label{RelayIRSsystem}
\end{figure}
\subsubsection{Relay-side ISs}
When deployed in wireless networks, passive IS and active relay differ in many aspects such as energy/spectral efficiency, hardware cost,  processing complexity, and coverage range \cite{Yildirim_2021,Basar_Yildirim_2021}.   
Specifically, passive IS can operate in full duplex and enhance the communication performance within its local coverage cost-effectively; while active relay typically operates in half duplex and is able to achieve broader coverage at the expense of more power consumption \cite{wu2019-intelligent,bjornson2020-intelligent,direnzo2020-reconfigurablea}. 
To harness their combined advantages, the relay-side IS deployment strategy has been proposed in some prior works. However, most of the existing works on relay-side IS simply add ISs to the relay system and deploy them separately \cite{abdullah2020-hybrid,abdullah2021-optimization,bie2022-deployment,wang2022-task,obeed2022-joint}, which thus incurs additional deployment cost and signaling overhead. 
In contrast, an alternative relay-side IS deployment scheme was proposed in \cite{zheng2021-irs} to integrate the active relay directly into the IS, by exploiting the new role of the IS controller as an active relay, as shown in Fig.~\ref{RelayIRSsystem}. Specifically, IS-integrated relay significantly enhances coverage performance compared to conventional ISs (with passive signal reflection/refraction only), by enabling the IS controller to actively relay information without additional deployment or hardware cost.
Moreover, similar to IS-integrated BS, IS-integrated relay also has a considerably shorter distance between IS REs and the active relay (see Fig.~\ref{RelayIRSsystem}), which yields several advantages such as reduced product-distance path loss (equivalently, increased effective channel gain), improved path diversity, expanded coverage range, and reduced signaling/information exchange overhead.

However, the channel estimation for the IS-integrated relay-aided communication system becomes more challenging than that for the conventional IS-aided or relay-aided communication systems due to the increased number of channel coefficients and the near-field channel response between IS elements and relay antennas. 
To address this issue, a practical channel estimation scheme was proposed in \cite{zheng2021-irs}, where the cascaded/direct CSI of the BS-relay link and the relay-user link is estimated in parallel at the BS and user, respectively, based on the pilot signals sent from the relay.
For practicality and to reduce training overhead, a codebook-based reflection design tailored for the IS-integrated relay may be considered, which needs to account for the mixed near- and far-field effect between relay/IS and BS/users (see Fig.~\ref{RelayIRSsystem}).
On the other hand, half-duplex relay is considered in \cite{zheng2021-irs} for ease of implementation, where two-phase relaying transmission is required and the IS can enhance both BS-relay and relay-user links by exploiting its time selectivity property over two phases. To boost transmission efficiency, the full-duplex relay could be considered for one-phase relaying transmission. However, this may introduce additional self-interference, necessitating sophisticated signal processing techniques to properly manage the interference, and the IS reflection would need to balance the transmission performance between BS-relay and relay-user links.

\subsubsection{Aerial ISs}
In addition to being deployed terrestrially, ISs can also be deployed or mounted on aerial platforms such as UAVs or balloons to achieve signal enhancement by providing additional LoS air-to-ground or air-to-air links, especially for post-disaster or scarcely covered areas \cite{qingqing2021JSAC-UAV,10266977,ning2023intelligent,You2021WC-IRS-UAV,m4}. 
To fulfill this application, the joint optimization of the aerial IS placement and the 3D passive beamforming was considered in \cite{HaiquanLu2021TWC,HaiquanLu2020ICC-AIRS}. 
Despite exploiting the additional design degree-of-freedom (DoF) provided by the vertical dimension, the enhanced performance resulting from aerial ISs is mainly constrained by the limited onboard energy and computational capabilities. 
To overcome these challenges, practical energy-efficient resource allocation and placement/trajectory optimization schemes have been proposed for designing aerial ISs \cite{ZhuohuiYao2021JSAC}, where DL-based algorithms exhibit potentials for solving non-convex optimization problems with coupled variables more efficiently as compared to conventional optimization algorithms \cite{BinDuo2023IOT}.
However, the signal transmitted on aerial links could suffer from high path loss, especially for high frequency signals under long-distance transmission.
To address this issue, aerial relaying architectures were studied in \cite{JAES2023SAG-IS} where UAV-mounted ISs relay the transmitted signals from aerial platforms at higher altitudes.

Furthermore, the application of IS to other types of airborne systems such as low earth orbit (LEO) satellite communication is an emerging direction of high interest. 
Specifically, LEO satellite communication systems suffer from the large Doppler effect due to the high mobility of satellites, while its unfavorable effects on signal quality can be effectively mitigated by ISs \cite{Basar2021Doppler,Toka2024arxiv}. 
In addition, the qualities of satellite-terrestrial and inter-satellite links can be improved by utilizing ISs.
For example, a new IS-aided satellite communication system with two-sided cooperative ISs has been proposed, where the communications between the LEO satellite and various ground nodes are enhanced by distributed ISs deployed near them \cite{Zheng2022Intelligent}. 
The authors of \cite{Zheng2022Intelligent} also proposed an efficient cooperative beamforming design and a practical transmission protocol to achieve distributed channel estimation and beam tracking, and demonstrated the substantial performance gain and great potential of the IS-aided satellite communication.
Nevertheless, to be more widely used in practice, the effects of aerial IS's mobility (e.g., trajectory drifts due to hardware imperfections) on system performance deserve further investigation.

\subsubsection{Roadside and Vehicle-side ISs}

Existing research on IS has primarily focused on static or quasi-static wireless channels, which may not be applicable to high-mobility scenarios with fast time-varying channels, such as vehicular communications.
In high-mobility communications, the transmitted signal often arrives at the receiver with rapidly time-varying phase shifts at different Doppler frequencies due to random scattering and high-speed motion, leading to superimposed multipath fast channel fading. 
Owing to its real-time control of signal reflection/refraction, IS provides a new means of converting wireless channels
from fast fading to slow fading in high-mobility scenarios, thus achieving more reliable
communications. Initial works on IS-aided high-mobility communications have considered deploying multiple ISs at the roadside to enhance the communication performance between the BS and high-mobility users moving on the road \cite{Huang2023Roadside,Sun2021Channel,Bora2023IRS-Assisted,Chao2023Channel}. 
However, this approach requires a large number of roadside ISs (due to the limited coverage of passive reflection/refraction of each roadside IS) to guarantee seamless coverage for high-mobility users moving on the road, and frequent handovers between the user/BS and nearby serving ISs are required, which may require sophisticated ISs' coordination and beam alignment/tracking.

\begin{table*}[!t]
\caption{Comparison of Roadside IS and Vehicle-side IS \cite{Huang2023Intelligent}}\label{table_comp}
\centering
\begin{tabular}{|c|cc|cc|c|}
    \hline
    \multirow{2}{*}{Deployment strategy} & \multicolumn{2}{c|}{Channel characteristics}     & \multicolumn{2}{c|}{Handover requirement}    & \multirow{2}{*}{Deployment cost} \\ \cline{2-5}
    & \multicolumn{1}{c|}{IS-BS channel} & User-IS channel & \multicolumn{1}{c|}{IS-BS handover} & User-IS handover &                         \\ \hline\hline
    Roadside IS & \multicolumn{1}{c|}{Quasi-static}  & Time-varying     & \multicolumn{1}{c|}{Frequent}        & Frequent          & Scales with road length \\ \hline
    Vehicle-side IS                     & \multicolumn{1}{c|}{Time-varying} & Quasi-static & \multicolumn{1}{c|}{Less Frequent} & No Need & Scales with vehicle number       \\ \hline
\end{tabular}
\end{table*}

An alternative strategy is to install a vehicle-side IS at each vehicle to constantly improve communication performance between the remote BS and the IS-aided user \cite{Huang2021Transforming}. Different from roadside IS deployment, vehicle-side IS deployment only requires one IS per vehicle. In this case, the vehicle-side IS can easily associate with the onboard users constantly and much less frequent
IS-BS handover is triggered. The comparisons between roadside and vehicle-side IS deployment are summarized in Table \ref{table_comp}.
Note that both roadside and vehicle-side ISs can be based on either signal reflection or refraction, depending on the deployment conditions. Moreover, to achieve high passive beamforming gain for both roadside and vehicle-side ISs, accurate CSI acquisition is essential yet practically challenging. This is due to the high-dimensional CSI associated with a large number of IS elements and the short channel coherence time due to high mobility. As such, instead of full channel estimation/acquisition,
it is practically appealing to perform beam alignment to efficiently track the fast time-varying cascaded user-IS-BS channel with low training overhead, which deserves further investigation.

{{\bf{Summary:}} In this subsection, we have reviewed the state-of-the-art IS deployment strategies, including BS-side ISs, multiple cooperative ISs, relay-side ISs, aerial ISs, roadside ISs and vehicle-side ISs, which are important for practical use and adaption to various wireless environments and application scenarios. 
From the perspective of network architecture, BS-side ISs can save the high cost of densely deploying user-side ISs and also provide better overall system performance. 
Apart from deploying ISs near BSs, the IS-integrated BS architecture can further decrease the product-distance path loss by integrating IS REs into the BS's antenna radome, while the increased channel estimation complexity can be addressed by codebook-based IS reflection design. 
In complicated wireless environment with dense obstacles and scatterers, multiple ISs can be deployed to establish cascaded virtual LoS links by utilizing graph-based system modeling and optimization methods to improve the communication performance. 
For further performance enhancement, ISs can be deployed close to or integrated into active relays to combine the advantages of both. 
Considering the vertical dimension of wireless networks, aerial ISs can provide additional design DoFs for coverage enhancement. 
To cope with the fast time-varying channels in high-mobility scenarios, ISs can be deployed along roadside or on/in vehicles to mitigate the unfavorable effects of fast fading. 
For more practical use of these IS deployment strategies, however, various issues such as cross-band interference, hardware imperfections and costly channel estimation should be further investigated.}

  \begin{figure*}[!t]
	\centering
	\includegraphics[width=0.9\textwidth]{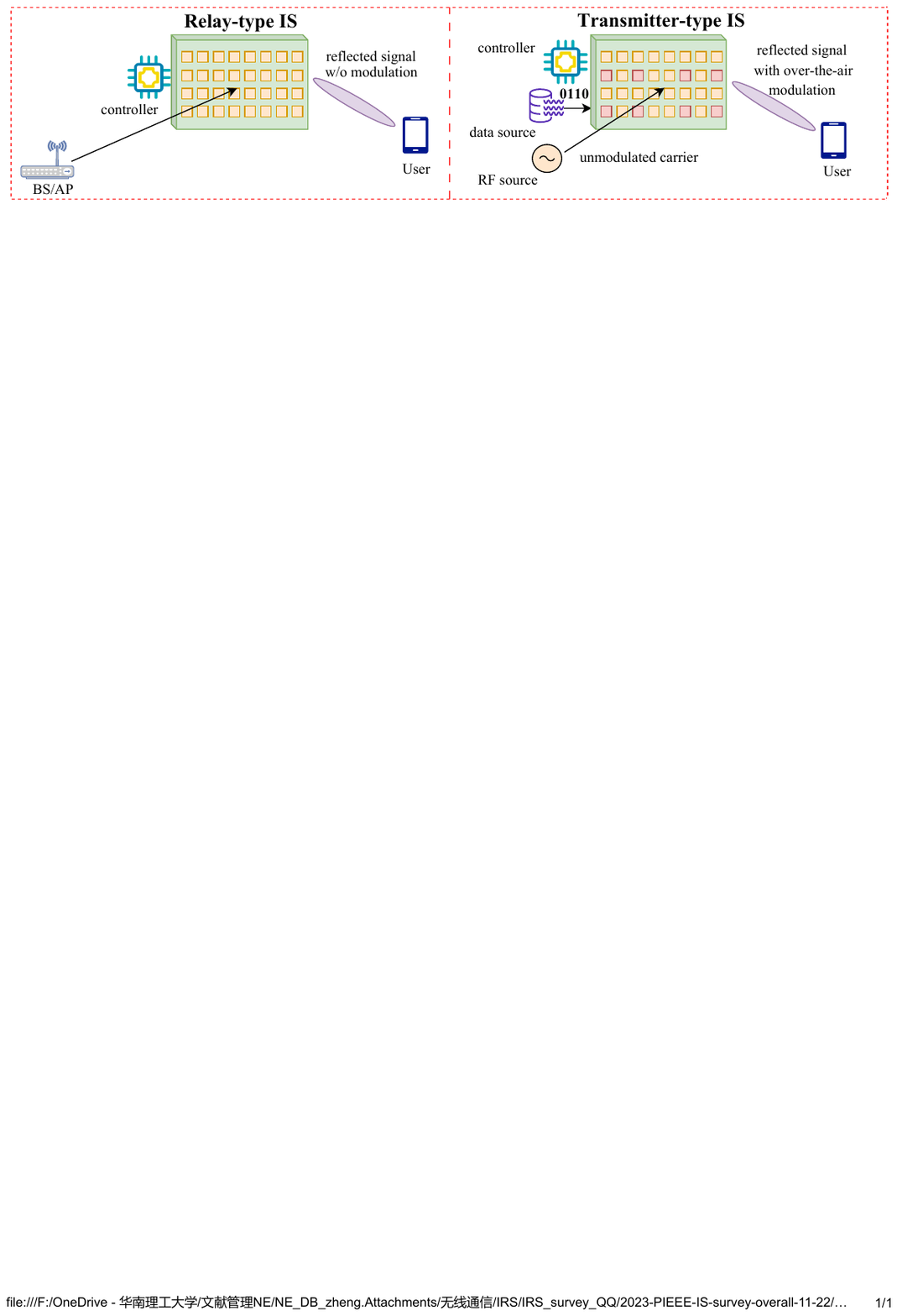}
	\caption{Comparison of transmitter-type and relay-type ISs.}
        \label{TransmitRelay-IS-Compare}
\end{figure*}

\subsection{Modulation}
    From a network functionality and information transmission perspective, we can classify ISs into two types depending on their roles in the network: relay-type IS and transmitter-type IS \cite{Basar_2021}. Specifically, an individual IS can function as either a relay or a transmitter, depending on its functionality and the specific use cases. Relay-type IS, as depicted in Fig.~\ref{TransmitRelay-IS-Compare} and discussed in the previous subsections, is strategically positioned between the source (e.g., BS) and the destination (e.g., users) to enhance communication performance between them, especially when the direct link is considerably weak or obstructed. 
    On the other hand, transmitter-type IS is an integral part of the transmitter, participating in its signaling and modulation processes, as illustrated in Fig. \ref{TransmitRelay-IS-Compare}. A key requirement is the capability for each IS element’s phase/amplitude change at the symbol level, which poses a critical challenge in hardware design and implementation. However, with typical symbol periods of current wireless systems like 5G/WiFi being in the order of microseconds, the switching speed of the IS element’s phase/amplitude can match the symbol transmission rate \cite{zhang2018space,tapio2021survey}. A practical approach for realizing transmitter-type IS is by connecting IS with its aided transmitter via a reliable wired link, thus enabling real-time control of IS reflection on a per-symbol basis.
    In this subsection, we focus on the transmitter-type IS and provide an overview of two types of transmitter-type IS-based modulation: amplitude/phase modulation (APM) and index modulation (IM).

    \subsubsection{IS for Amplitude/Phase Modulation}
    In a setup where the IS operates in transmitter mode, the incoming unmodulated carrier signal can be manipulated to encode information bits. 
    In simpler terms, feeding the IS with an unmodulated carrier allows it to generate virtual signal constellations for transmitting information.
    %(beyond the information from the BS or users). 
    By adjusting IS reflection amplitudes and phases appropriately, one can generate virtual APM constellations. In this case, virtual phase shift keying (PSK) constellations can be more easily created by altering outgoing signal phases. 
    
    The application of transmitter-type IS for APM offers two significant advantages. First, the hardware architecture of an IS-based transmitter can be simpler compared to traditional setups with upconverters and filters. Generating an unmodulated cosine carrier signal is straightforward, typically using an RF digital-to-analog converter with internal memory and a power amplifier. Second, the proximity of the IS to the signal source mitigates the multiplicative path loss effect. It is even feasible to position the IS in the transmitter's near-field to further enhance received signal power, provided near-field effects are carefully considered. In the following, we briefly overview different APM schemes based on transmitter-type IS in the existing literature.
    
    The early work of \cite{Basar_2019_LIS} envisioned using the IS as an RF chain-free BS/AP by creating virtual PSK constellations over-the-air to perform data transmission. This study also provided a theoretical framework to analyze the bit error probability of the underlying system. Independently, the authors of \cite{Tang_2019} developed an RF chain-free transmitter that employs metasurfaces to enable over-the-air 8-PSK modulation. This innovative design simplifies hardware, improves signal quality, and enhances the efficiency of wireless communication systems. In \cite{Dai_2020_2}, another strategy was considered to implement different modulation schemes, including quadrature amplitude modulation (QAM) and PSK, thanks to the use of harmonic frequencies. Specifically, IS has shown excellent capability to manipulate EM waves in different domains, such as space, time, and frequency, and this property is exploited for information modulation. Using this flexibility, the study of \cite{Zhao_2018} implemented an IS-based frequency shift keying (FSK) scheme. Interested readers can refer to \cite{Dai_2021} and references therein for detailed description of information metasurface-enabled transmitter architectures.
    
    Another distinctive attribute of transmitter-type IS is its ability to function as a virtual MIMO terminal. In this setup, an IS exposed to an unmodulated carrier signal can be configured to act as a virtual MIMO transmitter, completely devoid of RF chains. Specifically, in \cite{Khaleel_2021}, considering IS elements partitioning, two novel IS-based MIMO solutions were proposed. First, dividing the IS into two subsurfaces and using the PSK modulation over two time slots, an IS-assisted Alamouti’s scheme \cite{Alamouti} is realized with the IS acting as the transmitter. Second, an IS is used to enhance the performance of the nulling and canceling-based suboptimal detection procedure in a spatial multiplexing-based MIMO system and additional information bits are also transmitted by the phases of the IS REs to boost spectral efficiency.
    Later, in \cite{Zheng2023Simultaneous}, a novel space-time code based on PSK modulation was designed for the single-antenna transmitter and IS, collectively acting as a virtual two-antenna transmitter to achieve transmit diversity without the need for any CSI.
     Moreover, it is demonstrated in \cite{Zheng2023Simultaneous} that in addition to transmit diversity, IS can also achieve passive beamforming simultaneously, thus fulfilling both transmitter-type and relay-type functions at the same time.
     In \cite{Tang_2020}, utilizing multiple digital-to-analog converters, a MIMO system that can transmit multiple streams was realized, and a $2\times 2$ MIMO system was experimentally tested. It was shown that two independent $16$-QAM data streams can be successfully received by a traditional receiver at a carrier frequency of $4.25$ GHz. The above studies paved the way for designing more advanced IS-based transmitters subsequently. Specifically, in \cite{Chen_2021}, a dual-polarized IS architecture was created to transmit two data streams simultaneously at a transmission rate of $ 20 $ Mbps.
     
     On the other hand, the principle of transmitter-type IS is also similarly applied in backscatter communication, where a backscatter device (BD) transmits messages to a backscatter receiver (BR) by reflecting an incident signal from an external RF emitter.
The adaptability of IS allows for its effective deployment in backscatter communication networks. Specifically, the fundamental principle of IS-aided backscatter communication was introduced in \cite{Liang2022Backscatter}, with various types of modulation considered. 
The authors of \cite{Basharat2022Reconfigurable} further analyzed cooperative IS and backscatter communication performance with different amplitude/phase modulations. 
The BER performance of IS-aided backscatter communication networks was examined in \cite{Chen2021Performance}, demonstrating the effectiveness of ISs in improving the channel conditions of both source-reader and source-tag links. 
The closed-form achievable rate expressions for IS-aided backscatter communication networks were provided in \cite{Khan2023Capacity}. 
Further investigation was made in \cite{Loku2023RIS}, where the achievable rate of IS-aided backscatter communication networks adopting $M$-PSK was maximized by jointly optimizing the phase shifts of IS. 
Unlike the aforementioned literature where an IS is used solely for link enhancement, \cite{Basar_2019_LIS} explored the function of an IS as an AP with an unmodulated carrier, showing that ISs can also perform backscatter communication. 
Building on this result, \cite{Yang2024Segmented} proposed dividing an IS into two parts, one for transmitting the primary data signal and the other for transmitting the backscatter signal.

    \subsubsection{IS for Index Modulation}
    IM, which leverages the indices of available transmit entities to encode digital information \cite{Basar_2017}, has also motivated another new line of research on transmitter-type IS-based communication systems. In the early work of \cite{Basar_2019_LIS_2}, an IS illuminated by an unmodulated carrier was utilized to perform IM by directing the signal to a terminal's receive antennas. This study also introduced the concept of IS-based IM as a potential solution beyond MIMO by considering IM at different parts of the system. In the following studies of \cite{Canbilen_2020,Canbilen_2022}, the performance of a space shift keying scheme was investigated in the presence of an IS.
    
    The concept of IS-based IM was further advanced in \cite{Guo_2020}, where a promising IM scheme was proposed for transmitter-type IS. Depending on joint or independent data mapping of the transmitter and the IS to deliver information, IM is further divided into two categories: jointly mapped and separately mapped IM. To enhance transmission reliability, the authors of \cite{Guo_2020} proposed a discrete optimization-based joint signal mapping, shaping, and reflecting design with a given transmit signal candidate set and a given reflecting pattern candidate set. By combining reflect beamforming and IM, a subset of the IS REs are activated to reflect a sharp beam towards the intended destination while exploiting the index combination of the on-state IS elements to implicitly convey additional information \cite{Lin_2021}. Given the potentially large number of REs at the IS, this leads to significant computational and storage overhead for mapping information bits to on-state elements. 
     The authors thus adopted a grouping strategy to solve this issue by dividing the IS into subgroups for activation. 
     To further enhance reflection power efficiency, the authors of \cite{Lin2022Reconfigurable} proposed quadrature IM for transmitter-type IS.  In this approach, the IS elements are partitioned into two subsets to reflect incident signals towards two orthogonal spaces, thus achieving simultaneous passive beamforming and quadrature IM-based information transfer. 
     Using a similar IS subgrouping strategy but considering a hybrid IS composed of both active (amplifying) and passive reflecting elements, hybrid IM was introduced in \cite{Yigit_2023}. In this scheme, an over-the-air IS modulation scheme is considered by turning active and passive IS subgroups on and off to create a virtual amplitude shift keying (ASK) constellation. The phase modulation concept has also been combined with space shift keying, where the IS embeds its own information by adjusting its phase shifts \cite{Singh_2022}.

    {\bf{Summary:}} In this subsection, we have mainly reviewed the works on transmitter-type IRS, which serves a unique role in network functionality. 
    A transmitter-type IS can simplify hardware architecture and enhance signal quality by participating in signaling and modulation processes, which are generally based on two types of modulation: APM and IM. 
    Specifically, ISs can implement various modulation schemes (e.g., PSK, QAM, and FSK) and act as virtual MIMO terminals, thus enabling transmit diversity. 
    Through IM techniques, ISs can selectively activate/deactivate reflecting elements to convey additional information. 
    Despite these promising recent advances in IS-based modulation designs, critical challenges remain to be tackled for fully unlocking their potential for future wireless networks. 
    As the changing speed of each IS element’s phase/amplitude at the symbol level poses stringent requirements, the resulted low data rate of reflection modulation is another practical issue to address in future work.

\begin{figure*}[!t]
   \centering
   \includegraphics[width=1.0\textwidth]{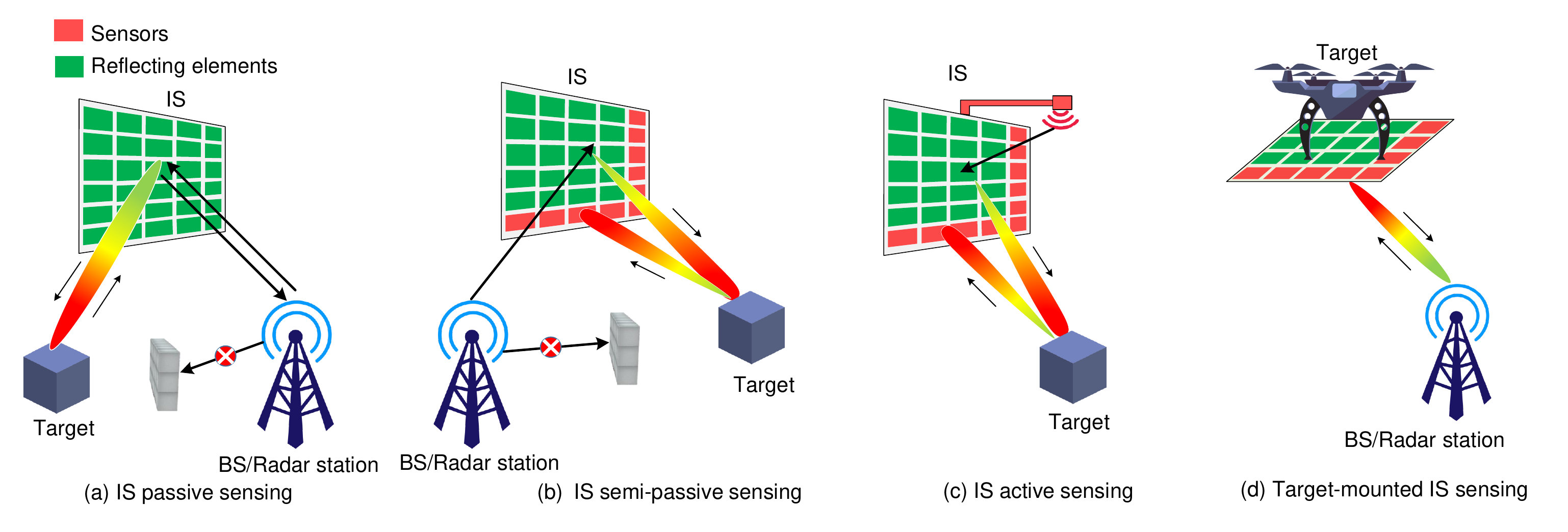}
   \caption{Schematics for four IS-aided wireless sensing architectures.}
   \label{sensing_arc}
   \end{figure*}

\subsection{Sensing and ISAC}

ISAC has been regarded as a promising technique for the 6G wireless network, where the sensing function and the communication function can be carried out
simultaneously by sharing the same spectrum and hardware
facilities \cite{conde,cral,9858656}. However, ISAC systems still face practical challenges. For example, sensing coverage is significantly smaller than communication coverage due to severe path loss in the BS-IS-target-IS-BS cascaded echo link. Although
the sensing coverage can be guaranteed by deploying more BSs, this method incurs high hardware cost and
energy consumption.
On the other hand, hardware sharing of communication and sensing transceivers increases the design and implementation complexity, while
spectrum sharing of communication and sensing signals results in interference and performance trade-off. Fortunately, the use of IS can achieve higher spectral/energy efficiency for wireless communication\cite{IRSMA, IRSCH}, as well as higher accuracy for wireless sensing, thus promising to enable ISAC for 6G \cite{imag,qingmulti,9769997}. Specifically, the IS can establish LoS channels to enhance both sensing/communication coverage, provide flexible passive beamforming to balance their performance trade-off, and reduce hardware cost and energy consumption compared to dense deployment of BSs \cite{10226306,10284917}. 
Also, equipping ISs with sensing capabilities enables an easier deployment of the ISs \cite{m5} and energy harvesting capabilities \cite{m6}.

\subsubsection{IS-enhanced Wireless Sensing}
Promising new system architectures are essential to overcome the sensing range bottleneck for wireless sensing systems. IS passive sensing is shown in Fig. \ref{sensing_arc}(a), where the sensing performance is improved
by establishing a virtual LoS link from the BS to the IS and then to the target \cite{xujiej,peng2024semi}.
However, the practical performance of IS passive sensing is limited by the severe path loss of the cascaded echo link. To address this issue, IS semi-passive sensing was proposed in \cite{activesM_shao} (see Fig. \ref{sensing_arc}(b)), where
dedicated low-cost active sensors are installed on the IS (i.e., semi-passive IS), allowing it not only to reflect signals from the BS but also to directly receive echo signals from the target for sensing. As such, IS semi-passive sensing incurs lower path loss compared with IS passive sensing. Nevertheless, if the BS
is far from the IS, the sensing accuracy of IS semi-passive
sensing remains limited. To overcome this issue, a new type of IS-aided sensing architecture, called  {\it IS active sensing}, has recently been proposed  \cite{activesens_shao} (see Fig. \ref{sensing_arc}(c)).
Specifically, in IS active sensing, the IS controller, typically responsible for reflection control and information exchange only, also acts as a transmitter to send probing signals for sensing. This allows the IS to autonomously send (via the IS controller) and receive (via sensors) sensing signals for target localization, thus avoiding the dependence on BSs for sensing signal transmission/reception. 
Although IS active sensing incurs higher hardware and energy cost than IS passive/semi-passive sensing, it achieves superior sensing performance due to the short signal propagation distance and also helps reduce the BS transmit power for sensing. 

For conventional IS-aided wireless sensing systems, the IS is deployed in the environment as an anchor node, and its sensing performance relies crucially on the strength of the reflected echo signal from the target to the receiver for detection/estimation. However, this approach may be practically inefficient in achieving reliable sensing when the radar cross section (RCS) of the target is limited. Recently, an interesting new approach called {\it target-mounted IS} was proposed in \cite{securesens_shao, tam_wang, targetmount_shao, controsens_shao}, where IS is mounted on the target (see Fig. \ref{sensing_arc}(d)) to provide enhanced/secure sensing by leveraging IS's controllable signal reflection and high spatial resolution thanks to its large aperture. For example, in \cite{tam_wang}, the authors considered mounting fully passive IS on the sensing target, thereby estimating the IS's location and orientation as that of the target via a tensor-based algorithm. Due to the improved target RCS and controllable signal reflection, this approach can significantly enhance the 6D sensing performance of the target (i.e., 3D position and 3D orientation) even when the number of receivers/receiver antennas is small.

Besides enhancing sensing accuracy, target-mounted IS has also been proposed to improve wireless sensing security. Specifically, the secure sensing aims to enhance sensing performance
of the legitimate radar station (LRS) while preventing the target
detection by any unauthorized radar station (URS). To achieve this goal, the authors in \cite{securesens_shao} proposed a general secure sensing protocol with target-mounted semi-passive IS, where each radar coherent-processing
interval is divided into two phases. In the first
phase, the IS sensors estimate the LRS/URS channel and waveform
parameters with all IS REs switched
off. After that, in the second phase, based on the estimated
parameters, IS reflection is designed to simultaneously
boost the received signal at the LRS receiver and suppress
that at the URS receiver. Benefiting from the flexible passive beamforming gain, such a new design with target-mounted IS can offer significant secure sensing performance gains as compared to
the conventional sensing without target-mounted IS. 
Moreover, target-mounted IS also gives rise to a novel application in electromagnetic stealth technology for evading radar detection \cite{zheng2023intelligent,Xiong2024Stealth},  which offers a flexible and adaptable solution to overcome the limitations of traditional stealth materials. In \cite{zheng2023intelligent,Xiong2024Stealth}, the authors proposed an optimization problem for designing the IRS’s reflection at the target to minimize the sum of received signal power over all adversary radars to achieve electromagnetic stealth. Low-complexity closed-form solutions based on reverse alignment/cancellation and minimum mean square error (MMSE) criteria were proposed in \cite{zheng2023intelligent} for the single-radar and multi-radar cases, respectively. This new application of target-mounted IS opens up new research avenues in anti-radar detection, providing a promising direction for future exploration in this field.

\begin{figure}[!t]
   \centering
   \includegraphics[width=0.48\textwidth]{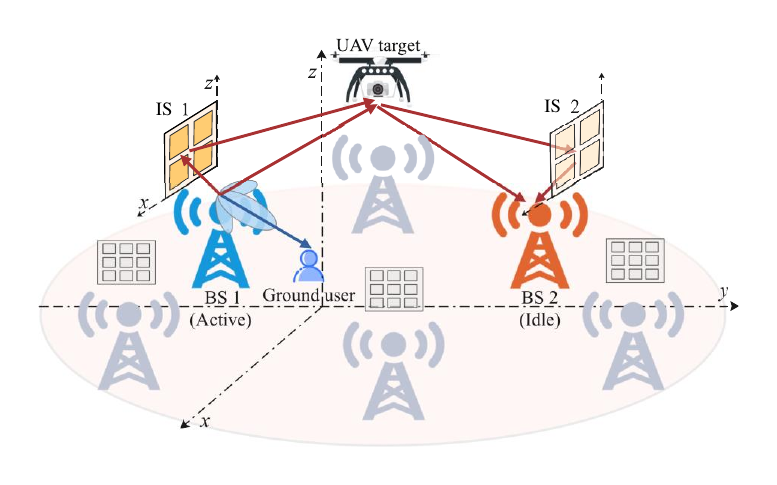}
   \caption{IS-aided ISAC for cellular-connected UAVs.}
   \label{UAV_ISAC}
\end{figure}
\subsubsection{IS-aided ISAC}
How to design IS-assisted ISAC systems to achieve optimal sensing and communication performance without significantly changing existing protocols in cellular systems is a key challenge \cite{meng2023intelligent,9913311,10143420}. To overcome this challenge, the authors in \cite{uavisac_pang} proposed to exploit cooperative ISs' passive beam searching and the communication signal from the BS (e.g., pilot signal sent for channel estimation) for cellular UAV target sensing. The associated sensing protocol is shown in Fig. \ref{UAV_ISAC}, where each channel
coherence interval is divided into three phases. In the first
phase, background channel is estimated without IS reflection. After that, in the second phase, the angle parameters between IS 1 and UAV are estimated with IS 2 OFF. Finally, in the
third phase, the angle parameters between IS 2 and UAV are estimated with IS 1's beam fixed as the best one obtained in the second phase. Note that by properly selecting the idle BS (i.e., without communicating with any user) and its sensing resource block, the communication interference from other active BSs (i.e., communicating with a user) to the sensing signal at idle BS can be ignored thanks to the inter-cell interference coordination (ICIC). The authors demonstrated that this cellular ISAC scheme not only requires no modifications to the existing BS beamforming design and transmission protocol, but also does not impact the communication performance of users in the existing network, thus achieving a superior performance tradeoff between sensing and communication. 

The IS is also particularly advantageous in 
mmWave ISAC to compensate for the high path loss, thus greatly improving the sensing
coverage and communications performance \cite{chen2023joint,10464353}. In mmWave communication,
beam scanning/training is widely applied\cite{gaommw} and can be exploited for simultaneous target sensing. In \cite{beamscan_li}, the authors considered a mmWave ISAC system where a multiple-antenna BS aims to communicate with a single-antenna user, and also to detect the angle of a sensing target from the semi-passive IS. They proposed a two-phase simultaneous beam training and sensing (STAS) protocol to enhance local communication and sensing performance at the same time via IS passive beam scanning. Specifically, in the first phase, the ISAC system finds the best beam for communication user and concurrently detects the target based on its echo signal received by IS sensors. Next, in the second phase, data are sent to the communication user with the best IS beam found in the first phase. Due to the effective utilization of the downlink beam training signal for simultaneous sensing, the STAS protocol outperforms the benchmark protocol with orthogonal beam training and sensing.

\begin{figure}[!t]
   \centering
   \includegraphics[width=0.43\textwidth]{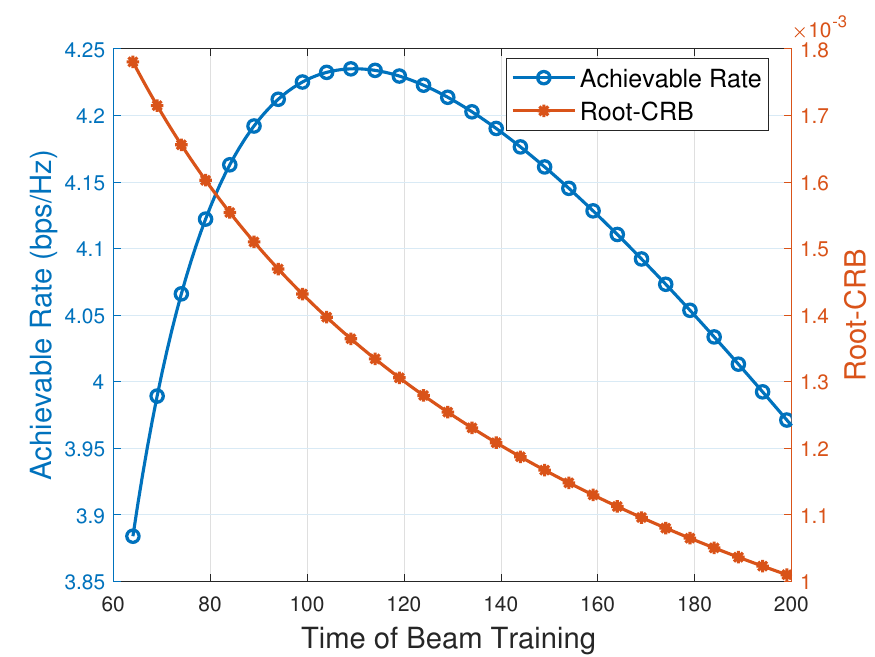}
   \caption{Achievable rate and RCRB versus
beam training time.}
   \label{isac_crb}
\end{figure}

To show the fundamental trade-off in the above
IS-aided mmWave ISAC, Fig. \ref{isac_crb} shows the achievable rate for communication and root CRB (RCRB) for sensing versus the IS beam training time with the STAS scheme \cite{beamscan_li}. The number of IS REs and sensors are set as 64 and 8, respectively, and the carrier frequency is 28 GHz.
As IS beam training time increases, IS beamforming gain increases at the expense of reduced data transmission time, which results in an initial increase and subsequent decrease in the achievable rate for the communication user. In contrast, sensing RCRB monotonically decreases with beam training time. Thus, beam training time needs to be properly set to balance the communication-sensing performance trade-off.

\begin{table*}[!t]
      \centering
      \begin{threeparttable}
      \renewcommand{\arraystretch}{1.2}
      {
      {\caption{A Summary of Representative Works on IS-Aided Sensing/ISAC Systems.}
      \small
      {\begin{tabular}{|m{1.2cm}|m{2.2cm}|m{1.9cm}|m{1.8cm}|m{1.85cm}
      |m{3.96cm}|m{2.4cm}|}\hline

      {\bf  Reference}  &{\bf Applications} &{\bf Link status}
       & {\bf  Type of IS} &{\bf Role of IS } & {\bf  Design objectives} & {\bf  Algorithms} \\ \hline
       Sensing \cite{activesM_shao}, \cite{activesens_shao}         & Target DoA estimation   & Only LoS      &Semi-passive IS   &    Sensing   & Improve the MSE of sensing    & MUSIC   \\ \hline

   Sensing  \cite{securesens_shao, controsens_shao}
          & Secure sensing   & Only LoS      &Semi-passive IS   &   Sensing and sensing security & Maximize
          IS reflected signal power to legitimate radar, while limiting it
          to unauthorized radar   & PDD-based algorithm   \\ \hline

       Sensing  \cite{zheng2023intelligent, Xiong2024Stealth}
          & Electromagnetic stealth    & Only LoS      &Semi-passive IS   &   Stealth and sensing security & Optimize the IS’s reflection at the target to minimize the received signal power of adversary radars
   & Reverse cancellation/MMSE and Lagrange multiplier \\ \hline
   
          Sensing  \cite{founbuz}
          & Far-field and near-field target detection   &  LoS and NLoS      &Passive IS   &   Sensing & Maximization of the probability of
detection under a fixed probability
of false alarm   & Forward-backward optimization via alternating
          maximization   \\ \hline

         Sensing  \cite{metahu}
          & 3D object detection   &  Only LoS       &Passive IS   &   Sensing & Minimizing cross-entropy loss of the beampattern  & Deep reinforcement learning   \\ \hline

        ISAC  \cite{Jointliu}
          &  DFRC with Rayleigh fading  & Communication LoS, target LoS      &Passive IS   &   Sensing and communication & Maximize the radar output
          SINR while satisfying the
          communication QoS
          requirement   & ADMM and MM  \\ \hline

          ISAC  \cite{deepliu}
          & DFRC with Rician fading   & Communication NLoS, target LoS      &Semi-passive IS   &  Communication & Improve the MSE of sensing and communication
          channels   & Train with the Adam optimizer and MSE loss function   \\ \hline

          ISAC  \cite{securecwu}
          & Secure ISAC    & Communication NLoS, target LoS      &Passive IS   &   Sensing, communication, and information security & Maximize the
            sensing beampattern gain while guaranteeing communication and information security requirements
          & Lagrange duality and MM   \\ \hline

          ISAC  \cite{widewei}
          & Wideband DFRC  & Communication NLoS, target LoS      &Passive IS   &   Sensing and communication & Maximize the average SINR of radar and the minimal communication
            SINR among all users  & Dinkelbach's method and ADMM   \\ \hline

               ISAC  \cite{beamscan_li}
          & Millimeter-Wave DFRC  & Communication LoS, target LoS      &Semi-passive IS   &   Sensing and communication & Use downlink beam scanning for simultaneous beam training and target sensing  & ISAC protocol design \\ \hline
      \end{tabular} } } }
      \begin{tablenotes}
        \footnotesize
        \item[1] PDD: penalty dual decomposition; MM: majorization-minimization; ADMM: alternating direction method of multipliers; SINR: signal-to-interference-plus-noise-ratio; MUSIC: multiple signal classification; CRB: Cramer-Rao bound.
      \end{tablenotes}
    \end{threeparttable}
      \end{table*}

%\subsubsection{Promising Directions for Future Work}
{\bf{Summary:} In this subsection, we have discussed IS-aided sensing schemes based on four approaches, namely, IS passive sensing, IS semi-passive sensing, IS active sensing, and target-mounted IS sensing. We compared their pros and cons in terms of performance, hardware cost, and implementation complexity. In addition, we discussed IS-aided ISAC for cellular-connected UAV and mmWave systems.}
In Table IV,
we summarize the up-to-date research works on
exploring IS-aided sensing and ISAC.
To motivate future research on IS-aided ISAC, several important directions are briefly discussed as follows which still need in-depth investigation. First, it is necessary to explore the connection between the IS-aided sensing and communication channels to enhance ISAC performance. Furthermore, new IS architectures, such as target-mounted IS\cite{targetmount_shao}, STAR-RIS \cite{stars}, and active IS\cite{r3_11} (to be discussed in the next section in detail),
open new opportunities for ISAC performance improvement.

\begin{figure*}[!t]
   \centering
   \includegraphics[width=0.8\textwidth]{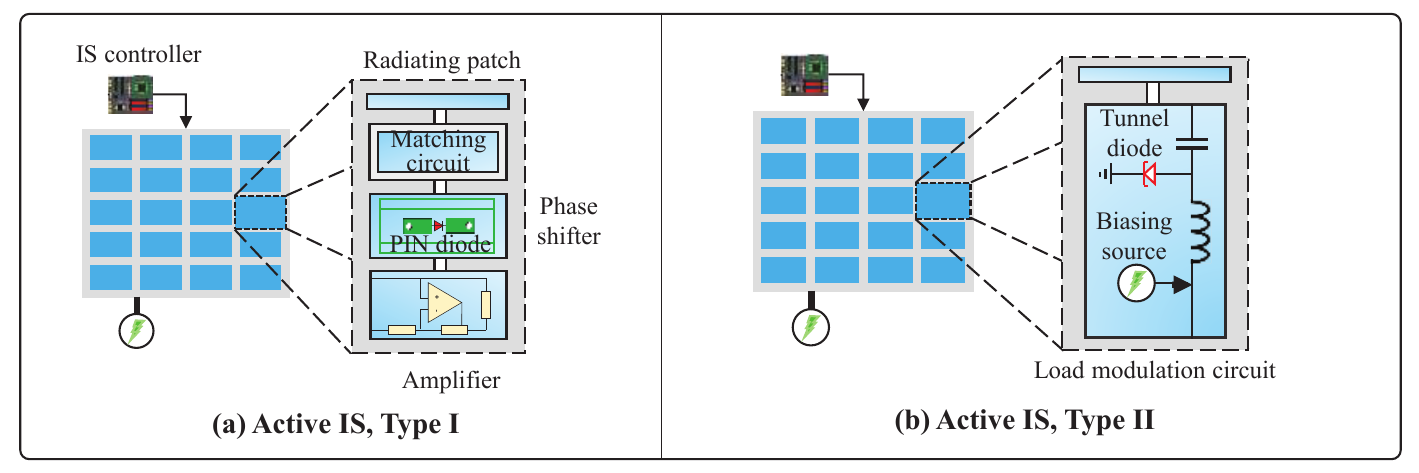}
   \caption{Two types of active IS hardware architectures.}
   \label{FigCSY-1}
\end{figure*}

\section{Advanced IS Architectures: Recent Progress and New Challenges}

Besides the passive reflection-based ISs discussed in Section II, several new IS architectures have been proposed to overcome their limitations and enable broader applications with improved performance, including active IS, omnidirectional IS, and holographic IS,  elaborated as follows.

\subsection{Active IS and Integrated Active/Passive IS}
\subsubsection{Active IS}
The communication performance of passive-IS-assisted  systems is practically constrained by the severe product-distance path-loss in signal reflections, which can be compensated for by equipping ISs with a massive number of passive reflecting elements or reduced by deploying the passive ISs closer to the transceivers. 
However, these methods may not be applicable in practice, due to the high design complexity and practical site  constraints. 
To address this issue, an efficient approach is by deploying a new type of IS for assisting wireless communication, called \emph{active} IS, which enables the signal reflection with \emph{power amplification} for effectively compensating for the product-distance path-loss \cite{zhang2022active,r3_1,9896755}. Generally speaking, active IS can be implemented by two types of hardware architectures as illustrated in Fig. \ref{FigCSY-1}, namely, 1) the cascaded amplifying and phase-shifting (APS) circuits (see Fig. \ref{FigCSY-1} (a)) \cite{zhang2022active} and 2) load modulation circuits (e.g., tunnel diodes) (see Fig. \ref{FigCSY-1} (b)) \cite{r3_1}. With the power amplification capability, active IS can be used for not only  improving the communication performance, but also enhancing the performance of other  wireless applications, such as wireless power transfer, physical layer security \cite{lv2022ris,9716895,9947328,zargari2022multiuser,10061167,9946415,10113784,10177727}.  
In this subsection, we provide an overview of  recent research results  on active IS from the communication design perspective, including active-IS reflection optimization, channel estimation, and deployment.

First, the active-IS reflection design is more complex than the passive-IS case, since it needs to jointly optimize the active-IS reflection amplitude and phase. Moreover, compared with passive IS, there is a new design tradeoff for active IS between maximizing the received signal power and minimizing the  amplification noise and residual self-interference power. 
Although the above issues usually render non-convex constraints over IS reflection coefficients and hence non-convex optimization problems, various methods have been proposed in the  literature to optimize the active-IS communication performance from different aspects. For example, the authors in \cite{r3_1} proposed an alternating optimization (AO)-based algorithm to maximize the SNR for an active-IS-aided SIMO system. In addition, the authors in \cite{r3_2} investigated a multi-user MISO system, where a joint
transmit beamforming and reflect precoding scheme was proposed to maximize the sum-rate by using fractional programming techniques.
Moreover, instead of the fully-connected active IS, a sub-connected active-IS architecture was proposed in \cite{sub_1} to reduce the power consumption of an active-IS aided multi-user MISO system, where the active IS elements are grouped into multiple sub-surfaces, each consisting of one power amplifier. This new architecture offers a flexible tradeoff between minimizing the power consumption and maximizing the communication rate for active IS. 
%Unlike \cite{sub_1} that simply assumed independent signal amplification at sub-surface elements, the authors in  \cite{sub_2} further improved the accuracy of the signal model for sub-connected active ISs, by taking into account the combination and re-distribution of incident signals on the elements of each IS sub-surface powered by the same amplifier. 
Furthermore, for wide-band active-IS systems, the authors in \cite{act_wide} 
proposed a novel hardware architecture for OFDM active-IS systems to address its  frequency non-selectivity issue. Specifically, each active-IS element/sub-surface is equipped with multiple amplifying and phase-shift circuits operating at different frequencies. Then the subcarriers are partitioned into several groups,  each configured by one amplifying and phase-shift circuit. 

Second, for active-IS channel estimation, the 
 conventional methods  designed for passive ISs cannot be directly applied, since the active-IS reflection design generally requires the CSI of the separate transmitter-IS and IS-receiver links (instead of the cascaded CSI for passive IS) due to the new consideration of amplification noise. 
Moreover, the amplification factors need to be properly chosen in the active-IS channel estimation for satisfying the maximum amplification power constrain.
%t, which, however, is determined by the unknown CSI of the transmitter-IS link. 
To address the issue in estimating the BS-IS channel, the authors in \cite{r3_4} proposed two practical channel estimation schemes with and without IS-mounted sensors, respectively.
%Specifically, with low-cost sensors mounted, active IS can estimate the BS$\to$IS channel based on the sensors' received signals, by exploiting the strong channel correlation between the sensors and IS elements. On the other hand, for active IS without sensors, the BS can estimate the BS-IS channel based on the reflected signal over the BS-IS-BS link, by leveraging  the wireless sensing capability of the BS. 
In addition, by assuming the BS-IS link CSI  known \emph{a priori} based on the fixed BS and IS locations, the authors in \cite{r3_3} proposed a new training reflection pattern for  active IS, by minimizing the estimation error variance of the IS-user channel in the presence of active-IS-induced noise.

Last, for the active-IS deployment design, several factors need to be considered, such as the physical environment, number of ISs and their REs, uplink or downlink communication, amplification power budget, etc. Specifically, for the single-active-IS aided communication system, it was shown in \cite{r3_11} that the active-IS should be deployed closer to the receiver to compensate for the amplification noise in the downlink/uplink. 
%However, for joint uplink and downlink communications, it needs to be placed between the transmitter and receiver to balance the uplink and downlink performance.
%To further boost the communication performance, the authors in \cite{r3_12,r3_13} considered the double-active-IS aided wireless systems and jointly optimized IS deployment with reflection optimization and elements allocation for maximizing the achievable rate over the double-reflection link under the \emph{per-element} and \emph{total} amplification power constraints, respectively. Specifically, under the practical per-element power constraint,  an AO-based algorithm was proposed in \cite{r3_12} to optimize the BS beamforming and the active-IS reflections, elements allocation, and placement  for maximizing the achievable rate.
 It was revealed in \cite{r3_12} that given the fixed per-element amplification power, the received SNR at the user increases asymptotically with the \emph{square} of the number of REs, while 
%; while given the fixed number of reflecting elements, the SNR does not increase with the per-element amplification power when it is asymptotically large  due to the linearly-increasing amplification noise power.
%   On the other hand, 
%    the authors in \cite{r3_13} shown that by optimizing the active-IS placement and elements allocation, 
the double-active-IS aided wireless system only achieves a \emph{linear} capacity scaling order with the number of active REs under the total amplification power constraint, due to the noise amplification over the double-reflection link \cite{li2023double}.

\subsubsection{Integrated Active/Passive IS}
\begin{figure*}[!t]
   \centering
   \includegraphics[width=0.8\textwidth]{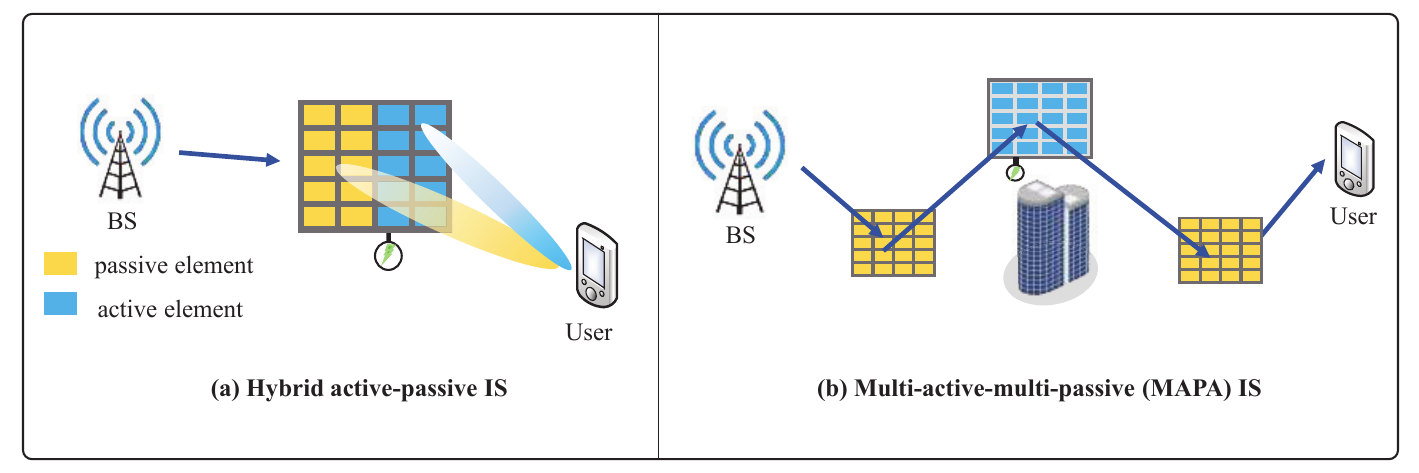}
   \caption{Illustration of integrated active and passive IS designs.}
   \label{FigCSY-2}
\end{figure*}

Besides the use of  active and passive ISs only,  recent research has also been dedicated to studying  efficient approaches to integrate both active or passive ISs in one system to harness their respective advantages. Specifically, one efficient approach is by deploying both active and passive elements in a single surface as shown in Fig. \ref{FigCSY-2} (a), which is called  \emph{hybrid} (active/passive)  IS \cite{r3_5,nguyen2022hybrid}. For this architecture, the authors in
\cite{r3_5} proposed to optimize the hybrid-IS elements allocation based on the statistical CSI under the constraints on the total deployment budget and maximum amplification power. It was shown that with optimized elements allocation, the hybrid-IS is able to achieve better communication performance than the conventional IS with active or passive elements only.
%, since it can flexibly balance between the  power amplification gain of active IS and the superior beamforming gain of passive IS. 
The same hybrid-IS architecture is further studied in \cite{r3_6_1} to maximize the system energy-efficiency.
On the other hand,  the authors in \cite{r3_6} considered a different hybrid-IS architecture, where each element is integrated with an amplifier and can be switched between the active and passive modes. For this  hybrid-IS architecture, it was demonstrated  that  the received SNR at the user can be further improved by optimizing the placement of active IS elements according to different channel gains at different elements' locations. 

In contrast, another design  of integrated active and passive ISs  deploys  both active and passive ISs in a distributed manner  to improve the communication performance of the wireless network  as illustrated in Fig. \ref{FigCSY-2} (b) \cite{r3_7,r3_8,r3_9}. For example, instead of deploying multiple passive ISs only in the wireless system for passive beam routing, a more effective approach is by adding several active ISs in the network to enable \emph{opportunistic} signal amplification over some  multi-reflection signal paths, thus more  effectively compensating   for the product-distance path-loss. For this new multi-active and multi-passive (MAMP) IS-aided wireless system, the authors in  \cite{r3_7}
proposed a new hierarchical beam routing design to solve the intricate MAMP-IS beam routing problem based on graph theory. 
Besides the beam routing design, the communication performance of MAMP-IS aided wireless system can be further improved by optimizing the placement of active/passive ISs. In \cite{r3_8}, the authors considered a wireless information and power transfer system aided by a single active-IS and multiple passive-ISs via multi-reflection path. It was shown that given fixed passive-IS locations, the optimal active-IS placement is generally different for wireless information transfer (WIT) versus wireless power transfer (WPT) due to the different roles of active-IS-induced amplification noise. 
Furthermore, the authors in \cite{r3_9} optimized the locations of all active/passive ISs in an MAMP-IS aided wireless system in a given region. An optimization problem was formulated and efficiently solved to minimize the total deployment cost under the SNR constraints.
%, by selecting an optimal subset of all candidate locations for deploying ISs and optimizing the number of passive/active reflecting elements deployed at each selected location.

%In Table \ref{airs_table}, we summarize the existing works on the active-IS and integrated  active/passive ISs, while
In future work, there are several key design issues worthy of investigation. For example, the existing works on active-IS have usually assumed ideal reflection coefficients with continuous and independent phase-shift and amplitude control, which may not be implementable in practice  and thus calls for efficient algorithms to handle their practical constraints.
In addition, for MAMP-IS aided systems, it is interesting to study how to optimize the density of active and passive ISs in a multi-cell network for maximizing the network throughput by using e.g.,  the  stochastic geometry theory \cite{lyu2021hybrid}. 
Furthermore, a research direction is the design of globally passive ISs based on the  optimization of surface waves, since no power amplifiers are required but more advanced optimization schemes are needed \cite{m7}.

\subsection{Omnidirectional IS}

Our discussions so far have mainly focused on reflection-based ISs, passive or active. However, reflection-based ISs have one fundamental limitation after they are deployed in practice: they can only \emph{reflect} incident waves from a restricted angular range, i.e., only users and the BS that are located in the same reflection half-space of the IS can benefit from the signal reflection by the IS. 
To overcome this restriction, a new concept of omnidirectional IS has been proposed~\cite{HB-IOS-2022}. 
The omnidirectional IS can enable a dual function of signal reflection and refraction simultaneously, i.e., some power of the incident signal is reflected to serve users located in the same reflection half-space of the IS, and the rest of power penetrates the IS to serve users located in the refraction half-space of the IS \cite{9820777,9961851}. 
As a result, a doubled 360$^\circ$ service coverage is ensured compared to reflection-based IS. This type of surfaces is also referred to as STAR-RIS as aforementioned \cite{YXJRYHL-STAR-2021}. 
In this subsection, we first introduce the working principles of the omnidirectional IS and its new characteristics. Then, two hardware implementations of the omnidirectional IS are presented. 
Finally, emerging applications of the omnidirectional IS are discussed.

\subsubsection{Working Principles}
\begin{figure}[!t]
	\centering
	\includegraphics[width=0.40\textwidth]{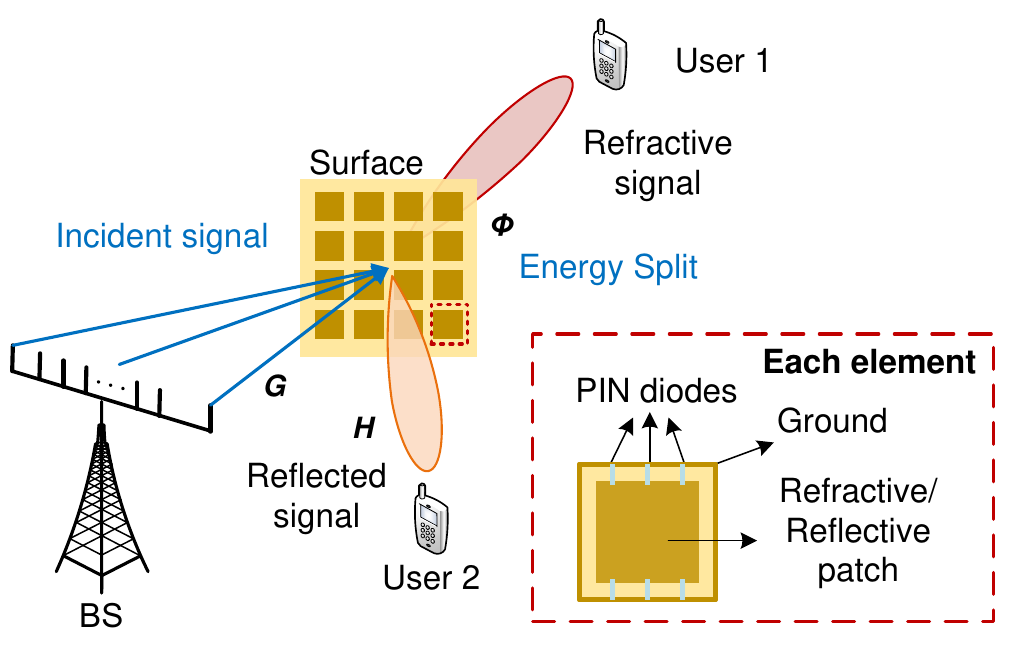}
	%\vspace{-3mm}
	\caption{An illustration of the omnidirectional IS-based hybrid beamforming for two users.}
	\vspace{-0.5cm}
	\label{angle}
\end{figure}

An omnidirectional IS is an engineered surface that comprises electrically-controllable scattering elements. Each element comprises multiple metallic patches and $N$ PIN diodes that are evenly distributed on a dielectric substrate, as shown in Fig. \ref{angle}. The metallic patches are connected to the ground via the PIN diodes that can be switched between their ON and OFF states according to predetermined bias voltages. The ON/OFF state of these PIN diodes determines the amplitude and phase responses applied by the omnidirectional IS to the incident signals. Each metallic patch can be configured in $2^{N}$ different states, which results in $P \le 2^N$ different amplitudes $\Gamma$ and phases $\theta$ (usually equally distributed in $[0, 2\pi]$), depending on the implementation of the omnidirectional IS. As illustrated in Fig. \ref{angle}, when a signal impinges, from either side of the surface, upon one element of the omnidirectional IS, for example, a fraction of the incident power is reflected and refracted towards the same side and the opposite side of the incident signal, respectively\cite{SHBYMZHL-Beamforming-2022}. Under the assumption of periodic boundary conditions, complex-valued reflection and refraction coefficients of a generic omnidirectional IS element can be defined. Mathematically, the received signals at each user can be expressed by
\vspace{-1.5mm}
\begin{equation}
	\bm{y} = \bm{H}\bm{\Phi}\bm{G}\bm{x},
        \vspace{-1.5mm}
\end{equation}
where $\bm{x}$ is the signal vector at the BS, $\bm{G}$ is the channel matrix between the BS and the omnidirectional IS, and $\bm{H}$ is the channel matrix between the omnidirectional IS and the user. $\bm{\Phi}$ is the phase shift matrix of the omnidirectional IS, which is different for reflection and refraction in general. $\bm{\Phi}$ is typically diagonal, and its diagonal elements correspond to the reflection/transmission coefficient of each omnidirectional IS element $\Gamma e^{-j \theta}$, respectively.

In general, the reflection and refraction/transmission coefficients may depend on the direction of incidence. The amplitude of the reflection and transmission coefficients of a generic omnidirectional IS element are determined by the size and shape of each element. It is worth mentioning that the proposed omnidirectional IS is flexibly designed to ensure that the amplitude of the reflected and refracted signals can be different. The power ratio between them is, in particular, a constant determined by the hardware implementation of the omnidirectional IS elements. When the states of the PIN diodes are changed, the reflection and transmission coefficients are changed simultaneously. For instance, the phase shifts for reflection and refraction satisfy a linear model as suggested in \cite{HB-IOS-2022}, i.e., a pair of reflection and refraction phase shifts satisfies $\theta_t(i) - \theta_r(i) = c(i)$, where the subscript~$i$ refers to the $i$-th omnidirectional IS element and $c(i)$ is a constant. This indicates the coupling of refraction and reflection, which needs to be considered in the omnidirectional IS beamformer design.

\subsubsection{New Characteristics}

Due to the coupling of refraction and reflection, the following key factors need to be revisited in order to achieve full-dimensional omnidirectional IS-aided communications.

\begin{itemize}
\item \textbf{Channel Characteristics:} As the omnidirectional IS splits the energy of the incident signals, one may raise a question: Does the uplink (UL) and downlink (DL) reciprocity hold for omnidirectional IS-aided wireless communications? 
According to the theoretical analysis and experimental results given in~\cite{SSHFLB-Reciprocity-2023}, the authors showed that the channels are reciprocal while the beams are not. This indicates that we may not need separate channel estimations for UL and DL, but we may need different phase configurations for UL and DL communications, which is different for reflection-based ISs. Channel estimation is, however, a critical issue for omnidirectional IS-aided wireless communication systems because of the coupling effect of reflection and refraction. An initial attempt to tackle this issue can be found in \cite{CCYXY-2022}, where the authors used the minimum mean square error (MMSE) metric to estimate the CSI, by jointly designing the pilot sequences of the users, the energy splitting ratio, and the training pattern matrices under a coupled phase shift model. 

\item \textbf{Phase Shift Optimization:} Due to the coupling between reflection and refraction, the phase optimization for an omnidirectional IS is different from that of reflection-based ISs, where only the users on one side of the surface are considered. Two cases were considered in the literature: 1) \emph{CSI is known.} The authors of \cite{SHBYMZHL-Beamforming-2022} proposed a hybrid beamforming framework for an omnidirectional IS-aided wireless system, where an omnidirectional IS is deployed at the cell edge to extend the coverage. Here, assuming that the CSI is perfectly known at the BS, the authors jointly optimized the digital beamformer at the BS and the analog beamformer enabled by the omnidirectional IS with discrete phase shifts to maximize the sum-rate. Their results showed that an omnidirectional IS enhances the data rate of users located on both sides of the surface, which asymptotically doubles the achievable sum-rate and the service coverage; 2) \emph{CSI is unknown.} In dynamic environments, it is costly to obtain real-time CSI. The work in \cite{YBHMLL-Codebook-2022} developed a codebook-based scheme to bypass the channel estimation. In this work, a phase shift configuration of the omnidirectional IS corresponds to a codebook, and the codebooks for the BS and the omnidirectional IS are pre-designed jointly. Then, a multi-layer beam training scheme was proposed to select a codeword with the maximum data rate. The work \cite{SHBZL-Location-2021} further analyzed the optimal codebook size to achieve a trade-off between the data rate and the training overhead for an omnidirectional IS-aided wireless system. The results showed that the optimal codebook size is influenced by the refraction-reflection ratio, which decreases as the ratio approaches to one. Moreover, as the reflection and transmission coefficients depend on the incident angle, the orientation and position of the omnidirectional IS also need to be considered in the phase shift optimization~\cite{SHBZL-Location-2021}.
	
\end{itemize}

\subsubsection{Hardware Implementation}
Two implementation methods can be used to realize simultaneous reflection and refraction of omnidirectional IS, elaborated as follows.

\begin{itemize}

\item \textbf{Power Domain:} A straightforward idea is to split the incident signal into two parts: one for reflection and the other for refraction, as shown in Fig. \ref{angle}. Such a function can be realized by a symmetric structure provided in~\cite{HSBYMMZHL-Implementation-2022}, for which the layout parameters can be optimized by modeling each element as a two-port network. With such a structure, the incident signals first induce currents on the patches, and then radiate reflective and refractive signals on both sides of the omnidirectional IS. By changing the equivalent impedance of each element through switching the ON/OFF states of embedded PIN diodes, the phases and amplitudes of the refracted and reflected signals are changed accordingly, thus enabling beamforming. Following this element design, the authors of \cite{SHBYMZHL-Circuit-2022} implemented the omnidirectional IS and built an omnidirectional IS-aided wireless communication prototype, as shown in Fig.~\ref{prototype}. Their experimental measurements verified the feasibility of full-dimensional communications provided by the omnidirectional IS. They also pointed out that the response of each omnidirectional IS element is a function of the angle of incidence, which should be considered in the beamformer optimization, as otherwise beam misalignment can be resulted. 

\begin{figure}[!t]
	\centering
	\includegraphics[width=0.45\textwidth]{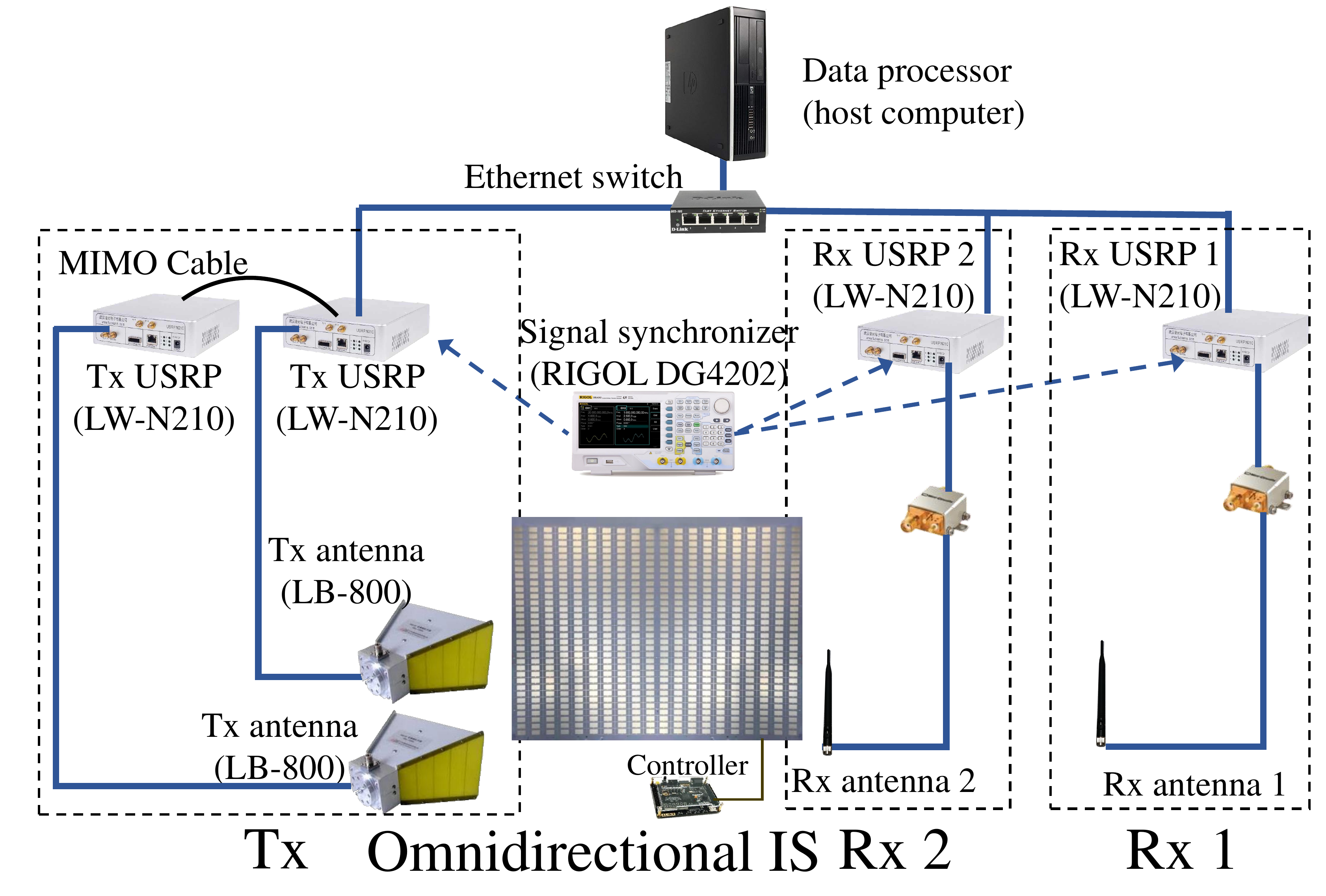}
	\vspace{-0.3cm}
	\caption{An omnidirectional IS-aided wireless communication prototype \cite{SHBYMZHL-Circuit-2022}.}
	\vspace{-0.5cm}
	\label{prototype}
\end{figure} 

\item \textbf{Polarization Domain:} Polarization is the property of an EM wave that determines the direction of the electric field oscillations. There could be different polarization states depending on the direction of these oscillations in the plane perpendicular to the propagation direction, such as linear (vertical and horizontal) and circular polarizations. Different polarizations can be utilized to control reflective and refractive waves simultaneously with a polarization-dependent characteristics provided by the omnidirectional IS. For example, an omnidirectional IS element can only reflect vertical-polarized waves while horizontal-polarized waves can only be refracted. An initial implementation for the omnidirectional IS in the polarization domain was reported in~\cite{ACFFPDJ-Polarization-2023}, where two orthogonal polarized components are integrated in one omnidirectional IS element with a decoupling structure to guarantee their orthogonality. However, with such a design, the effective aperture for either reflection or refraction is reduced to only half of that of the surface. Therefore, how to fully utilize the whole surface with different polarizations is still a challenging unsolved problem.

\end{itemize}

{It is worth noting that some existing designs utilize coupling among elements to realize simultaneous reflection and refraction \cite{HSB-Architectures-2023}. In other words, the incident signals on an element will also be reflected or refracted from the neighboring elements. As a result, the phase shift matrix $\bm{\Phi}$ is not diagonal any more for either reflection or refraction, and this type of surface is also referred to as BD-RIS as mentioned in Section I-B. Under the BD-RIS framework, the reflective and refractive patches for an element are grouped into a cell, and the cell is modeled as a two-port impedance network, which is similar to the equivalent circuit discussed in \cite{SHBYMZHL-Circuit-2022}. The BD-RIS considers a more general concept that the IS can be divided into several sub-surfaces and each sub-surface is modeled as an $M$-port impedance network, where $M$ is the number of reflective/refractive patches within the sub-surface. This implies that the incident signal on a sub-surface can be reflected or refracted by any patch within each sub-surface. This can be achieved by forming coupling currents among elements within the sub-surface. Such a coupling can provide another dimension to optimize, i.e., the equivalent amplitude of the coefficient for each element can be adjusted, leading to additional performance improvement. However, the coupling poses an increasing complexity on the hardware design of the omnidirectional IS. Therefore, it is necessary to strike a trade-off between the performance and complexity in such a framework. }

\subsubsection{New Applications}
The new characteristic of simultaneous reflection and refraction  has unlocked various new applications \cite{HB-IOS-2022}. For example, the omnidirectional IS has been introduced for self-interference cancellation in full-duplex communication systems. In particular, the omnidirectional IS can be deployed between the Tx and the Rx, and we can configure the phase shifts of the omnidirectional IS to enhance the received signal strength by reflection and to null the self-interference to the Rx on the refraction side. 

The integration of the omnidirectional IS with existing communication systems also faces its own challenges as the coupling of refraction and reflection further complicates the joint optimization, which requires further investigation.

\subsection{Holographic IS}

\begin{figure}[t]
\centering
\includegraphics[width=3.5in]{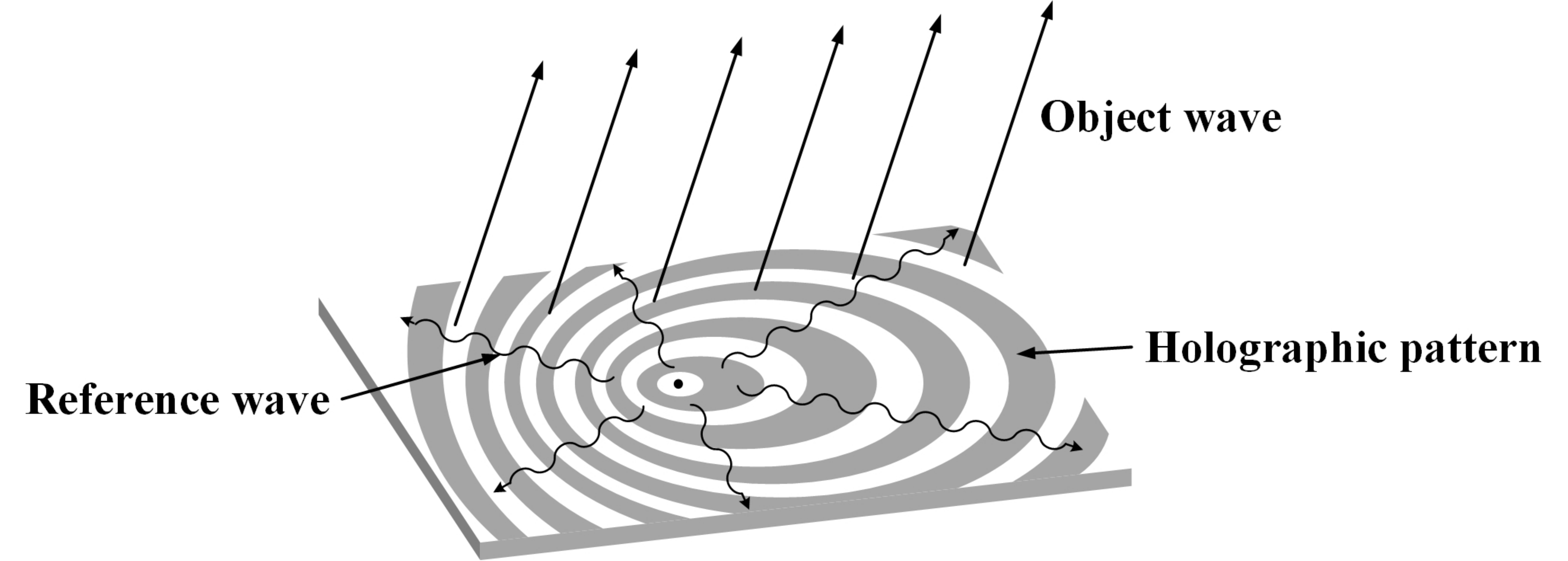}
%\vspace{-0.5cm}
\caption{Illustration of a holographic IS.}
\vspace{-0.5cm}
\label{URHS}
\end{figure}

Wireless communications enabled by holographic ISs strive to overcome the limitations of massive MIMO systems \cite{huang2020-holographic}. Specifically, holographic ISs refer to spatially-continuous EM  apertures incorporating densely-packed large number of radiation elements~\cite{LCC-2023}. By leveraging the holographic principle, as shown in Fig.~\ref{URHS}, the radiation characteristics of the EM wave (which is also called reference wave) propagating on the surface can be changed by the holographic pattern to generate the desired object wave~\cite{RD-2021}. Hence, holographic ISs are capable of achieving amplitude and phase modulation to enable signal processing within the EM domain, while maintaining a low hardware cost and power consumption. In this subsection, we first introduce the theoretical aspects and transmission schemes of holographic IS. Then, we present the hardware implementation of holographic IS. Finally, we discuss its potential applications.

\subsubsection{Theoretical Aspects}

The theoretical foundations of holographic IS such as channel model, degrees of freedom~(DoF), and system capacity require new and efficient approaches due to the dense packing of nearly infinite radiation elements and spatially-continuous EM apertures. In particular, one of the most prominent design shifts is from the traditional digital domain to the EM domain~\cite{JMMMH-2023}. This new domain provides a promising space to unveil the fundamental limitations of holographic IS-aided communications. In the following, we present the recent advances in theoretical aspects on holographic IS in terms of channel modeling, DoF, and system capacity.

\begin{itemize}
 \item \emph{Channel modeling:} Channel modeling of holographic IS systems differs from that of conventional MIMO systems due to the compact distribution of radiation elements in a holographic IS, where strong spatial correlations and mutual coupling effect need to be considered. To tackle this issue, a spatial correlation based channel model was proposed~\cite{OEL-2022}. Mathematically, the channel model can be denoted by $\bm{h}\sim \mathcal{CN}(0,\bm{R})$, where $\bm{R}=\mathbb{E}(\bm{h}\bm{h}^H)$ is given by
     \begin{equation}
     \begin{split}
     \bm{R}=\beta\int\int & f(\phi+\delta_{\phi},\theta+\delta_{\theta})\bm{a}(\phi+\delta_{\phi},\theta+\delta_{\theta})\\
     & \bm{a}^H(\phi+\delta_{\phi},\theta+\delta_{\theta})
     d\delta_{\phi}d\delta_{\theta},
     \end{split}
	\end{equation}
     $\beta$ is the average channel gain, $\phi$ and $\theta$ are the azimuth and elevation angles, respectively, $\delta_{\phi}$ and $\delta_{\theta}$ are the angular deviations corresponding to $\phi$ and $\theta$, $f(\phi,\theta)$ is the normalized spatial scattering function, and $\bm{a}(\phi,\theta)$ is the array steering vector. The authors of~\cite{SH-2021} further developed a joint spatial-temporal correlation model characterized by a four-dimensional sinc function by considering both spatial and temporal correlations. 
     
 \item \emph{DoF:} The DoF reveals the number of independent data streams that can be transmitted simultaneously. The DoF of holographic IS has been fully studied to explore the optimal communication performance. Specifically, the study in~\cite{D-2020} explored the DoF of a point-to-point holographic IS-aided system. An approximate closed-form expression of the DoF achieving performance limits was derived, which indicates that the DoF of LoS channel is only determined by geometric factors normalized to the wavelength, and can be higher than one. More recently, the authors of \cite{JMMMH-2023} have comprehensively analyzed the DoF and the optimal waveforms of holographic MIMO under general network deployments.
 
 \item \emph{System capacity:} As one of the most critical metrics for the system performance, the capacity of holographic IS-aided communication has also been investigated. The authors in~\cite{HDMA} presented pioneer studies on the system capacity of holographic IS considering LoS channels. It is proved that the normalized capacity will converge to a constant value as the terminal density grows and the wavelength approaches zero. Such an asymptotic limit is expressed as $\frac{P}{2N_0}$, where $P$ denotes the transmit power per volume unit and $N_0$ is the noise spatial power spectral density. In addition, the capacity with non-LoS (NLoS) channels was studied in~\cite{LCG-2022}. It is revealed that as the element spacing decreases, strong mutual coupling effects among elements will deteriorate the system capacity for a given number of radiation elements.

\end{itemize}

\subsubsection{Transmission Scheme}
The differences in transmission schemes between holographic ISs and conventional MIMO mainly involve in two aspects. On one hand, due to the large aperture of holographic ISs, the communication distances fall within the near-field region. Hence, transmission schemes applicable to near-field communication need to be developed for performance enhancement. On the other hand, in addition to conventional phase-controlled beamforming, holographic ISs can also apply amplitude modulation for holographic beamforming. Therefore, novel transmission schemes tailored for amplitude-controlled beamforming are also needed. In the following, we present the recent advances in transmission schemes in terms of near-field holographic IS and amplitude-controlled holographic IS aided communications.

\begin{itemize}
\item \emph{Near-field transmission scheme:} In near-field communications, the phases and distances observed by the user with respect to the elements are different, and thus the conventional uniform plane wave approximation in the far-field region cannot be applied anymore. Therefore, transmission scheme designs need to consider this characteristic for the system design. A mathematical model was presented to characterize the near-field channels and transmission patterns in~\cite{HNF-2022}. Also, the authors of~\cite{MLR-2021} developed effective beam-focusing transmission scheme by exploiting the distance information and phase variants to achieve favorable performance gains in the near-field region. Moreover, considering the randomness of user distribution, near-far field beamforming for holographic IS was proposed in for sum rate maximization.

\item \emph{Amplitude-controlled transmission scheme:} In the amplitude-controlled transmission, the amplitude of the EM wave at each radiation element is controlled by the holographic beamformer to generate desired directional beams~\cite{RD-2021}. Hence, the sum rate maximization problem can be modeled as a complex-domain optimization problem subject to unconventional real-domain amplitude constraints. The authors of~\cite{RBS-2021} introduced a novel hybrid beamforming scheme for a general amplitude-controlled holographic IS system. Specifically, the signal processing at baseband is conducted at the BS. The holographic IS utilizes an amplitude-controlled beamforming technique to generate desired directional beams and transmit the processed signals to each user.

\end{itemize}

\subsubsection{Hardware Implementation}
The primary challenge in the hardware implementation of holographic ISs is to design tunable radiation elements which are capable of optimizing the current distribution on the surface. Specifically, the tunability of radiation elements enables the construction of any holographic pattern on the surface, such that dynamic beamforming can be achieved. In the following we will list the state-of-the-art tuning mechanisms for the realization of holographic ISs, which rely on PIN diodes, liquid crystal, and photosensitive devices, respectively. {A summary of these hardware implementations is listed in Table \ref{RHS-compt}.}

\begin{figure*}[t]
    \centering
	\begin{minipage}[t]{0.5\linewidth}
		\centering
		\includegraphics[width=6cm]{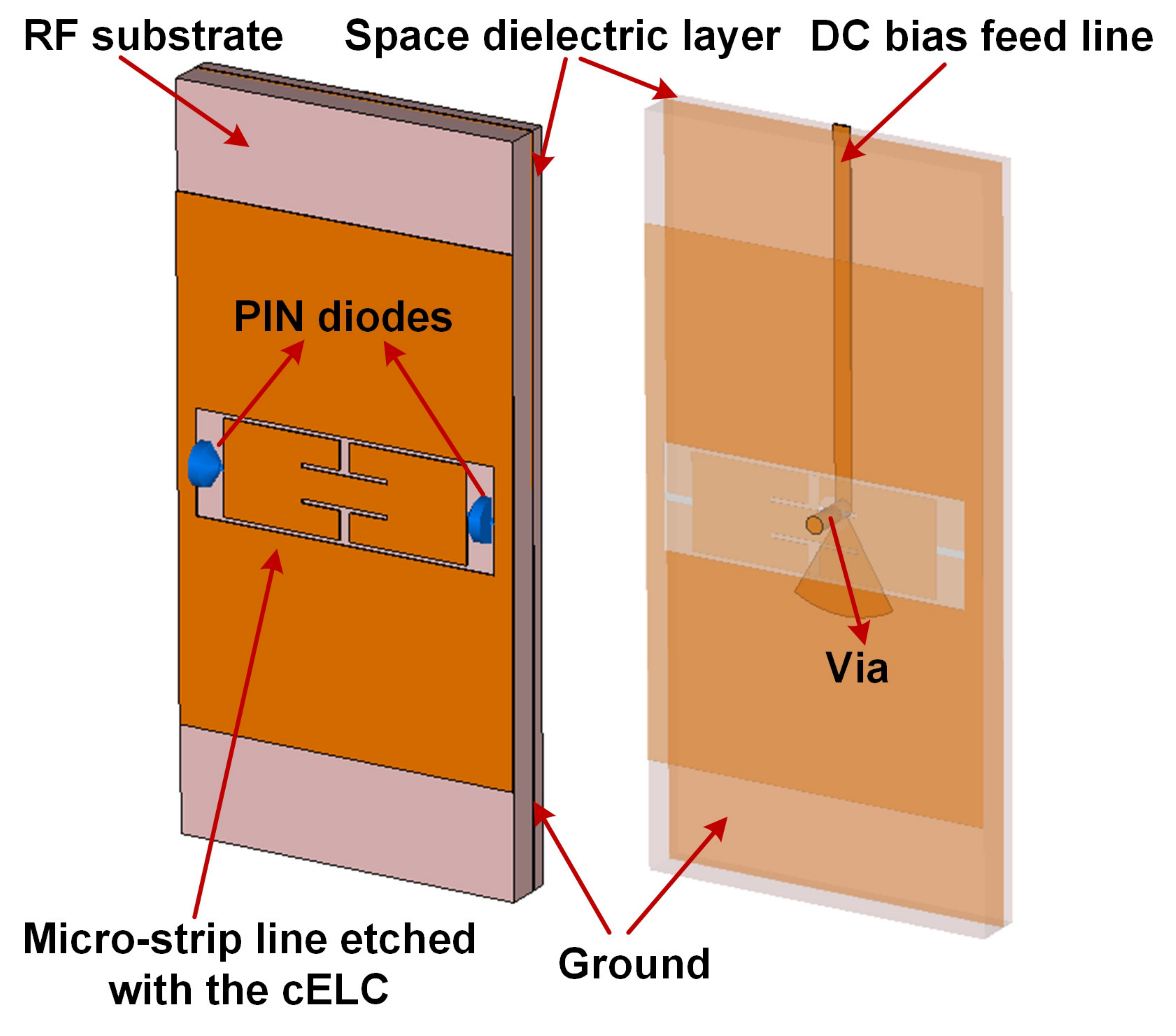}
		\label{2a}
	\end{minipage}%
	\begin{minipage}[t]{0.5\linewidth}
		\centering
		\includegraphics[width=6cm]{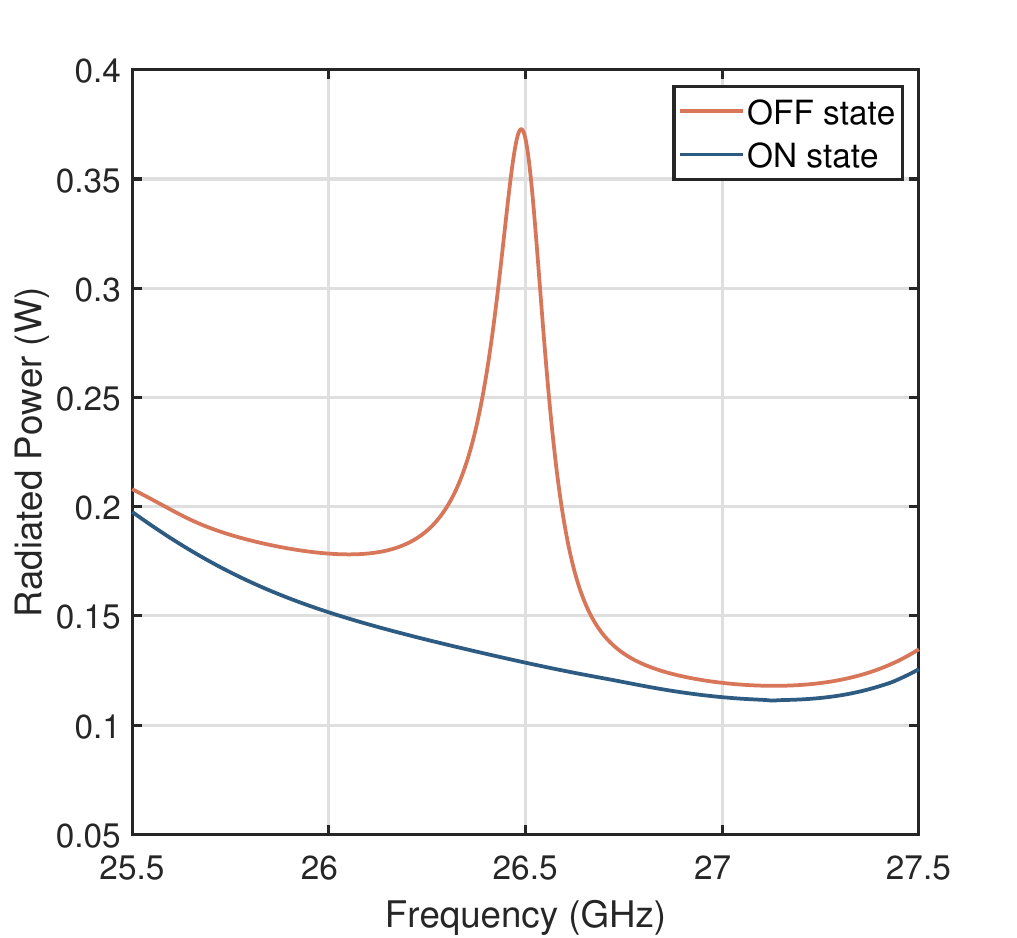}
		\label{2b}
	\end{minipage}
    \caption{Hardware design of a radiation element: (a) cELC-based structure; (b) Radiation power with different states of PIN diodes.}
    \label{comp}
\end{figure*}

\begin{table*}[!t]
	\centering 
	\renewcommand\arraystretch{1.6}
        {
	\caption{Comparison Between Different Hardware Implementations} \label{RHS-compt}
	\begin{tabular}{|m{2.6cm}<{\centering}|m{2.6cm}<{\centering}|m{3.5cm}<{\centering}|m{3cm}<{\centering}|m{3.5cm}<{\centering}|}
		\hline
		Tuning mechanism & Working frequency & Advantages & Limitation & Typical Applications \\
		\hline
		PIN diodes & (Sub-) mmWave & Low cost, fast response time and easy implementation & Nonlinearities in high-frequency bands  & Mobile communication scenarios\\
		\hline
		Liquid crystal & (Sub-) mmWave & Low power consumption, linear properties in high-frequency bands & Slow response time &(Quasi-) Static communication scenarios\\
		\hline
		Photosensitive devices & GHz-optical & Ultra-fast response time, protecting signals from EM interference & Difficult to maintain high phase stability & Long-distance deployments with optical fiber technology \\
		\hline
	\end{tabular}
        }
\end{table*}

{
\begin{itemize}
	\item \emph{PIN diodes:} A radiation element capable of controlling the radiation amplitude of the EM wave can be achieved through the integration of a complementary electric-LC (cELC) resonator, which incorporates PIN diodes for modulation purposes~\cite{R-2023}. The cELC resonator, as depicted in Fig.~\ref{comp}, is a micro-strip line etched with a circle of annular slot to form a metal patch combined with a closed loop. The operational states of PIN diodes are actively controlled by the application of distinct bias voltages. Such control holds significant influence over the mutual inductance properties of the cELC resonator, consequently dictating the radiation power of the EM wave.  Owing to their cost-effectiveness and easy implementation, PIN diodes find extensive applications predominantly in sub-mmWave and mmWave frequency ranges.
	
	\item \emph{Liquid crystal:} Although PIN diodes are capable of achieving high beampattern gains, the nonlinearities of PIN diodes in high-frequency bands deteriorate the performance of holographic ISs. Thanks to the linear properties of liquid crystal in high-frequency bands, the above limitation of PIN diodes can be overcome. The liquid crystal-based holographic ISs were proposed in~\cite{HS-2022}, where the liquid crystal is filled in cavity boxes between slots and radiating patches. The permittivity of liquid crystal is changed with the applied bias voltages, such that the EM wave can be regulated to achieve beamforming. Liquid crystal-controlled holographic ISs offer numerous benefits, such as their versatility in integration into diverse structures due to their fluid nature and efficiency in power usage, attributing to the predominantly capacitive functions that involve minimal currents. Nonetheless, a significant limitation is the relatively slow response time, which constrains their utility to applications that do not demand rapid tunability.
	
	\item \emph{Photosensitive devices:} In addition to the above electrical control methods, the tunability of radiation elements can also be realized by optical control. Specifically, different EM responses of each radiation element are controlled by properly illuminating the corresponding photosensitive semiconductors. Such an optical control method is capable of achieving a rapid tuning speed, while also safeguarding signals against EM interference. Moreover, when integrated with optical fiber, it becomes feasible to deploy holographic ISs and signal processing units at separate locations for long-distance deployments. However, a critical challenge associated with these optical tuning systems is the necessity for high phase stability, that is, the preservation of a constant phase difference among parallel optical signals~\cite{TG-2023}. Since optical fibers are highly susceptible to environmental influences, the above requirement is challenging to be met in practice.
\end{itemize}
}

\subsubsection{Potential Applications}
As an ultra-thin and lightweight antenna, the holographic IS has shown its potential in various wireless applications, including, e.g.,
\begin{itemize}
\item \emph{LEO satellite communications:} LEO satellite communications are one promising solution to deliver high-capacity backhaul or data relay services for terrestrial networks. Nonetheless, the high mobility of LEO satellites and the severe path loss pose challenges on antenna technologies in terms of precise beam steering and high antenna gain~\cite{RBH-2022}. Traditional antennas integrated with user terminals such as dish antennas and phased arrays either require heavy mechanics or costly phase shifters, rendering their adoption in practical systems economically unviable. As a new paradigm to achieve beamforming with low hardware cost and low power consumption, holographic ISs can be adopted in integrated terrestrial-satellite networks to overcome the above limitations. In particular, Kymeta together with Microsoft has developed a commercial holographic IS for satellite communications in the event of natural disasters.

\item \emph{Wireless Simultaneous Localization and Mapping (SLAM):} Wireless SLAM, as a technique for user positioning and mapping in uncharted environments, holds promise for enhancing location-based services. This method utilizes antennas to estimate both the time of arrival and the angle of arrival by analyzing the amplitudes of received multi-path components~\cite{HZY-2023}. Consequently, the precision of the SLAM system is determined by the directive gain of the antennas. Given that holographic ISs consist of numerous elements, offering enhanced beam-steering capabilities and high directive gain, they present a viable option to be applied in wireless SLAM systems. {It also shows great potential in radar detection \cite{XHHB-2022}.}

{
\item \emph{ISAC:} In 6G, it is envisaged that the functions of sensing and communication are fully integrated in the same system by sharing resources, waveforms, hardwares, etc. In ISAC systems, wideband waveforms are required to achieve high data rate for the communication function and high-resolution detection or estimation for the sensing function. However, wideband transmission with traditional phase controlled schemes typically suffers from beam squint. Holographic ISs, with the linear additivity of holographic patterns, can be used to effectively alleviate the beam squint issue \cite{zhang-isac-2022}.
}
\end{itemize}

\vspace{-2mm}

\section{Conclusions}

In this paper, we provided a comprehensive overview of the state-of-the-art results on IS-aided wireless systems and networks by focusing on the main design issues of commonly adopted reflection-based ISs such as reflection optimization, deployment, modulation, sensing and ISAC, as well as new and emerging IS architectures and their new design issues. Particularly, we pointed out several new trends in the research on ISs including the shifts from a single surface to multiple and cooperative-routing surfaces, from passive surfaces to active and integrated active/passive surfaces, from reflective surfaces to refractive and omni-directional surfaces, from terrestrial surfaces to integrated air-ground surfaces, from static surfaces to mobile surfaces, etc. Since there are still many open issues in the design and implementation of ISs-empowered wireless networks, we hope that this paper will serve as a timely and useful resource for future research on ISs to unleash their full potential for 6G and beyond.

%\section{references (maximum 5 pages)}

\bibliographystyle{IEEEtran}     
\bibliography{ref/IEEEabrv,ref/Overall}%SecII-B
%\bibliography{ref/IEEEabrv}
%\bibliography{ref/bib_2023}%SecII-C

\end{document}